\documentclass[apj]{emulateapj}

\usepackage{amsmath}
\usepackage{txfonts}
\usepackage{graphicx}

\def\mst{\,\,(*)}
\def\pst{\phantom{\mst}}

\def\***#1{\textsf{***#1***}}

\def\asca{\emph{ASCA}}
\def\sax{\emph{Beppo-SAX}}

\begin{document}

\shorttitle{CLUSTER TEMPERATURE PROFILES}
\shortauthors{VIKHLININ ET AL.}
\slugcomment{Submitted to ApJ 12/13/04}

\title{\emph{Chandra} temperature profiles for a sample of
  nearby relaxed galaxy clusters}

\author{A.~Vikhlinin\altaffilmark{1}\altaffilmark{,2},
M.~Markevitch\altaffilmark{1}\altaffilmark{,2},
S.~S.~Murray\altaffilmark{1}, C.~Jones\altaffilmark{1},
W.~Forman\altaffilmark{1},
L.~Van~Speybroeck\altaffilmark{1}\altaffilmark{,3}}

\altaffiltext{1}{Harvard-Smithsonian Center for Astrophysics, 60
  Garden St., Cambridge, MA 02138; avikhlinin@cfa.harvard.edu}
\altaffiltext{2}{Space Research Institute, Profsoyuznaya 84/32, Moscow,
  Russia.}
\altaffiltext{3}{This paper heavily uses the GTO program of our late
  colleague.}

\begin{abstract}

  We present \emph{Chandra} gas temperature profiles at large radii for
  a sample of 13 nearby, relaxed galaxy clusters and groups, which
  includes A133, A262, A383, A478, A907, A1413, A1795, A1991, A2029,
  A2390, MKW4, RXJ1159+5531, and USGC S152. The sample covers a range of
  average temperatures from 1 to 10 keV. The clusters are selected from
  the archive or observed by us to have sufficient exposures and
  off-center area coverage to enable accurate background subtraction and
  reach the temperature accuracy of better than 20--30\% at least to
  $r=0.4-0.5 r_{180}$, and for the three best clusters, to $0.6-0.7
  r_{180}$.  For all clusters, we find cool gas in the cores, outside of
  which the temperature reaches a peak at $r\sim 0.15\,r_{180}$
  and then declines to $\sim 0.5$ of its peak value at $r\simeq 0.5
  r_{180}$.  When the profiles are scaled by the cluster average
  temperature (excluding cool cores) and the estimated virial radius,
  they show large scatter at small radii, but remarkable similarity at
  $r>0.1-0.2 r_{180}$ for all but one cluster (A2390). Our
  results are in good agreement with previous measurements from
  \emph{ASCA} by Markevitch et al.\ and from \emph{Beppo-SAX} by
  DeGrandi \& Molendi. Four clusters have recent \emph{XMM-Newton}
  temperature profiles, two of which agree with our results, and we
  discuss reasons for disagreement for the other two.  The overall shape
  of temperature profiles at large radii is reproduced in recent
  cosmological simulations.

\end{abstract}


\section{Introduction}

Radial temperature profiles of the hot intracluster medium (ICM) in
galaxy clusters and groups is one of the prime tools to study the
gravitational processes responsible for large-scale structure formation
and non-gravitational energy input into the ICM. The temperature profile
is also an important cosmological measurement because in dynamically
relaxed systems, it is the basic ingredient in estimating the total
cluster mass distribution assuming hydrostatic equilibrium of the
ICM. \mbox{X-ray} mass measurements at radius $r$ are only as accurate
as $T$ and $dT/dr$ at that $r$ (e.g., Sarazin 1988). 

Temperature measurements at large cluster radii are technically challenging. 
Cluster brightness at the virial radius is only 10\% of the Cosmic X-ray
Background (CXB) in the soft X-ray band, and a smaller fraction of the total
(detector + CXB) background. The background spectrum is usually much harder
than that of the cluster, leading to even lower surface brightness contrast
at higher energies. The practical implication is that accurate spectral
analysis near the virial radius is not possible, even with very long
exposures, unless the background can be subtracted with better than 1\%
accuracy. Such accuracy is unachievable with past or present X-ray
telescopes. The surface brightness contrast near 0.5 of the virial radius is
higher, and the required accuracy of the background subtraction is $\sim
3\%$. This is easily achievable with \emph{Chandra} (Markevitch et al.\ 
2003) and marginally feasible with \emph{XMM-Newton} (Nevalainen,
Markevitch, \& Lumb~2005). 

%/home/maxim/TEX/jukka_bg_subm.ps.gz

%  The table
% below gives ratios of the quiescent detector background
% surface brightness to the cluster brightness (the same 6 kev
% cluster observed with the respective instrument, arbitrary
% units), in the 2-7 kev band where the background matters
% most:
% 
% ACIS-S3  ACIS-FI     EPIC-MOS1  EPIC-pn
%   1.5     0.92         0.47      0.44

Cluster observations with telescopes operating below the Earth's radiation
belts, such as \emph{ASCA} and \emph{Beppo-SAX}, are less affected by the
background. However, poor point spread functions (PSF) of \emph{ASCA} and
\emph{Beppo-SAX}, 1\arcmin--2\arcmin, was a major problem for temperature
profile measurements. This problem is non-existent for \emph{Chandra} and
not a particular concern for \emph{XMM-Newton} except in the cores of
clusters with peaked surface brightness profiles (e.g., Markevitch 2002). 

\begin{deluxetable*}{lcclccccl}
\tablecaption{Cluster sample and \emph{Chandra} observations\label{tab:chandra:obs}}
\tablehead{
& & & & \multicolumn{3}{c}{\emph{Chandra} observations}\\[2pt]
\cline{5-7}\\[-5pt]
\colhead{Cluster\hspace*{20mm}} &
\colhead{$\langle T\rangle$\tablenotemark{a}} &
\colhead{$z$} &
\colhead{$N_H$\tablenotemark{b}\pst} &
\colhead{Aim point} &
\colhead{Exposure, ksec} &
\colhead{ACIS Mode} &
\colhead{$r_{\text{max}}/r_{180}$} &
Comments
}
\startdata
A133\dotfill      & 4.2 & 0.057 & $1.53\times10^{20}\pst$ &  S+I   & 40+90       & F+VF & 0.67 & $14'$ offset in ACIS-I\\
A262\dotfill      & 2.1 & 0.016 & $8.10\times10^{20}\mst$ &  S     & 30          & VF   & 0.30 & \nodata\\
A383\dotfill      & 4.9 & 0.188 & $3.92\times10^{20}\pst$ &  I+I+S & 20+10+20    & VF   & 0.42 & \nodata\\
A478\dotfill      & 7.9 & 0.088 & variable\mst            &  S     & 43          & F    & 0.60 & \nodata\\
A907\dotfill      & 5.9 & 0.160 & $3.87\times10^{20}\mst$ &  I     & 49+35+11    & VF   & 0.58 & \nodata\\
A1413\dotfill     & 7.3 & 0.143 & $2.19\times10^{20}\pst$ &  I     & 10+113      & VF   & 0.70 & \nodata\\
A1795\dotfill     & 6.1 & 0.062 & $1.19\times10^{20}\pst$ &  S+I   & $7\times15$ & VF   & 0.53 & Multiple offsets \\
A1991\dotfill     & 2.6 & 0.059 & $2.45\times10^{20}\pst$ &  S     & 36          & VF   & 0.48 & \nodata\\
A2029\dotfill     & 8.5 & 0.078 & $3.04\times10^{21}\pst$ &  S+I   & 87+9        & F+VF & 0.60 & \nodata\\
A2390\dotfill     & 8.9 & 0.230 & $1.07\times10^{21}\mst$ &  S     & 95          & VF   & 0.90 & \nodata\\
MKW4\dotfill      & 1.6 & 0.020 & variable\mst            &  S     & 30          & VF   & 0.39 & \nodata\\
RXJ\,1159+5531\dotfill & 1.9 & 0.081 & $1.20\times10^{20}\pst$ &  S     & 70          & VF   & 0.38 & \nodata\\
USGC S152\dotfill & 0.7 & 0.015 & $1.55\times10^{21}\pst$ &  S     & 30          & VF   & 0.45 & \nodata
\enddata
\tablenotetext{a}{--- Emission-weighted temperature (keV), excluding central
  70~kpc (\S\ref{sec:self:sim}).} 
\tablenotetext{b}{--- Galactic absorption column density (cm$^{-2}$)
  adopted in this paper (\S\ref{sec:results:individ}). Stars mark those
  clusters with different radio and X-ray values of $N_H$.} 
\end{deluxetable*}

\begin{figure*}
\vspace{0.3\baselineskip}
\centerline{%
\framebox{\includegraphics[width=0.3\linewidth]{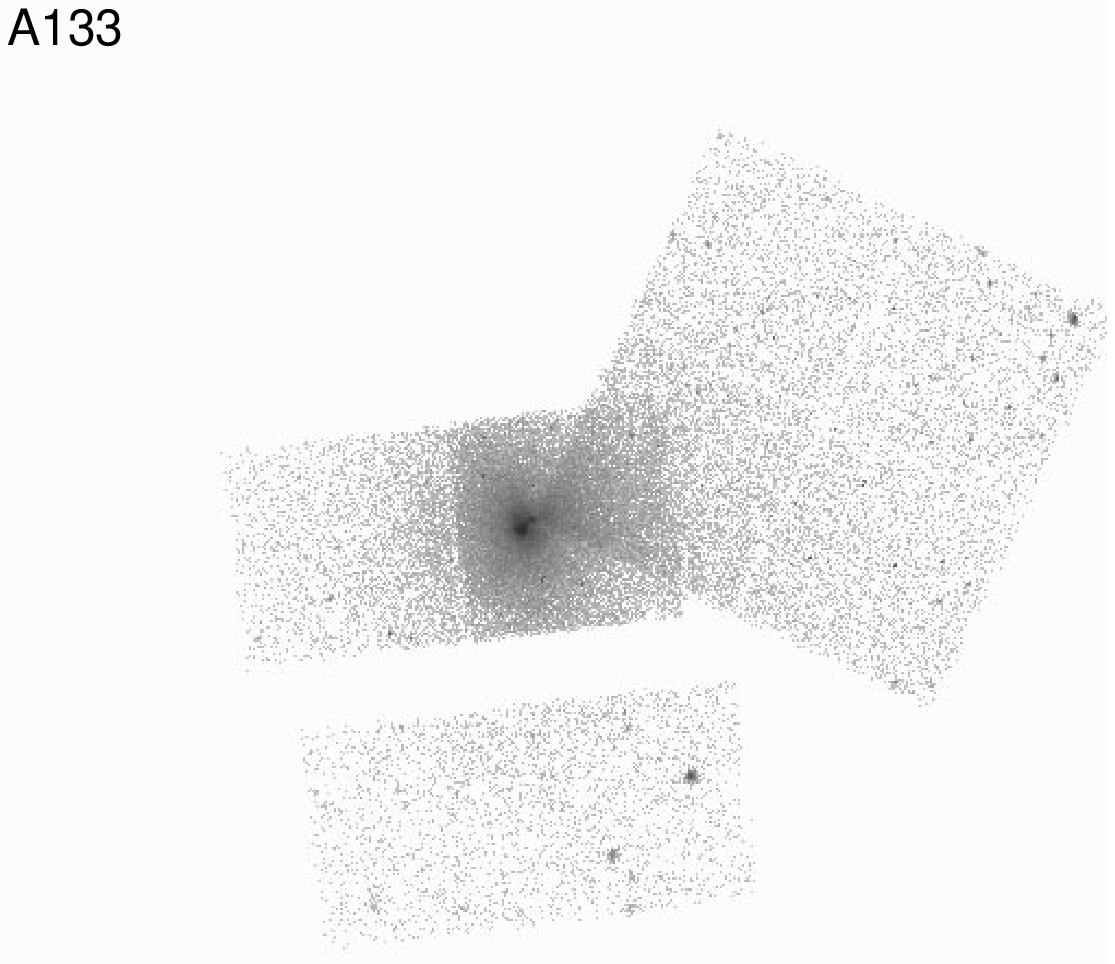}}~~~%
\framebox{\includegraphics[width=0.3\linewidth]{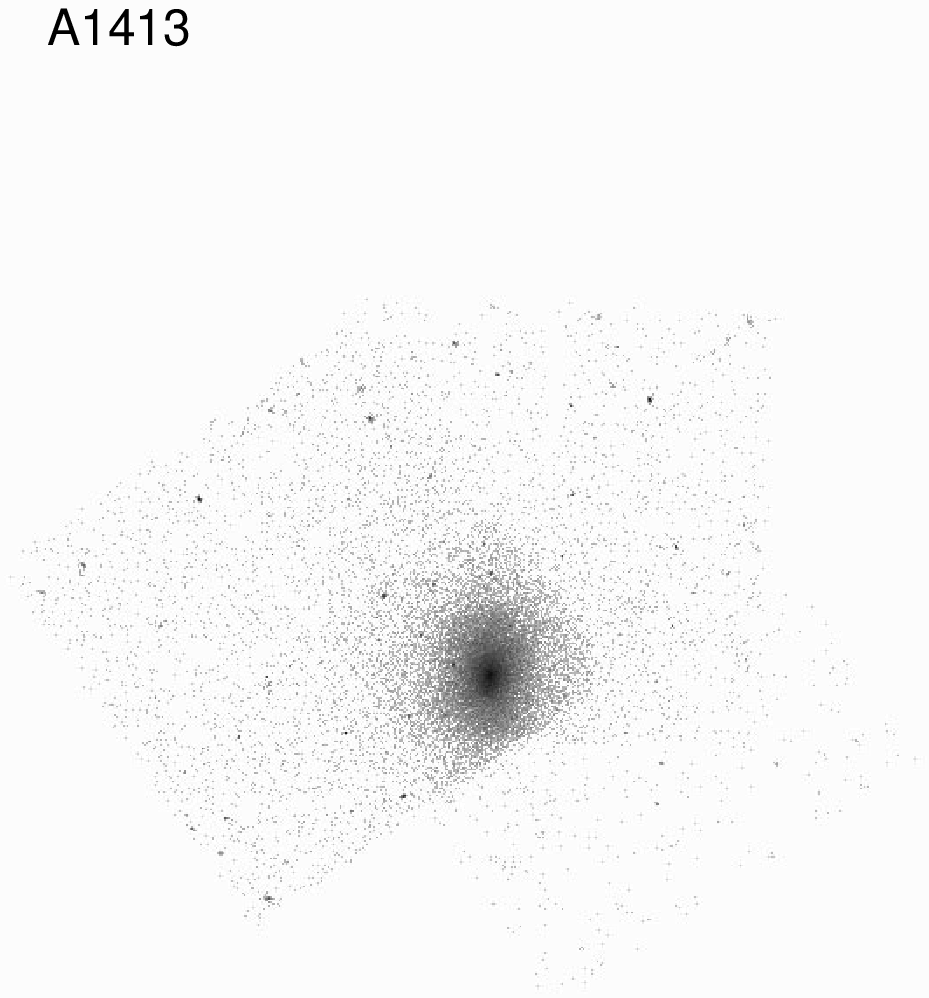}}~~~%
\framebox{\includegraphics[width=0.3\linewidth]{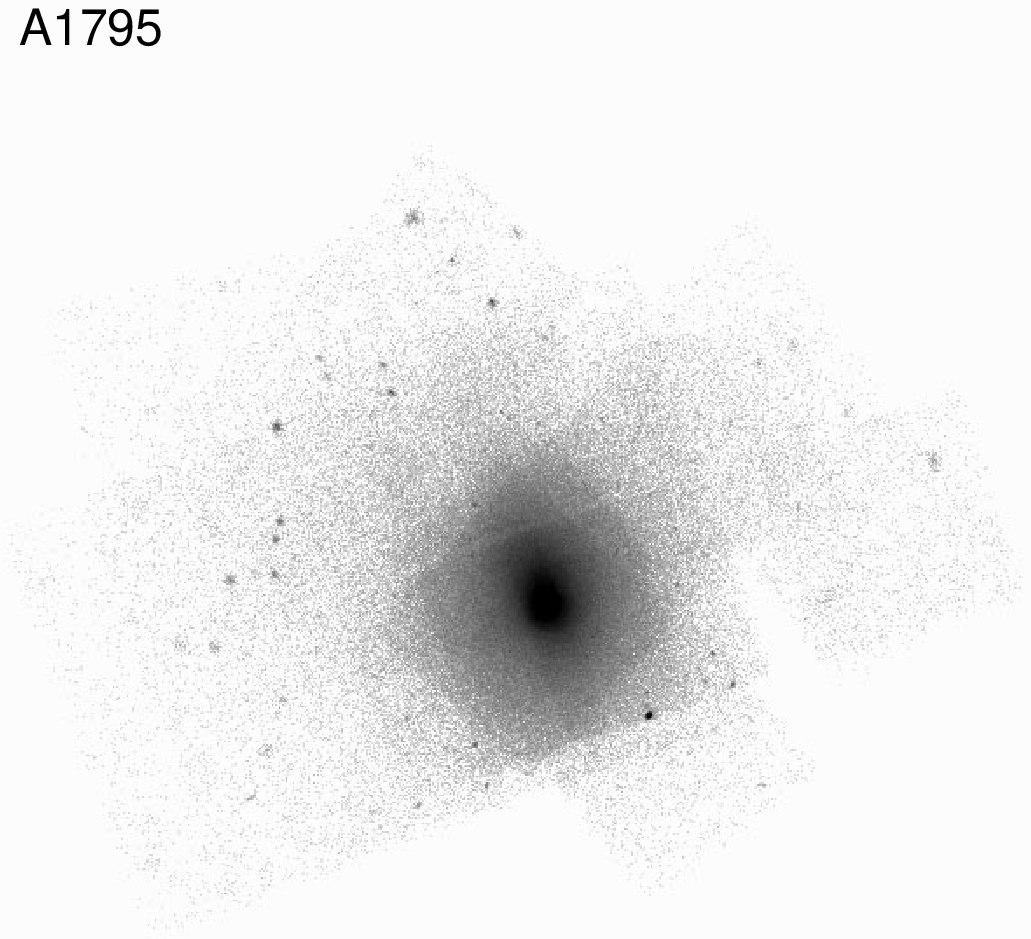}}
}
\medskip
\centerline{%
\framebox{\includegraphics[width=0.3\linewidth]{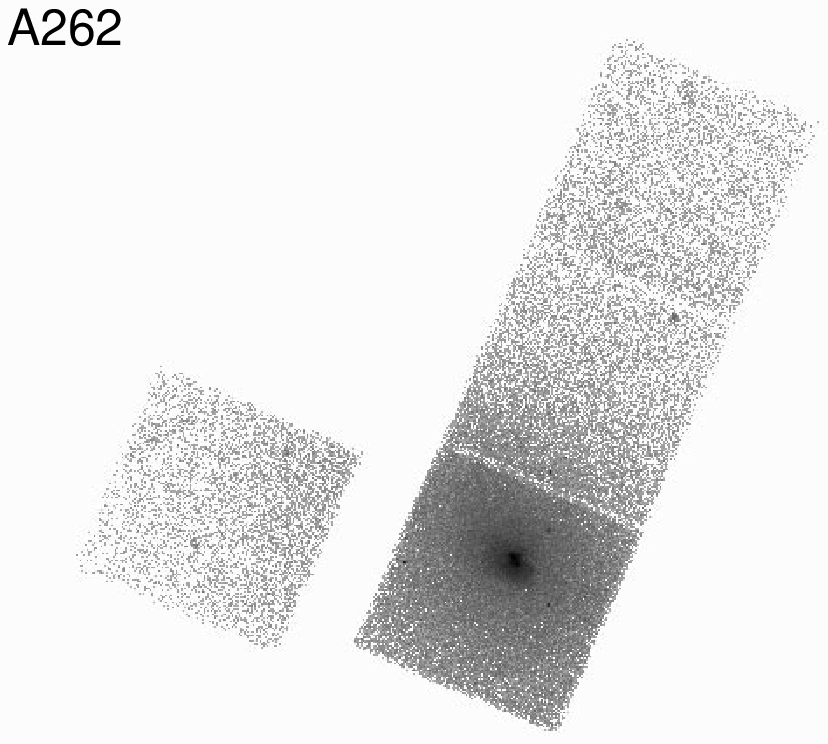}}~~~%
\framebox{\includegraphics[width=0.3\linewidth]{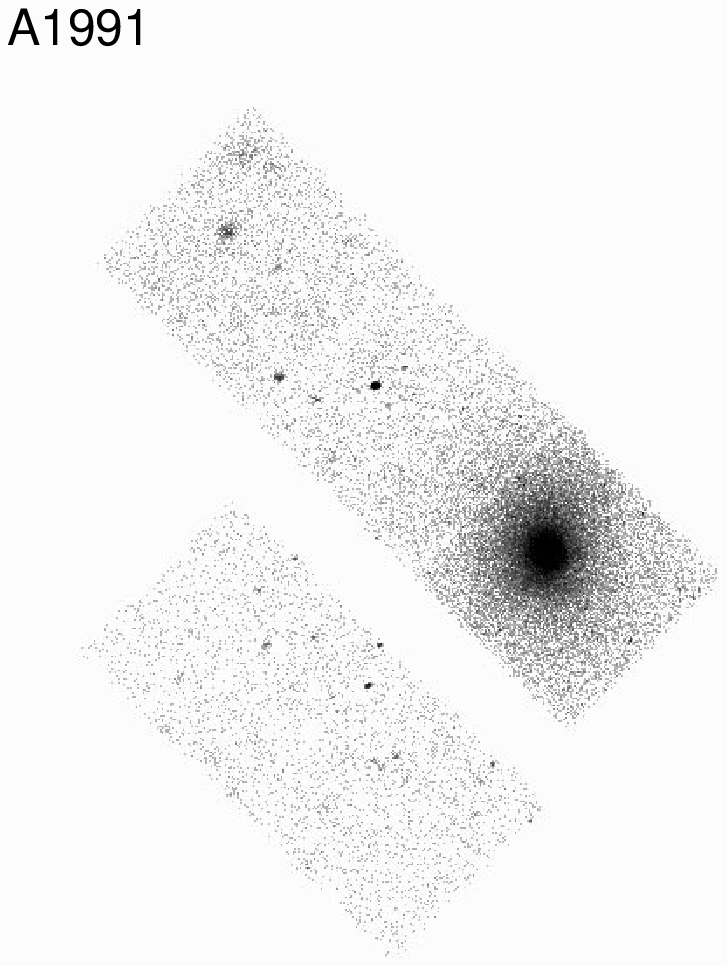}}~~~%
\framebox{\includegraphics[width=0.3\linewidth]{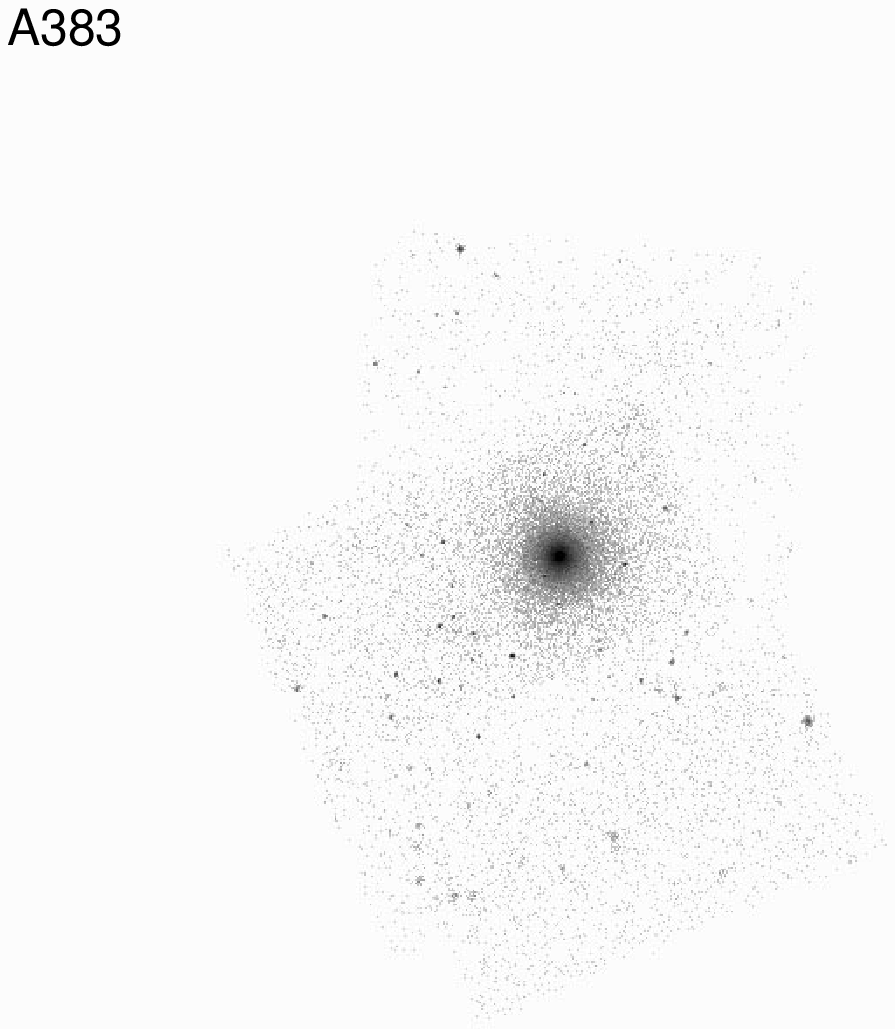}}
}
\medskip
\centerline{%
\framebox{\includegraphics[width=0.3\linewidth]{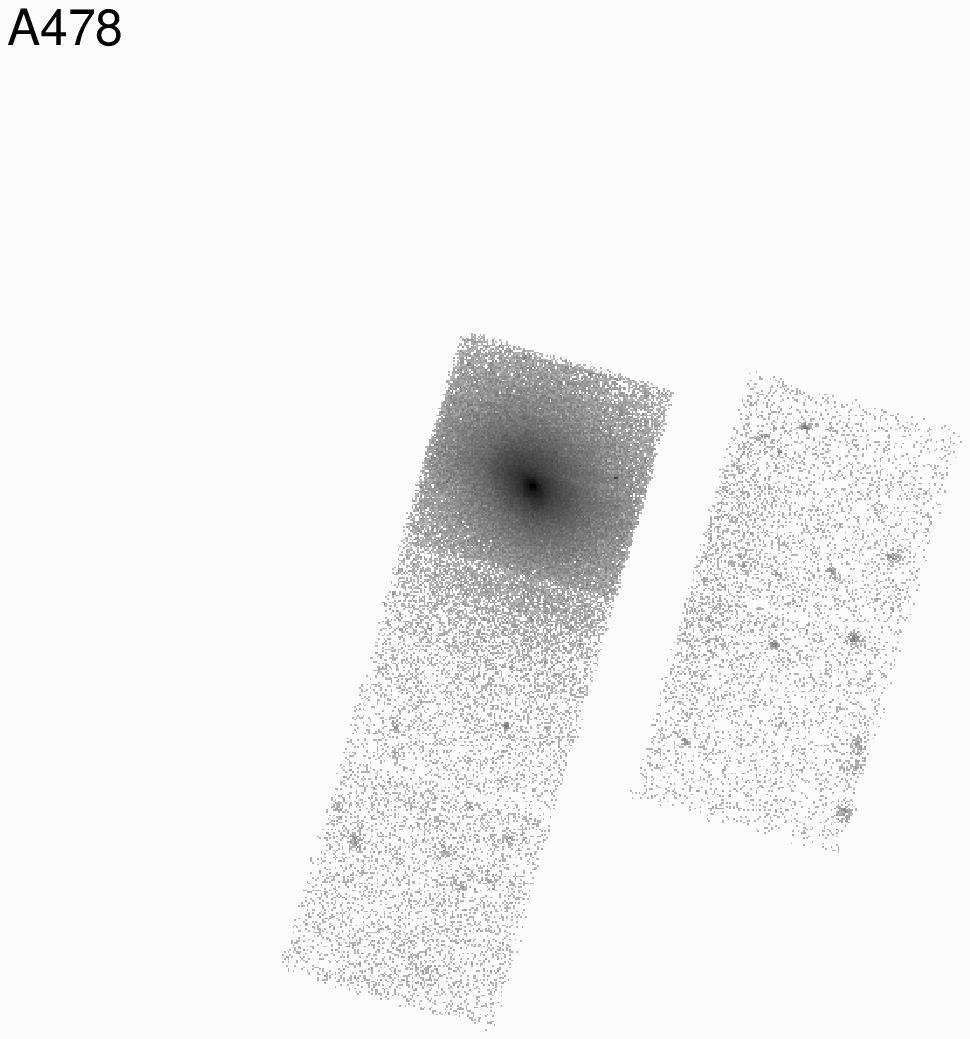}}~~~%
\framebox{\includegraphics[width=0.3\linewidth]{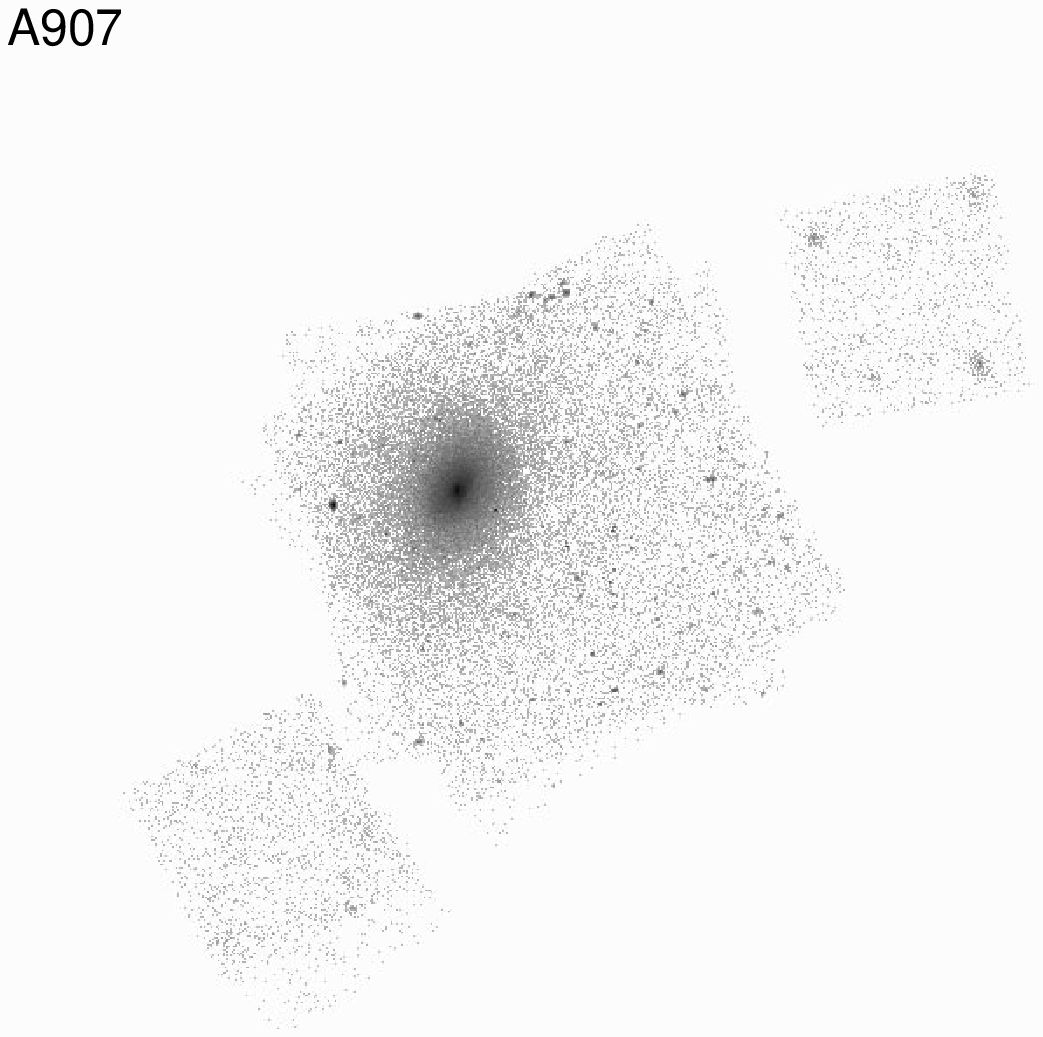}}~~~%
\framebox{\includegraphics[width=0.3\linewidth]{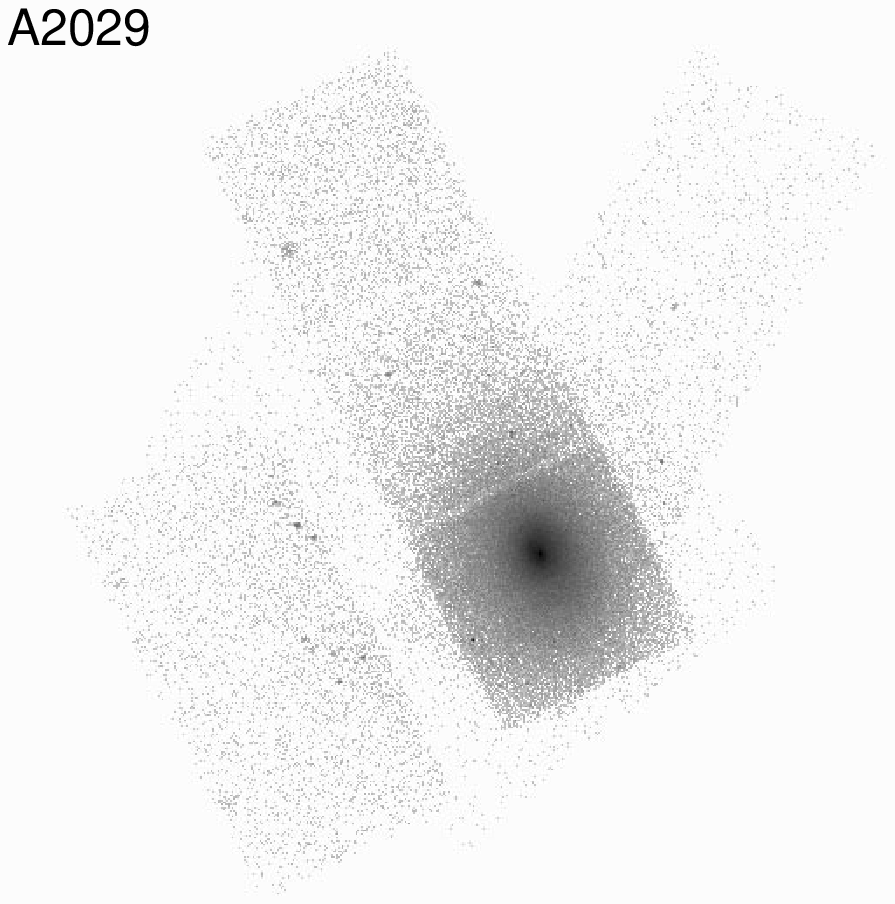}}%
}
\caption{Composite \emph{Chandra} images in the 0.7--2~keV
  band. The images were not flat-fielded to indicate the difference in
  exposure coverage for different locations within the cluster. Each CCD
  chip is $\simeq 8'$ on a side. North is up and East is to the left.} 
\label{fig:mosaic:img}
\end{figure*}

\begin{figure*}
\vspace{0.3\baselineskip}
\addtocounter{figure}{-1}
\centerline{%
\framebox{\includegraphics[width=0.3\linewidth]{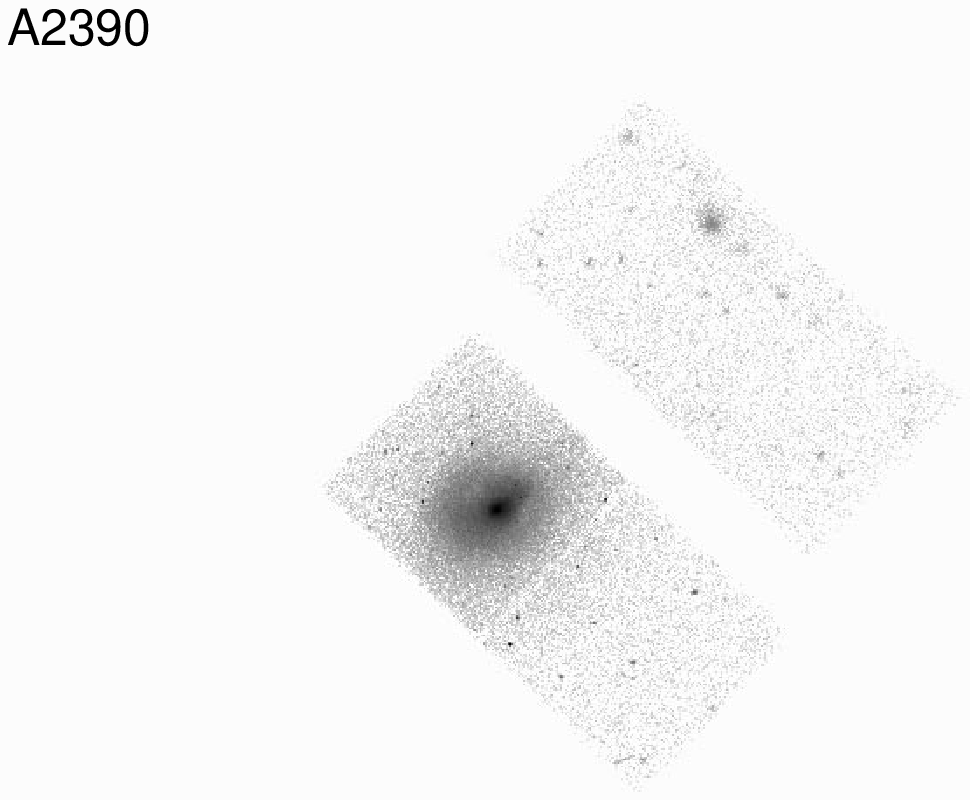}}~~~%
\framebox{\includegraphics[width=0.3\linewidth]{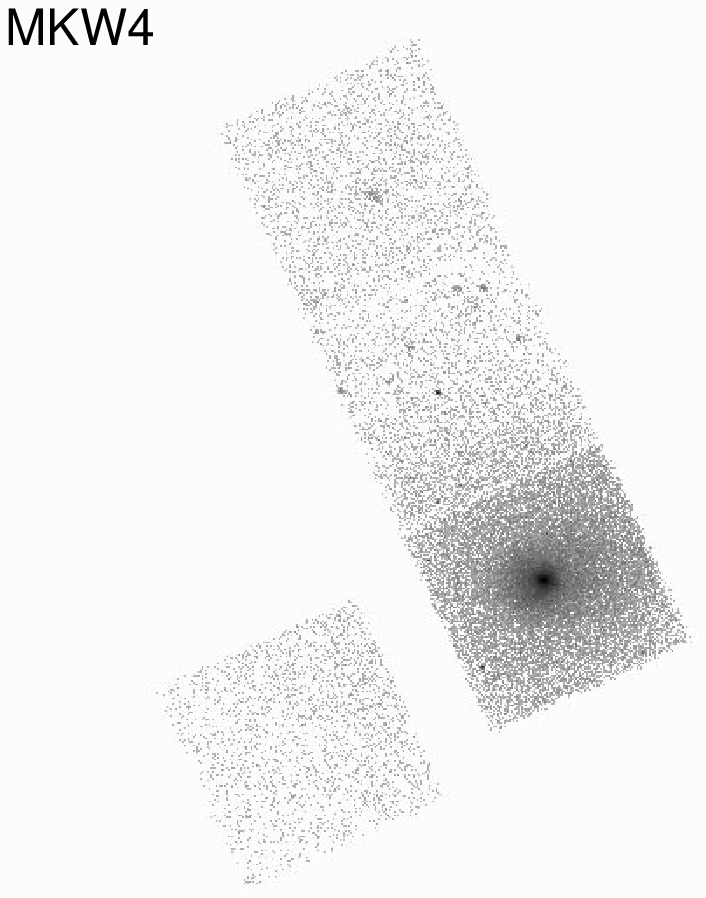}}%
}
\centerline{%
\framebox{\includegraphics[width=0.3\linewidth]{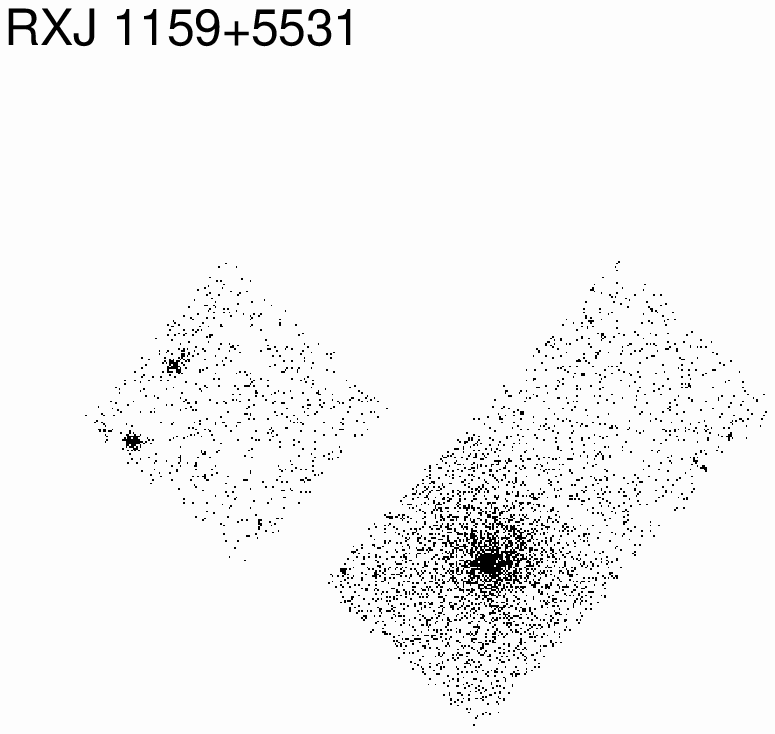}}~~~%
\framebox{\includegraphics[width=0.3\linewidth]{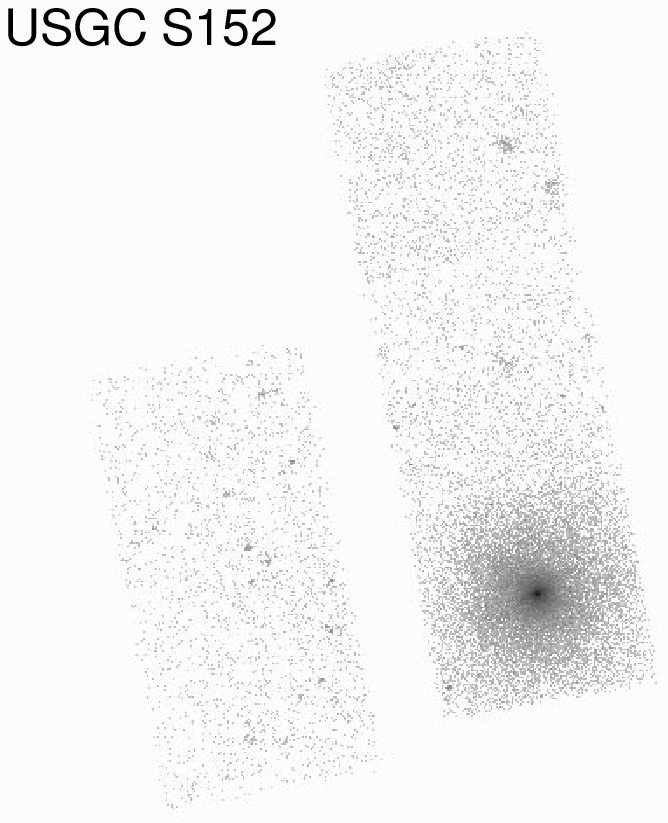}}%
}
\caption{\emph{continued}.} 
\end{figure*}

Given the technical difficulties, early measurements of cluster temperatures
at large radii have been controversial. \asca\ was the first instrument with
the necessary spectro-imaging capability.  Cluster analysis was complicated
by the complex PSF of the \asca\ telescope, which if not taken into account,
resulted in radially increasing apparent temperatures. Using \asca\ data,
Markevitch et al.\ (1996; 1998, hereafter M98; 1999) obtained temperature
profiles for a sample of 32 nearby clusters, which showed significant
declines with radius between $r=0.1-0.6\, r_{\rm vir}$ (hereafter,
$r_{180}\equiv r_{180}$ from Evrard, Metzler \& Navarro 1996). In
clusters without obvious mergers, the radial temperature profiles outside
the cool cores were similar when normalized to the virial radius. Consistent
results were obtained, e.g., by Ikebe et al.\ (1997), Cannon et al.\ (1999)
and Finoguenov et al.\ (2001). However, White (2000) found that most
clusters in their large \asca\ sample ``are consistent with isothermality at
the $3\sigma$ confidence level.''  Some of the differences between White
(2000) and other works can be attributed to White's use of a PSF model that
overestimated scattering at low energies. Also, White typically did not
extend measurements to large radii because of large uncertainty inherent in
his image deconvolution method, and in the overlapping radial range most of
his temperature profiles are in fact consistent with M98. 
% 
% We also note that most of White's
% temperature profiles were in fact consistent with M98 in the radial range
% where direct comparison was possible (i.e., outside the cool cores). 
% Because of the large uncertainty inherent in his image deconvolution method,
% White typically did not extend measurements to large radii where significant
% temperature declines were reported by M98. 

\sax\ was another instrument capable of spatially resolved spectroscopy;
compared to \asca, it had better PSF. Irwin \& Bregman (2000) analyzed
11 clusters and reported isothermal or even increasing radial
temperature profiles.  However, De Grandi \& Molendi (2002) pointed out
a technical error in that work.  Instead, De Grandi \& Molendi found, in
their analysis of 21 clusters, declining temperature profiles, in good
agreement with M98 outside the cores ($r>0.15-0.2\, r_{\rm vir}$),
although less peaked closer to the center. Similar profiles using \sax\
data were obtained by Ettori et al.\ (2000) and Nevalainen et al.\
(2001).

Temperature profiles for individual clusters have been derived in the
past few years using both \emph{Chandra} and \emph{XMM-Newton}
observations.  Previous \emph{Chandra} results were mostly confined to
the central region, typically within 0.2--0.3~virial radii
\citep{2001ApJ...557..546D,2001MNRAS.327.1057S,2002ApJ...567L..37M,2003ApJ...587..619S,2002MNRAS.336..299J,2003ApJ...586..135L,2004ApJ...604..116B,2003ApJ...598..250S}
although attempts were made to go to larger radii
\citep{astro-ph/0012215,2001ApJ...563...95M}.  These measurements cannot
be used to test the temperature decline at large radii observed by
\emph{ASCA} and \emph{Beppo-SAX}. Published \emph{XMM-Newton}
temperature profiles extend to a larger fraction of the virial
radius. Declining temperature profiles are observed in some clusters
\citep{2002A&A...394..375P,2003PASJ...55.1105T,2004A&A...413...49Z,astro-ph/0412233,2005A&A...430..385B}. 
However, there are also clusters reported to be isothermal at large
radii
\citep{2002A&A...394...77M,2004A&A...423...33P,2004A&A...415..821B,2004MNRAS.351.1439S,PrattArnaud2004}. 
These lists include results for both relaxed clusters and major mergers,
so some of the differences could be explained by the differences in the
cluster dynamical state.

\emph{Chandra} is well-suited for measurements of the temperature profiles
to 0.5--0.6 of the virial radius thanks to its stable detector background,
and fine angular resolution. 
%\emph{XMM-Newton} has an
% advantage of a larger effective area, but this can be compensated by
% longer exposure times of \emph{Chandra} observations. 
In this paper, we present measurements of the temperature profiles to at
least $\sim 0.4$ of the virial radius for a sample of 13 low-redshift
clusters observed by \emph{Chandra} with sufficiently long exposures
that the statistical temperature uncertainties at this radius are
reasonably small. The clusters are listed in
Table~\ref{tab:chandra:obs}. All these objects have a very regular
overall X-ray morphology and show only weak signs of dynamical activity,
if any. The main goal these observations was mass determination at large
radii from the hydrostatic equilibrium equation. All of them have
sufficient off-center area coverage to enable accurate background
modeling and subtraction, which is the critical element of our
analysis. Three clusters (A133, A907, A1413) were observed by us with a
specific setup optimized for studying the outer regions. Even though the
present cluster sample is not unbiased, it represents an essential step
towards reliable measurements of the gas density and temperature
profiles to a large fraction of the virial radius. 

We assume $h=0.72$, $\Omega_M=0.3$, $\Omega_\Lambda=0.7$. Measurement
uncertainties correspond to 68\% CL.

\section{Chandra data analysis}

The nominal aim points (ACIS-I or ACIS-S) and exposure time for
\emph{Chandra} observations analyzed in this paper are listed in
Table~\ref{tab:chandra:obs}.  Most of the observations were telemetered
in VFAINT mode which provides for better rejection of the
particle-induced background. In the case of ACIS-I pointings, we used
the data from four ACIS-I chips for the temperature analysis, and S2
chip was generally used for monitoring the background. In the ACIS-S
pointings, we used S3, S2, and available ACIS-I chips. The S1 chip was
used mainly to monitor the background, but in low-redshift cool clusters
(A262, MKW4, USGC~S152), we used the S1 data also for the temperature
profiles. In the rest of this section, we outline our \emph{Chandra}
data reduction procedures. 

\subsection{Calibration Corrections to Individual Photons}

We start with ``level 1'' photon lists and apply standard processing
using the CIAO tool \texttt{acis\_process\_events}. This includes
correction for Charge Transfer Inefficiency (CTI; Townsley et al.\ 2000,
Grant et al.\ 2004), re-computation of event grades, and detection of
so-called afterglow events. We then remove photons detected in bad CCD
columns and pixels and also those with bad ASCA grades (1,5,7). For
observations telemetered in VFAINT mode, we apply additional
background screening by removing events with significantly positive
pixels at the border of the $5\times5$ event island. 

The next step is to compute calibrated photon energies. We apply updated
ACIS gain maps and correct for its time dependence. Both corrections are
available in the CIAO~3.1. The new gain maps are required for analyzing
S1 data; improvements are very small for other chips. 

The final step is to examine background light curves during each
observation in order to detect and remove the flaring episodes. The
flare detection was performed following the recommendations given in
Markevitch et al.\ (2003). Clean exposure times are listed in
Table~\ref{tab:chandra:obs}. 

\subsection{Spectral Response Calibration}

To take into account the spatial dependence of effective area and
detector energy resolution, our analysis uses several important recent
calibration advances outlined below. 

ACIS optical blocking filters have been contaminated in-flight by a
substance containing C, O, and F. Photoelectric absorption in the
contaminant strongly reduces the low-energy effective area. Thanks to a
number of calibration analyses, the properties of the contaminant are now
known sufficiently accurately. The absorption spectrum of the contaminant
has been measured using grating observations of bright AGNs with continuum
spectra \citep{2004SPIE.5165..497M}. Time dependence of its thickness in
ACIS-S was determined from the flux ratios of the calibration source lines. 
These corrections are built into CIAO since v3.0. New to our analysis is the
inclusion of the spatial distribution of the contaminant. The center-to-edge
difference of its optical depth is $\Delta\tau\simeq 0.5$ around 0.7~keV in
both ACIS-I and ACIS-S arrays. Such variations, if unaccounted for, may have
a major impact on the derived ICM temperatures. Recently, the spatial
distribution of the contaminant has been measured accurately, so that the
residual variations of the effective area are less than 3\% at energies
above 0.6~keV \citep{vikhlinin2004a}. This correction
was released to the \emph{Chandra} users in October, 2004. 

An additional non-uniformity of the CCD quantum efficiency arises
because for a certain fraction of photons, the trailing charge produced
by CTI makes them appear as charged particle-induced events, and they
are screened out. Earlier calibration of this effect was based on a
limited dataset and lacked an accurate model for its energy
dependence. The resulting quantum efficiency residuals are $\simeq 5\%$
near $E=6\,$keV, marginally significant for the ICM temperature
measurements. The new calibration released with CIAO~3.1 eliminates
these residuals \citep{vikhlinin2004b}. 

The same CIAO release corrected an error in the CCD average quantum
efficiency (QE) for the backside illuminated chips (S3 and S1) both at
low and high energies \citep{edgar2004}. This error, if neglected,
results in a $\simeq 10\%$ mismatch in the temperature values determined
from the back- (BI) and frontside illuminated (FI) chips.  Temperature
profiles spanning both types of chips would be affected.

Finally, the released calibration underestimates the effective area of
the \emph{Chandra} mirror by $\sim 10\%$ just above the Ir M edge,
probably because the mirror surface is contaminated by a thin
hydro-carbon layer \citep{marshall2003}. We find that the corresponding
effective area correction can be approximated by a ``positive absorption
edge'',
\begin{equation}
  C=\exp(0.15 (E/E_0)^{-3}) \quad \text{for $E>E_0$},
\end{equation}
where $E_0=2.07$~keV. This correction reduces the best-fit temperatures by
5\%. The expected effect of mirror contamination on the vignetting is
negligible (D.~Jerius, private communication). Corrections of the mirror
area and updates to the QE of BI chips supersede the fudge factor of 0.93
that was suggested for the FI QE \citep{2001ApJ...563...95M}.

With all these recent calibration updates applied, the variations of the
effective area within the \emph{Chandra} field of view are modeled
accurately to within 3\%. This is verified directly by comparison of the
A1795 spectra observed at different locations \citep{vikhlinin2004a},
and indirectly by the agreement of the temperature profiles derived in
different CCD chips.

\subsection{Background Subtraction}

As was discussed above, correct background subtraction is crucial for
temperature measurements at large radii. The ACIS detector background can be
modeled very accurately, to within 2\% \emph{rms}, which is one of the major
advantages of \emph{Chandra} over \emph{XMM-Newton} for studying cluster
outskirts. 

The baseline background model can be obtained by using a compilation of the
blank-field observations, processed identically to the cluster data, and
``re-projected'' onto the sky using the aspect information from the cluster
pointing (see Markevitch et al.\ 2000 for a detailed description). This
procedure results in background residuals of $\lesssim10\%-20\%$ outside the
obvious flaring periods. The accuracy can be significantly improved by small
adjustments to the baseline model. 

\subsubsection{Quiescent Background}
\label{sec:bg:adj}

ACIS background above 2 keV is dominated by events from the charged
particles. There are secular and short-term variations of the intensity
of this component by as much as 30\%, but its spectrum is very
stable. Therefore, these trends can be accounted for by simply changing
the background normalization. The renormalization factor can be derived
for each observation using the data in the 9.5--12~keV band where the
\emph{Chandra} effective area is nearly zero and so all the observed
flux is due to the background. Such a renormalization reduces the
uncertainties in the modeling of the quiescent background to 2\%
\emph{rms}, provided that flaring periods were screened out properly
(Markevitch et al.\ 2003). 

We extract and fit cluster spectra separately in the BI and FI chips,
and flare filtering is also specific for each CCD type. Therefore, the
background renormalizations were determined independently for these chip
sets. A 2\% scatter in the overall normalization per set per pointing
was included in the overall temperature uncertainty budget. The accuracy
of the quiescent background subtraction was verified using the spectra
in the regions far from the cluster centers.

\subsubsection{Soft Diffuse X-ray Background}

In addition to the particle-induced background, the blank-field datasets
contain the diffuse X-ray background. This component makes a large
contribution to the total background below $\sim 1$~keV. The soft X-ray
background was studied with \emph{Chandra} by Markevitch et al.\ (2003)
and \emph{XMM-Newton} by Lumb et al.\ (2002).  For most of the sky above
the Galactic plane, the background spectrum is well represented by the
MEKAL model with the Solar metallicity and $T\approx 0.2$~keV. A second
component with $T\approx 0.4$~keV appears in locations where the diffuse
flux is high. A fraction of the soft background is geocoronal in origin
and variable in time (Wargelin et al.\ 2004). This component is
dominated by the O~VII and O~VIII lines and its spectrum also can be
approximated with sufficient accuracy by a thermal plasma model. 

The blank-field datasets contain a typical mixture of the Galactic and
geocoronal backgrounds. Since these components vary with location or
time, the soft background in individual pointings is usually slightly
different. Therefore, appropriate adjustments are needed, even though
they rarely lead to qualitative changes in the derived temperature
profile. 

\begin{figure}
\vspace*{-1.7\baselineskip}
\centerline{\hspace*{5mm}\includegraphics[width=0.485\textwidth]{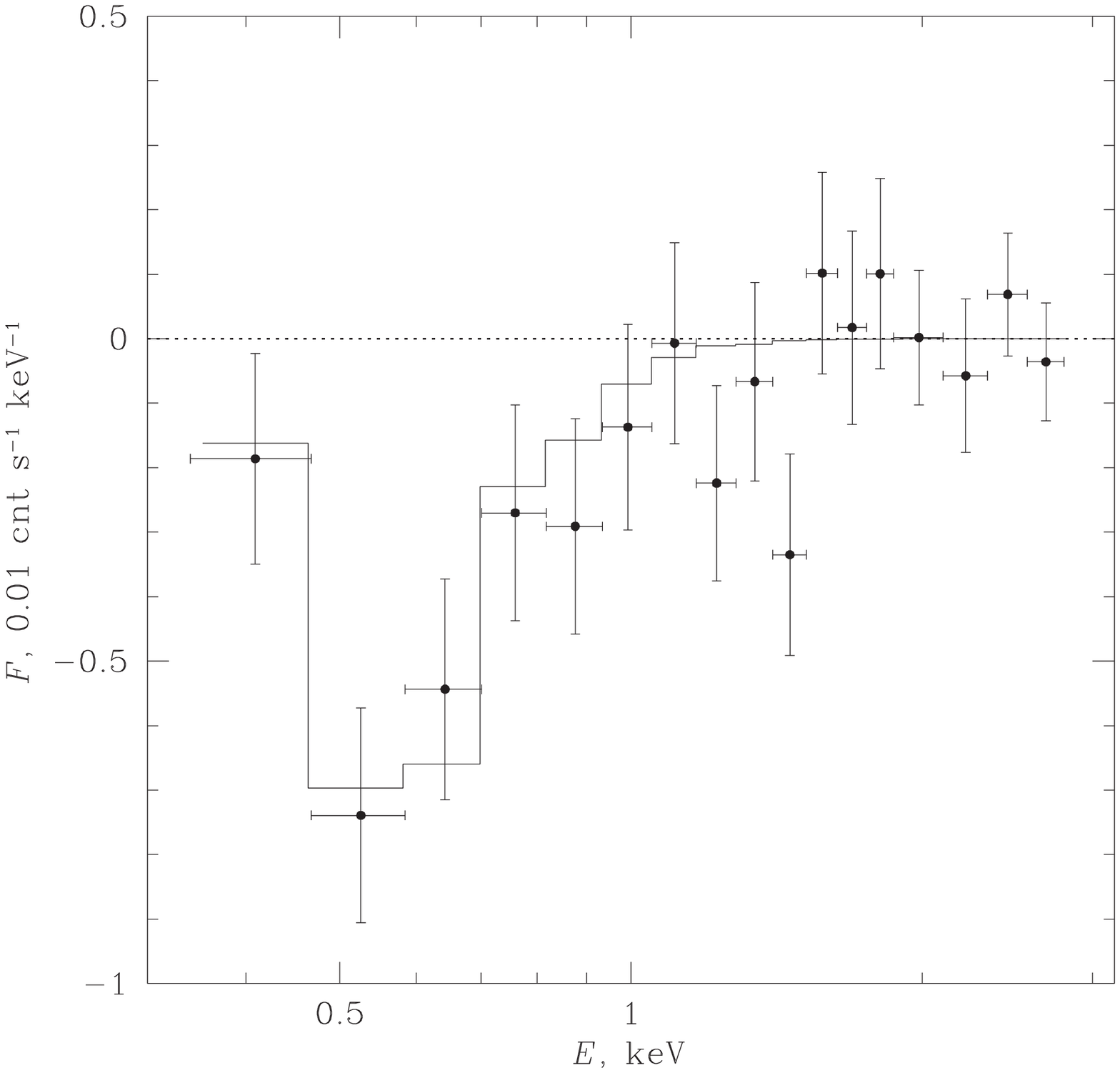}}
\vspace*{-1.0\baselineskip}
\caption{The spectrum of the offset A133 pointing in the S2 chip with
  the nominal background subtracted. Residuals give an example of
  oversubtraction of the sky soft background. The fit is MEKAL model
  with negative normalization, Solar metallicity, and $T=0.18$~keV. It
  is included in the final background model.} 
\label{fig:a133:soft:spec}
\end{figure}

A certain fraction of the detector area in most of our observations was
essentially free from the cluster emission and so we were able to
determine the soft background adjustments \emph{in situ}. We extracted
spectra in those regions, subtracted the adjusted blank-field background
(\S\ref{sec:bg:adj}), and fit the residuals with an unabsorbed MEKAL
model, whose normalization was allowed to be negative. The fit was
performed in the 0.4--1~keV band because the soft background is usually
dominated by the oxygen lines near 0.6~keV
(Fig.~\ref{fig:a133:soft:spec}). The derived adjustment is the real sky
X-ray emission and is subject to the spatial variations in the effective
area. Therefore, it must be included as an additional component in the
spectral fits, with its normalization scaled by the region area. This
approach was used before by, e.g., \cite{2001ApJ...563...95M}, and a
similar method is used for \emph{XMM-Newton} (e.g., Pratt et al.\ 2001,
Majerowicz et al.\ 2002).

The soft-band adjustments are within 20\% of the nominal background flux
in the 0.6--1~keV band in all our clusters, and do not qualitatively
change the behavior of the temperature profiles in the radial range we
consider. The only exceptions are A1991 and A2029 which are projected on
the North Polar Spur. The soft background, if neglected, would have a
serious impact on the derived temperature profiles in these cases. The
statistical uncertainty in the normalization of this component is
included in the uncertainties for the cluster temperatures\footnote{We
  fitted the cluster spectra with the background normalizations fixed at
  $\pm 1\sigma$ from the nominal value. The changes in the best-fit
  cluster $T$ were treated as the corresponding uncertainty. They were
  added to the statistical uncertainties in quadrature. Identical
  technique was used to estimate the uncertainties related to
  normalization of the quiescent background.}. 

Our adjustment procedure implicitly assumes that there are no
significant fluctuations of the soft background on $\sim 15'$ scales. We
cannot test this assumption internally in each observation because
cluster emission fills large fraction of the field of view. This is a
limitation which can be addressed only by accurate mapping of the soft
background in the vicinity of each cluster. However, we do not expect
this to be a serious problem because the adjustments are usually small. 
In those cases with the high Galactic foreground flux, we have used the
\emph{ROSAT} All-Sky Survey maps to verify the absence of surface
brightness gradients on relevant spatial scales. 

\begin{figure*}
\vspace*{-\baselineskip}
\centerline{%
\includegraphics[width=0.485\linewidth]{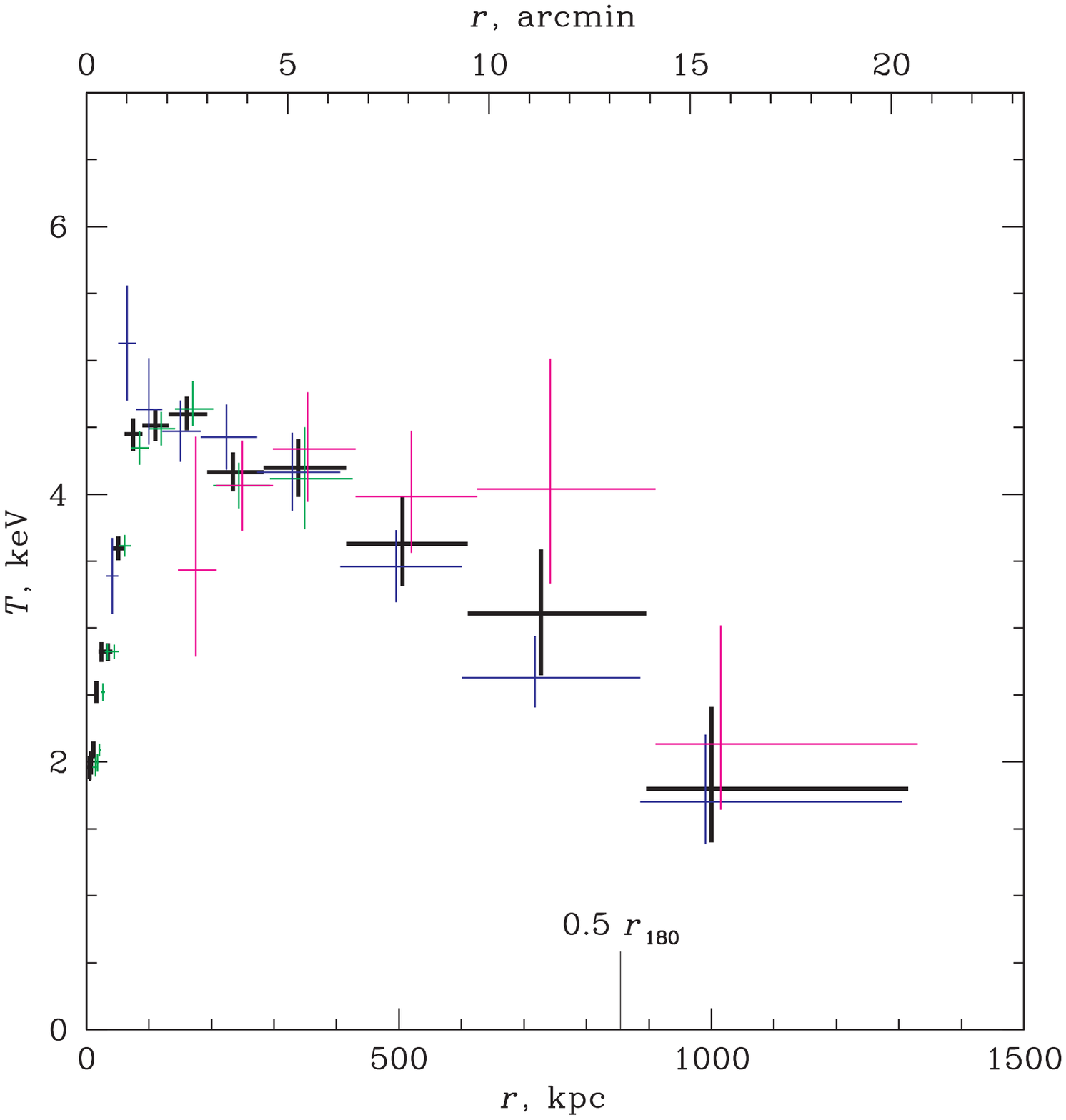}\hfill%
\includegraphics[width=0.485\linewidth]{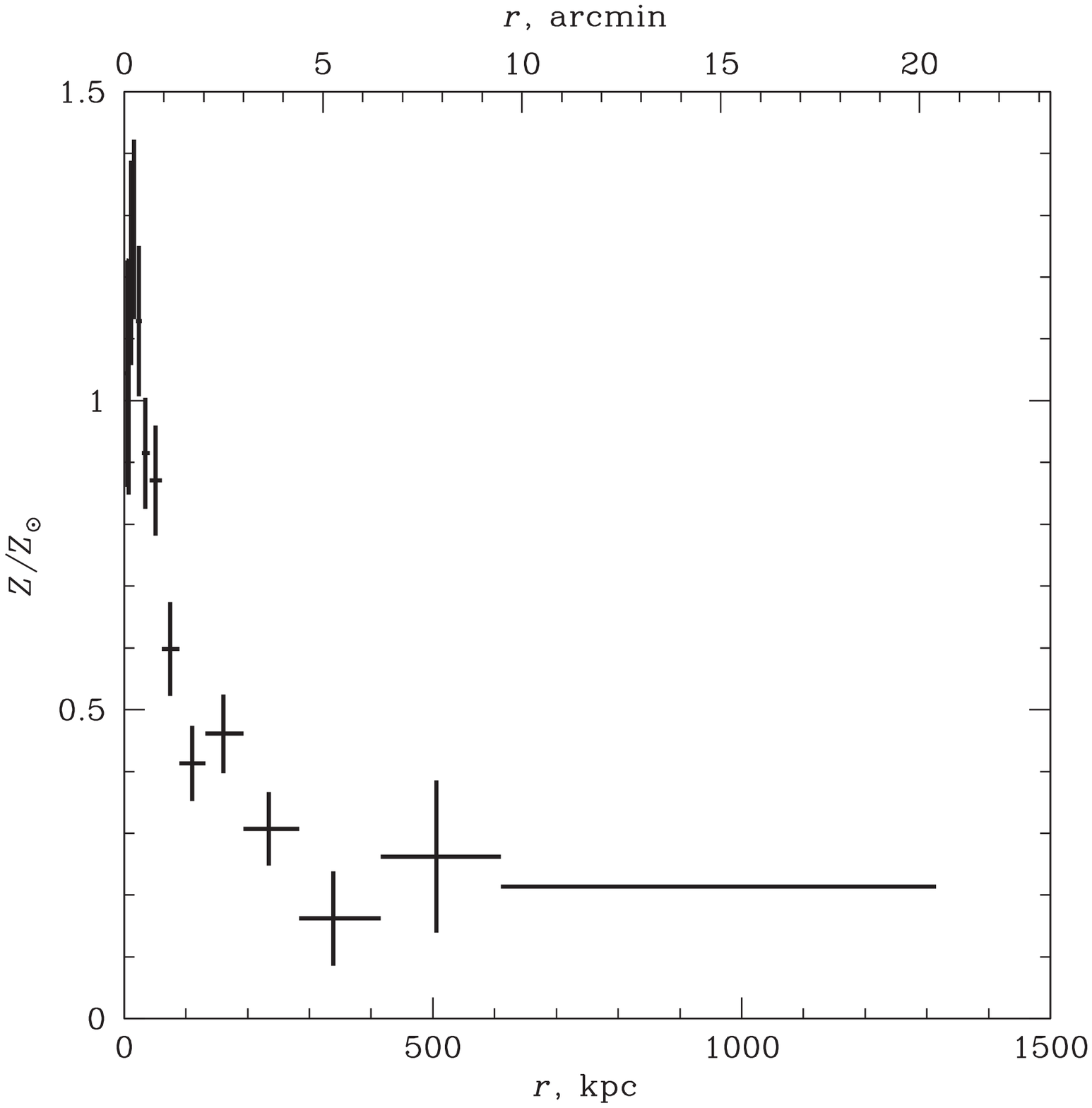}
}
\vspace*{-0.2\baselineskip}
\caption{Projected temperature and metallicity profiles for A133. The vertical
bars are at the emission-weighted radius within each annulus. The
independent temperatures measurements from S3 and FI chips in the ACIS-S
pointing, and from ACIS-I pointing are shown in green, magenta, and
blue, respectively. The values from the joint fit to these three
datasets are shown in black. The metallicity at $r>600$~kpc was fixed at
the average value from $r=270-600$~kpc, $Z=0.21$, and hence shown
without the vertical error bars.} 
\label{fig:a133}
\end{figure*}

\subsubsection{Subtraction of Readout Artifact}

ACIS CCDs are exposed to the source during the readout cycle, which
takes 41~ms or 1.3\% of the nominal exposure time (3.2~s). This results
in 1.3\% of the source flux being uniformly re-distributed into a strip
spanning the entire CCD along the readout direction, creating a
``readout artifact''. This effect must be accounted when the readout
artifact from the bright, central region contaminates the low surface
brightness cluster outskirts. Fortunately, the readout artifact for
sources without pile-up can be subtracted almost precisely, using a
technique proposed by \cite{2000ApJ...541..542M}. A new dataset is
generated with the \texttt{CHIPY}-coordinate of all the photons in the
original observation randomized, and their sky coordinates and energies
recalculated as if it were a normal observation. The obtained dataset is
normalized by the ratio $41\,\text{ms}/3.2\,\text{s}$ and treated as an
additional background observation. The normalization of the blank-field
background is reduced by the same 1.3\% to account for this subtraction. 
This correction was applied for all our clusters. 

\subsection{Spectral Fitting}

The cluster spectra were extracted in annuli centered on the X-ray
surface brightness peak, separately for each pointing and for FI and BI
CCDs. The radial boundaries were chosen so that
$r_{\text{out}}/r_{\text{in}}=1.5$ except for A262 where the statistical
accuracy was sufficient to measure temperatures in narrower annuli,
$r_{\text{out}}/r_{\text{in}}=1.33$. The spectra were fit in the
0.6--10~keV band to a single-temperature MEKAL model. The metallicity
was allowed to be free in each annulus in the central region, where the
statistical uncertainties were $\Delta Z<0.2$. At larger radii,
metallicity was fixed at the average value in the last two radial bins
where it was actually measured. If several datasets (pointings or FI/BI
chips) contributed to the same annulus, they were fit jointly, with the
spectral parameters (temperature, metallicity, Galactic absorption) tied
and normalizations free for each dataset. 

\begin{figure*}
\vspace*{-\baselineskip}
\centerline{%
\includegraphics[width=0.485\linewidth]{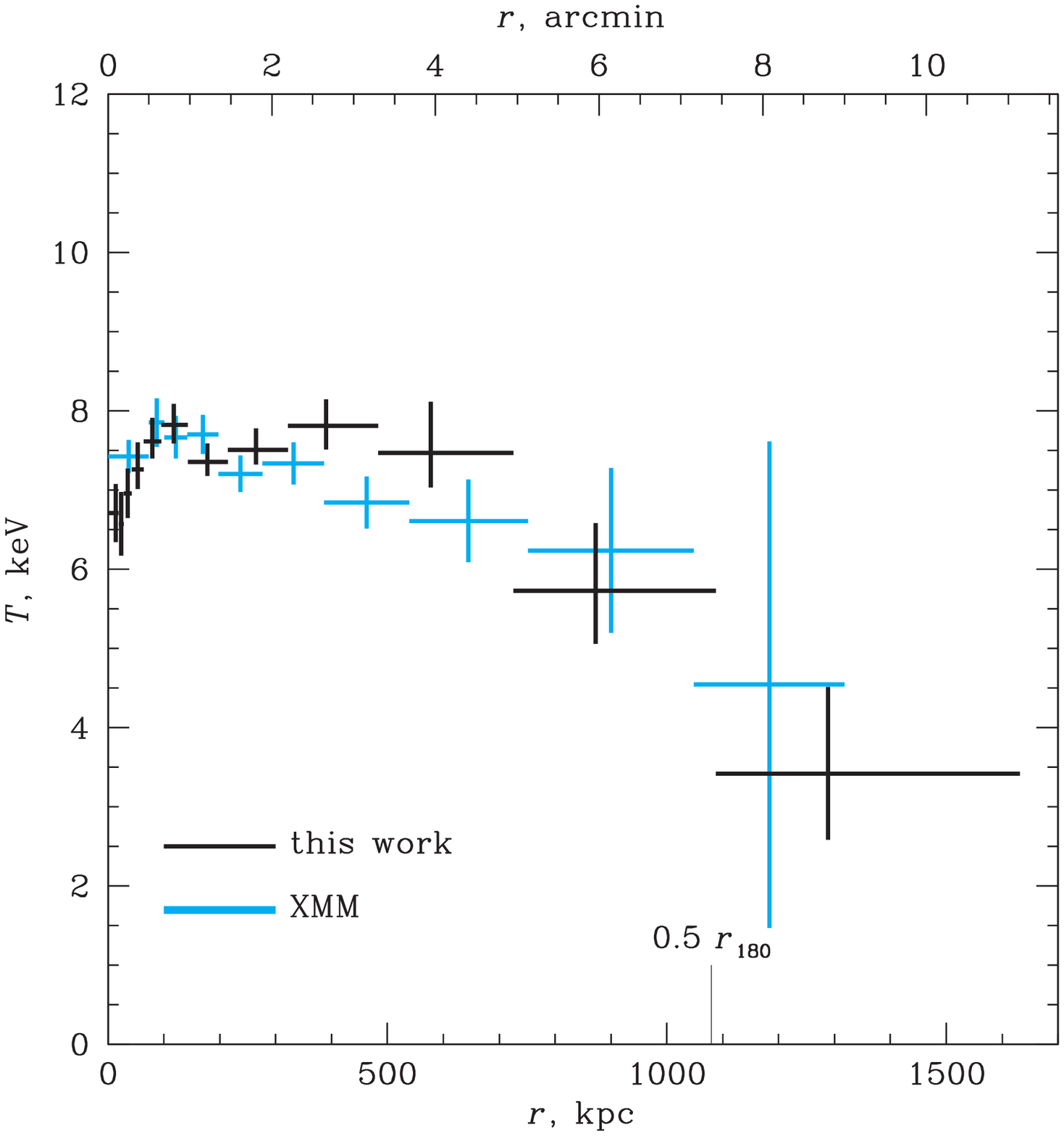}\hfill%
\includegraphics[width=0.485\linewidth]{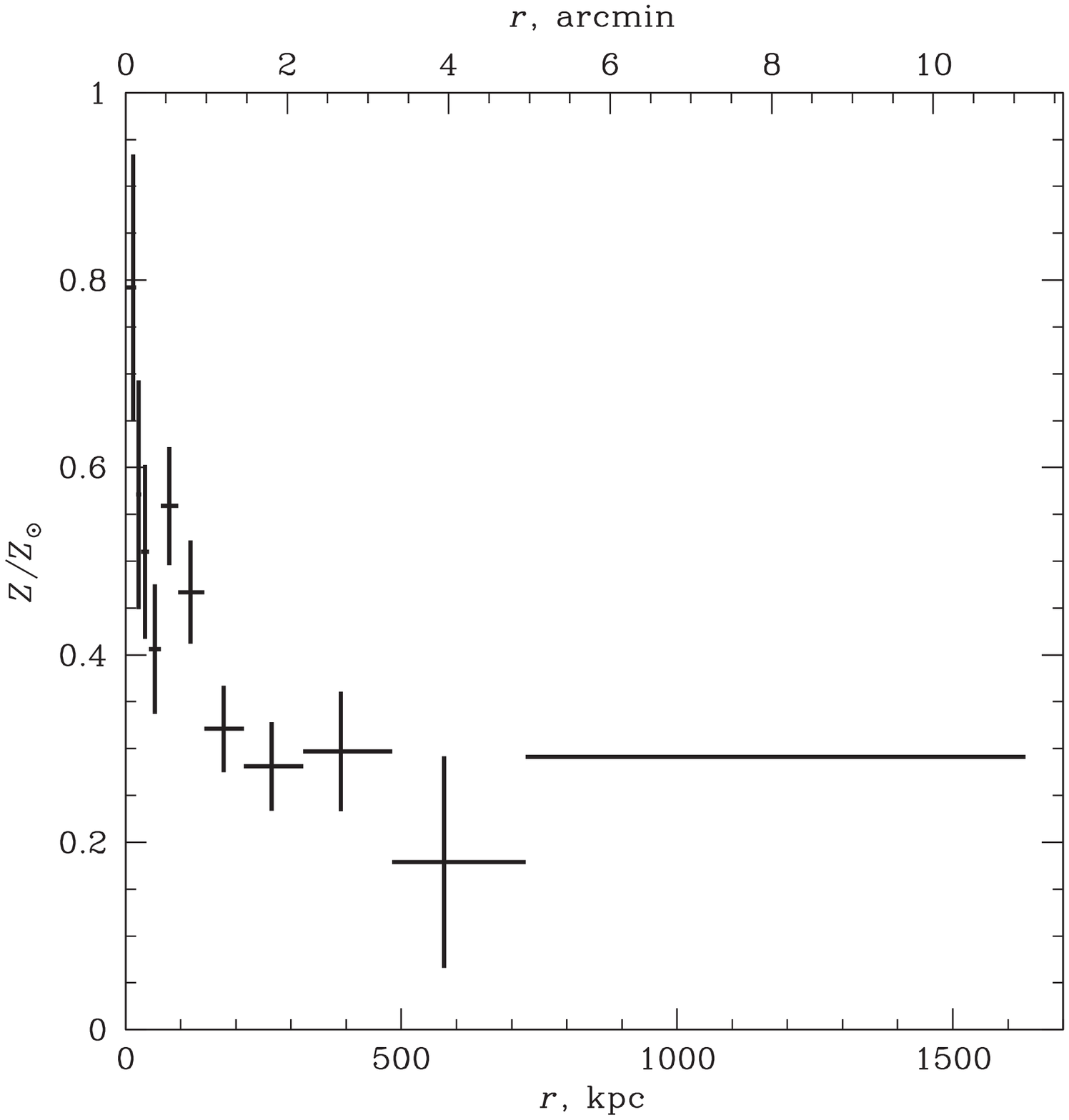}%
}
\vspace*{-0.2\baselineskip}
\caption{Temperature and metallicity profiles for A1413. The blue error
  bars in the left panel show \emph{XMM-Newton} results from
  \citep{2002A&A...394..375P}, renormalized by $+10\%$ to account for
  \emph{Chandra}--\emph{XMM-Newton} cross-calibration (see text).} 
\label{fig:a1413}
\end{figure*}

Absorption was fixed unless a significant variation was found between
the best-fit values in three cluster-centric annuli with radii 0--200,
200--400, and 400--800~kpc, where it usually could be measured
accurately. Significant variations were detected only in A478 and
MKW4. Furthermore, in most cases, the best-fit value of $N_H$ from the
X-ray fit was consistent with the value from the radio surveys
\citep{1990ARA&A..28..215D}, and we used the latter in those cases. 

\section{Results for individual clusters}
\label{sec:results:individ}

\subsection{Abell 133}

Abell~133 is a relaxed cluster at $z=0.057$ with the average temperature
$T\simeq 4.5$~keV. It was observed by \emph{Chandra} for 40~ksec in
ACIS-S and for 90~ksec in the offset pointing in ACIS-I, which we have
designed specifically for this measurement
(Fig.~\ref{fig:mosaic:img}). The X-ray image of A133 is nearly axially
symmetric on large scales. The only detectable substructures are located
within the central 40~kpc and likely caused by activity of the central
AGN (Fujita et al.\ 2002)

The quiescent background had to be adjusted by $+5\%$ in the ACIS-I
pointings, any by $-11\%$ and $-8\%$ in the \mbox{ACIS-S} pointing in
the FI chips and BI chips, respectively. The spectrum extracted outside
$20'$ of the cluster center shows significant negative residuals around
0.6--0.7~keV, indicating that the soft X-ray background is
over-subtracted. The residuals can be fit by the MEKAL model with
$T=0.18$~keV and normalization corresponding to the 0.7--2~keV count
rate $(-1.9\pm0.3)\times10^{-5}$~cnt~s$^{-1}$~arcmin$^{-2}$. This
component, with the normalization scaled by the area of the spectrum
extraction region, is included in the cluster spectral fits.

Galactic absorption shows no statistically significant deviations and
the cluster-average value from the X-ray fit,
$N_H=(1.6\pm0.4)\times10^{20}$~cm$^{-2}$, agrees well with the radio
value $1.53\times10^{20}$~cm$^{-2}$, which was used thereafter. 

The temperature and metallicity profiles are shown in
Fig.~\ref{fig:a133}. There is excellent agreement between the
temperatures measured independently from the S3 and FI chips in the
ACIS-S pointing (green and magenta, respectively), and from the off-set
ACIS-I pointing (blue). We find similar agreement between different
pointings and chips types in all our clusters. 

The temperature values from the joint fits are shown by thick black
error bars. There is a significant temperature decline at large radii,
from $T=4.6\pm0.15$~keV at $r=170$~kpc to $T=1.8\pm0.6$~keV in the
outermost annulus at $r\simeq 1000$~kpc. The upper limit on the
temperature in this bin is 2.7~keV at the 90\% CL. The temperature
uncertainties quoted here and shown in Fig.~\ref{fig:a133} include
uncertainties in the normalizations of the high-energy and soft
background components, contributing $\pm 0.35$ and $\pm 0.1$~keV,
respectively in the outermost bin. 

The ICM metallicity is nearly Solar in the center and decreases to
$Z\simeq0.2$ at large radii. Similar metallicity profiles were derived
from \emph{Beppo-SAX} observations of several ``cooling flow'' clusters
\citep{2001ApJ...551..153D}. \emph{Chandra} data cannot constrain the
abundance outside $r=600$~kpc and so temperatures in this region are
derived with the abundance fixed at $Z=0.21$, the average value for
$r=270-600$~kpc.  If, instead, one assumes that the observed abundance
gradient continues at $r>600$~kpc, the best-fit temperatures will be
only $\sim 3\%$ lower.  The temperature variation with metallicity is
much smaller than the statistical uncertainties for any $Z<0.5$, and we
will ignore it for this and all other clusters. 

Inside the central 70~kpc, we observe a sharp drop in the projected
temperature, which is most likely caused by radiative
cooling. Interestingly, there is also a spike in metallicity within this
radius. Such behavior is observed in all our clusters. However, we note
that a reliable metallicity analysis in the central regions, where the
temperature gradients are strong, requires spectral deprojection. This
is beyond the scope of this work, since we are primarily interested in
the temperature profiles at large radii. Our values of the metallicity
in the very centers should be treated with caution.

\subsection{Abell 1413}
\label{sec:a1413}

Abell~1413 ($z=0.1429$) was observed in three pointings. One observation
(OBSID 537) was almost entirely affected by a background flare and it
was discarded from the analysis. The composite image from the remaining
two observations is shown in Fig.~\ref{fig:mosaic:img}. The cluster
center was placed in the I3 chip in the shorter observation (OBSID~1661,
10~ksec). In the longer observation (OBSID~5003, 75~ksec), the cluster
center was deliberately placed off-axis in the I2 chip to obtain better
coverage for the outer regions. Since the exposure times are very
different, our results are dominated by the data from OBSID~5003. The
X-ray image is elongated in the North--South direction. However, no
obvious signatures of a merger are observed, except for a weak edge in
the surface brightness at 400~kpc North of the cluster center. 

We do not detect any radial variations in the Galactic absorption, and
the averaged value from the X-ray spectral fit,
$N_H=(2.4\pm0.8)\times10^{20}$~cm$^{-2}$, is consistent with the radio
value, $2.19\times10^{20}$~cm$^{-2}$. The spectrum from $r>1700$~kpc
shows negative residuals near 0.6~keV, therefore the soft background is
over-subtracted. The corresponding background correction can be modeled
with the MEKAL spectrum with $T=0.20$~keV and normalization
corresponding to the 0.7--2~keV count rate
$(-2.7\pm0.2)\times10^{-5}$~cnt~s$^{-1}$~arcmin$^{-2}$. 

The projected temperature and metallicity profiles are shown in
Fig.~\ref{fig:a1413}. The results are qualitatively similar to A133 ---
there is a temperature drop by a factor of $\sim 2$ in the outermost
bin. There is also a cool region within the central 100~kpc associated
with a spike in the metallicity profile. 

A1413 is one of the clusters with the most accurately measured
\emph{XMM-Newton} temperature profiles \citep{2002A&A...394..375P}. 
Comparison with our results is obviously in order. The average
\emph{XMM-Newton} temperature of the cluster is $\approx 10\%$ lower
than our measurements. A similar difference is observed in distant
clusters (Kotov \& Vikhlinin, in preparation), and therefore most likely
reflects cross-calibration problems. This difference is unimportant for
comparison of the temperature gradients and we uniformly renormalized
the \emph{XMM-Newton} values by $+10\%$. The corrected \emph{XMM-Newton}
temperature profile is shown by blue error bars in Fig.~\ref{fig:a1413}. 
There is excellent agreement between the \emph{Chandra} and
\emph{XMM-Newton} measurements, except in the very center where
\emph{XMM-Newton} can be affected by poorer angular resolution. 

\begin{figure*}
\vspace*{-\baselineskip}
\centerline{%
  \includegraphics[width=0.485\linewidth]{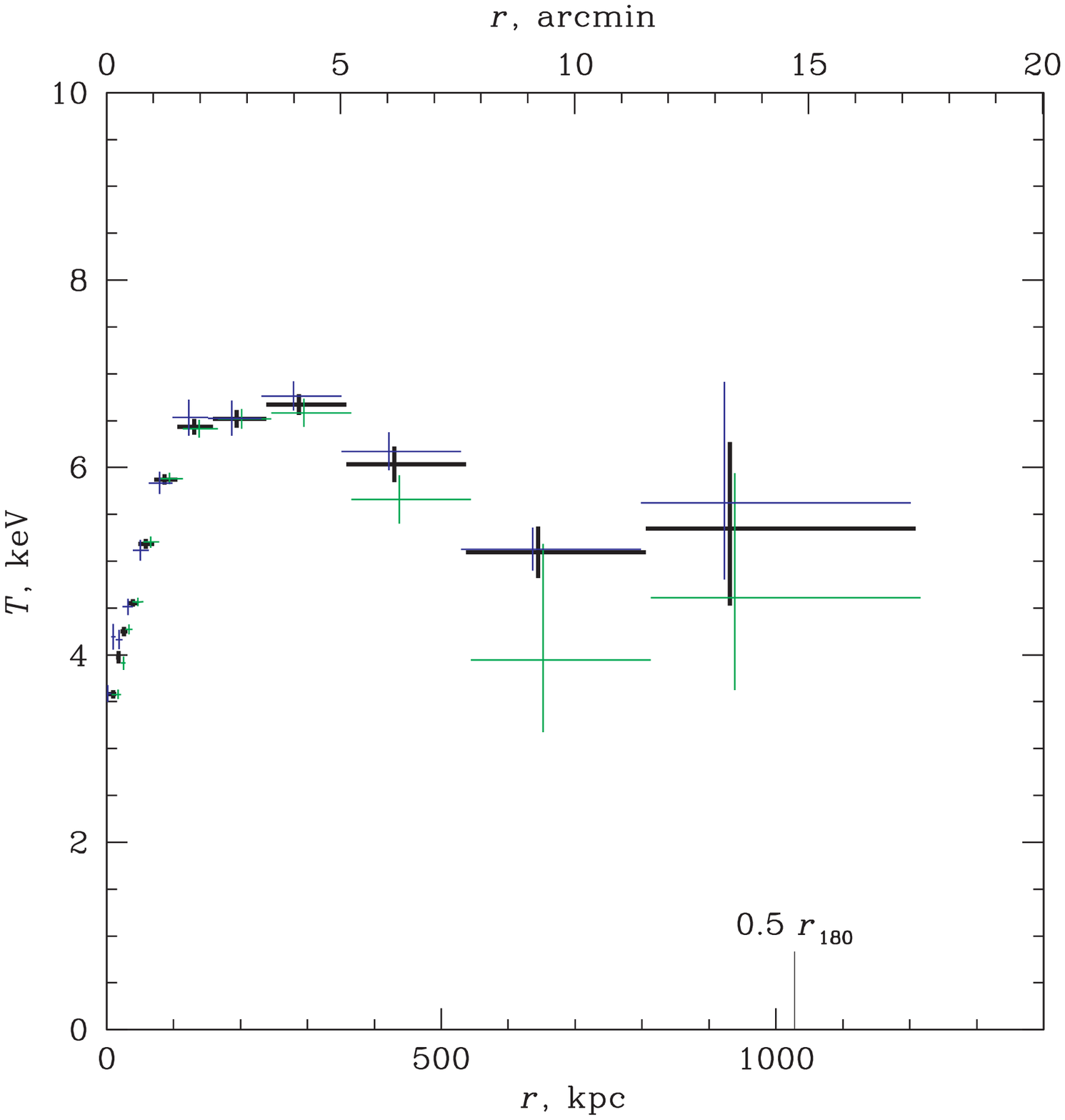}\hfill%
  \includegraphics[width=0.485\linewidth]{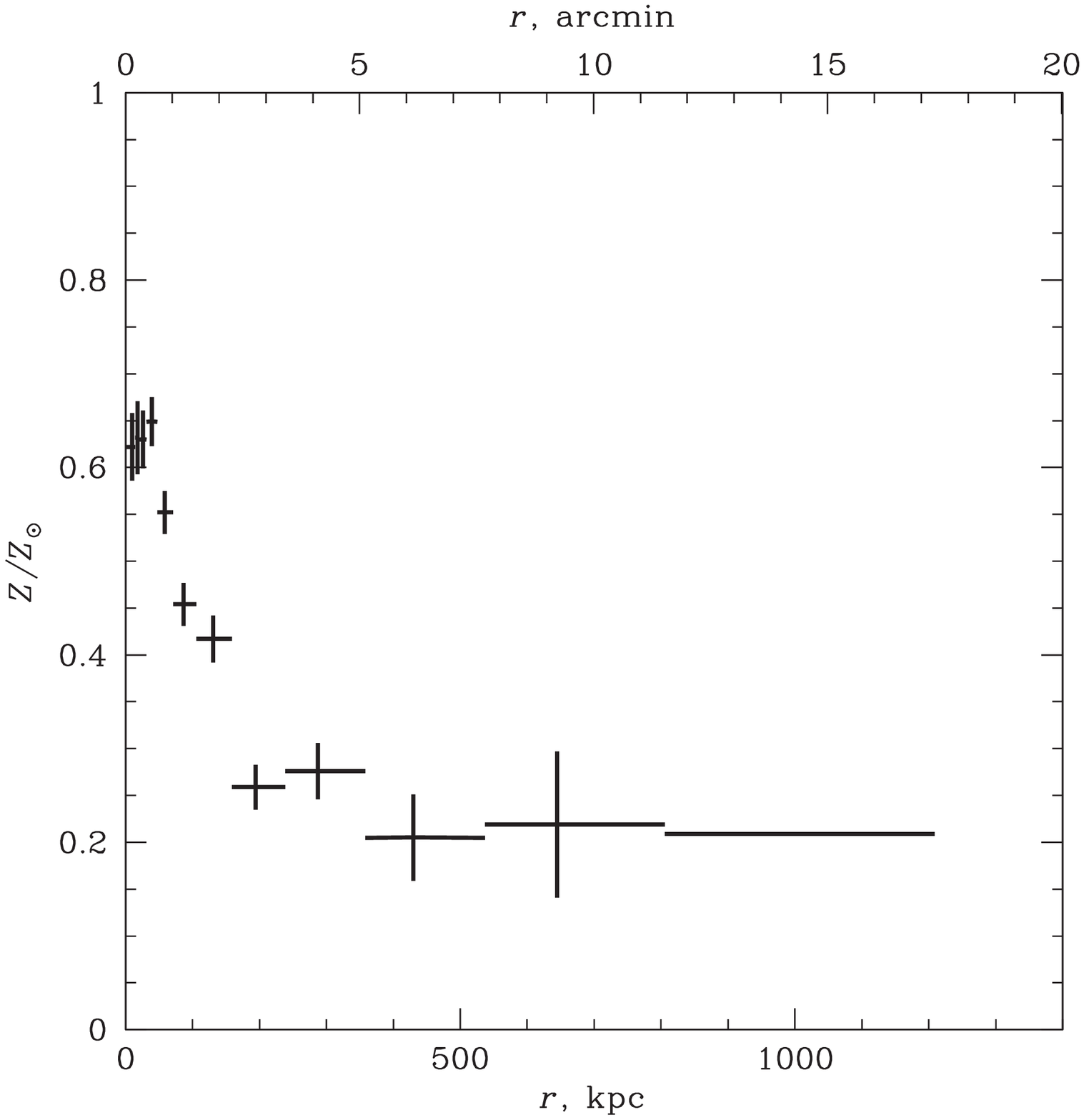}%
}
\vspace*{-0.2\baselineskip}
\caption{Temperature and metallicity profiles for A1795. Independent
  $T$ measurements from the BI and FI chips are shown by green and
  blue, respectively.} 
\label{fig:a1795}
\end{figure*}

\begin{figure}[t]
\vspace*{-\baselineskip}
\centerline{\includegraphics[width=0.485\textwidth]{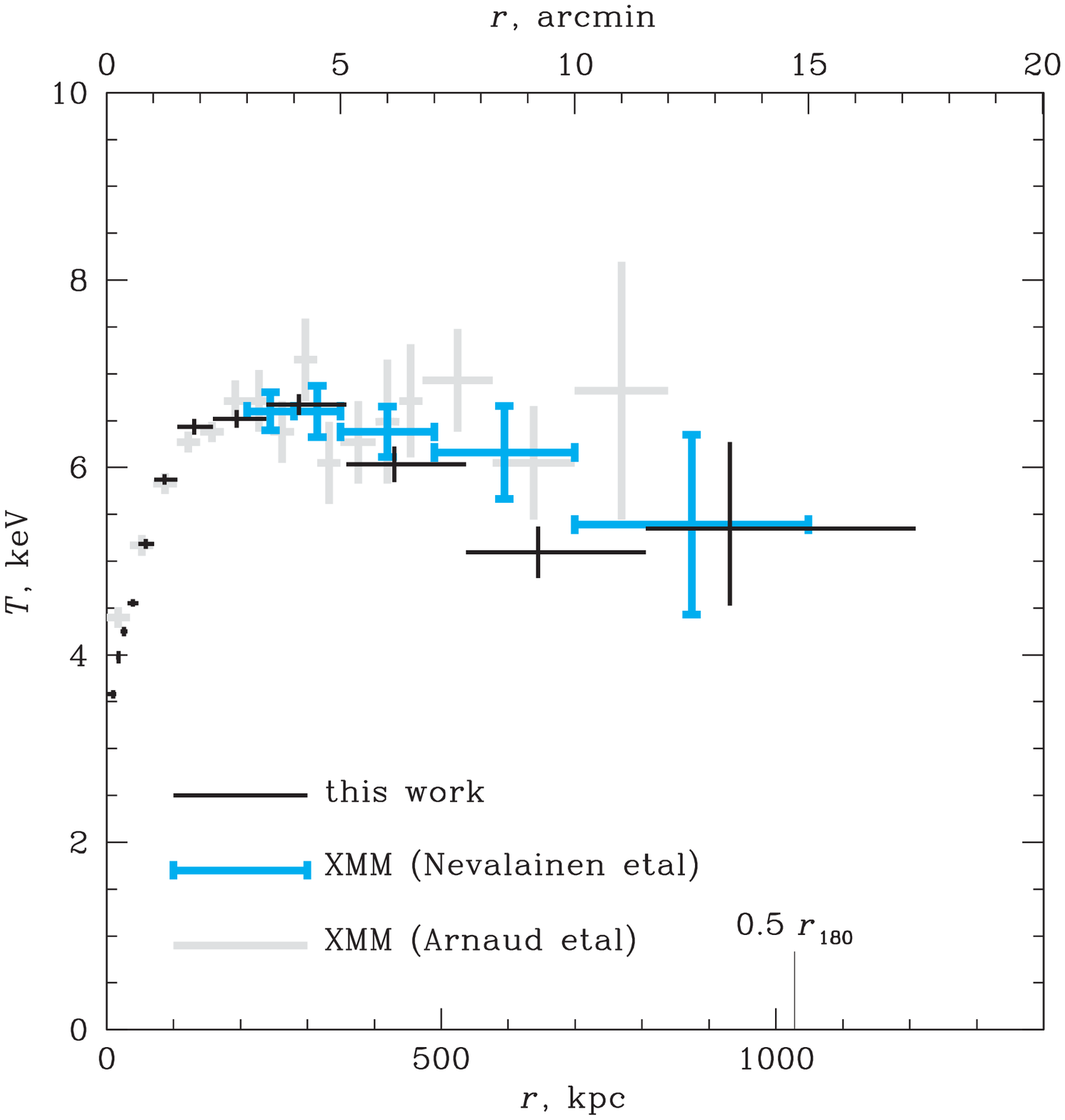}}
\vspace*{-0.2\baselineskip}
\caption{Comparison of \emph{Chandra} and \emph{XMM-Newton} temperature
  profiles for A1795. The \emph{XMM-Newton} results from
  \cite{2001A&A...365L..80A} are shown in grey. The same observation was
  also analyzed by Nevalainen et al.\ (2004; blue error bars with caps). 
  Note that \emph{XMM-Newton} values were renormalized by $+10\%$ to
  account for cross-calibration (see \S\,\ref{sec:a1413}).} 
\label{fig:a1795:prof:XMM-Newton}
\end{figure}

\begin{figure*}
\vspace*{-\baselineskip}
\centerline{%
\includegraphics[width=0.485\linewidth]{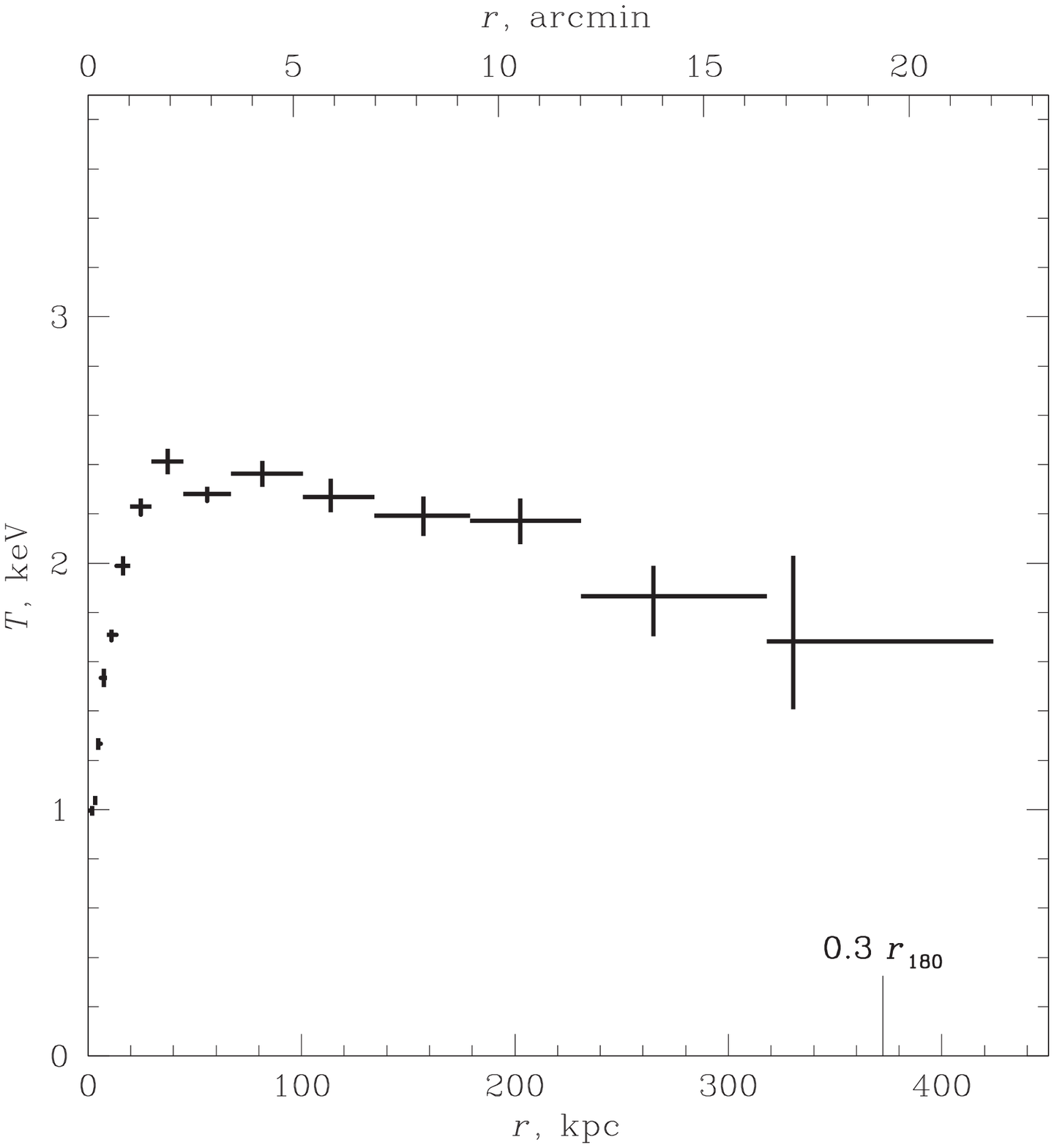}\hfill%
\includegraphics[width=0.485\linewidth]{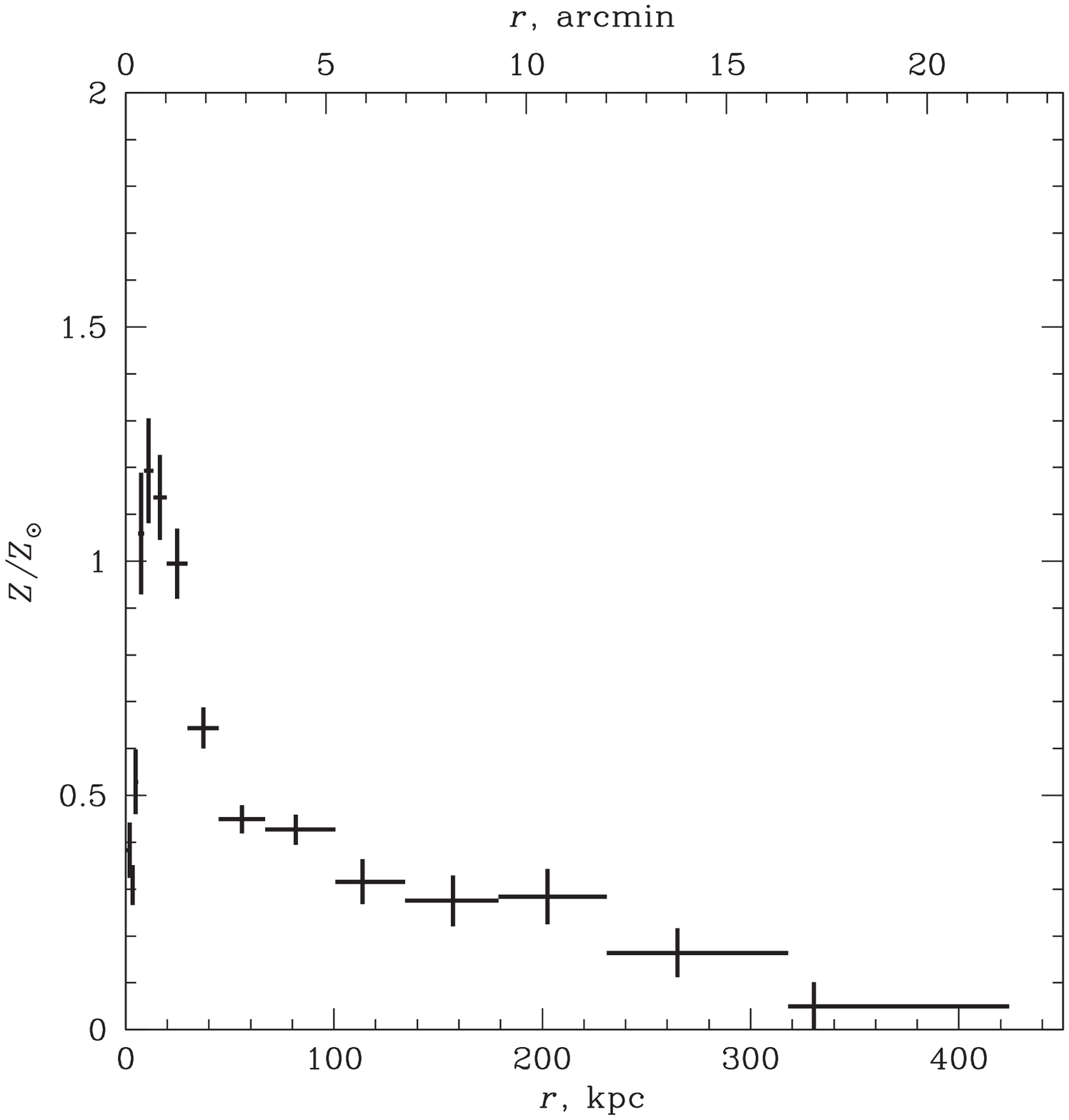}%
}
\vspace*{-0.2\baselineskip}
\caption{Temperature and metallicity profiles for A262.} 
\label{fig:a262}
\end{figure*}

\subsection{Abell 1795}
\label{sec:a1795}

Abell~1795 ($z=0.0622$), was observed by \emph{Chandra} multiple times,
within the Guaranteed Time program in 1999 and early 2000, and
subsequently in 6 pointings as a calibration target. We do not use the
1999 observation (OBSID 494) because accurate gain calibration is still
unavailable for that time period. The remaining 7 pointings, in which
the cluster was placed at different locations within the S3 and I3
chips, have a combined exposure of more than 100~ksec
(Fig.~\ref{fig:mosaic:img}). 

There is no significant variation of Galactic absorption with radius,
and the cluster-average value,
$N_H=(1.35\pm0.12)\times10^{20}$~cm$^{-2}$, is consistent with the value
from the radio surveys, $1.19\times10^{20}$~cm$^{-2}$. We do not detect
any signatures of incorrect subtraction of the soft background at the
largest radius covered by \emph{Chandra} pointings. Therefore, no
additional background correction is done, but a typical uncertainty in
normalization of such a component (see discussion on A133 and A1413
above) is still included in the final error budget. 

The temperature and metallicity profiles are shown in
Fig.~\ref{fig:a1795}. There is a significant temperature decrease at
large radii relative to the temperature just outside the central cool
region. The temperature drop is weaker than that in A133 and A1413, but
we note that \emph{Chandra} observations of A1795 cover a smaller
fraction of the cluster virial radius.

Multiple observations of A1795 at different locations on the detector
make this cluster an ideal case for testing the calibration
uncertainties. The temperatures derived independently from the data in
the BI and FI chips are shown in Fig.~\ref{fig:a1795} by green and blue
error bars, respectively. There is excellent agreement, even though
these datasets are from CCDs with substantially different quantum
efficiencies, and the data at the same distance from the cluster center
sample different detector regions and therefore different contamination
depths. 

Temperature profile for A1795 was the first one published from
\emph{XMM-Newton} \citep{2001A&A...365L..80A}. \emph{XMM-Newton} results
are compared with our measurements in
Fig.~\ref{fig:a1795:prof:XMM-Newton}. No obvious trend for the
temperature to decline with radius is suggested by the Arnaud et al.\
analysis (grey error bars), but the difference with our results is
within their statistical uncertainties. A subsequent reanalysis of the
same \emph{XMM-Newton} data by \cite{Nevalainen2004}, who performed a
more rigorous background flare exclusion, produced a temperature profile
fully consistent with our results (blue error bars with caps). 

\begin{figure*}
\vspace*{-\baselineskip}
\centerline{%
\includegraphics[width=0.485\linewidth]{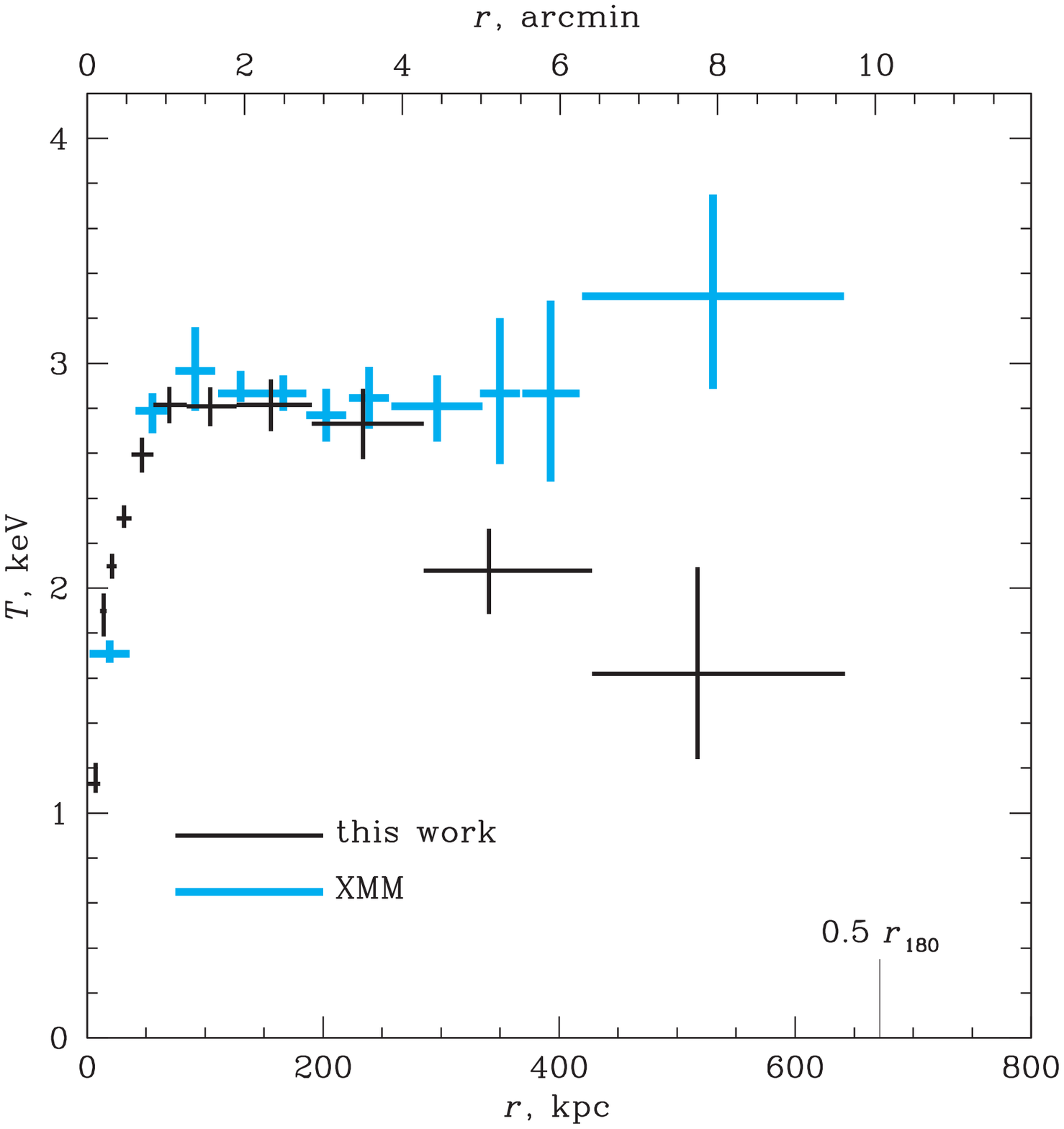}\hfill%
\includegraphics[width=0.485\linewidth]{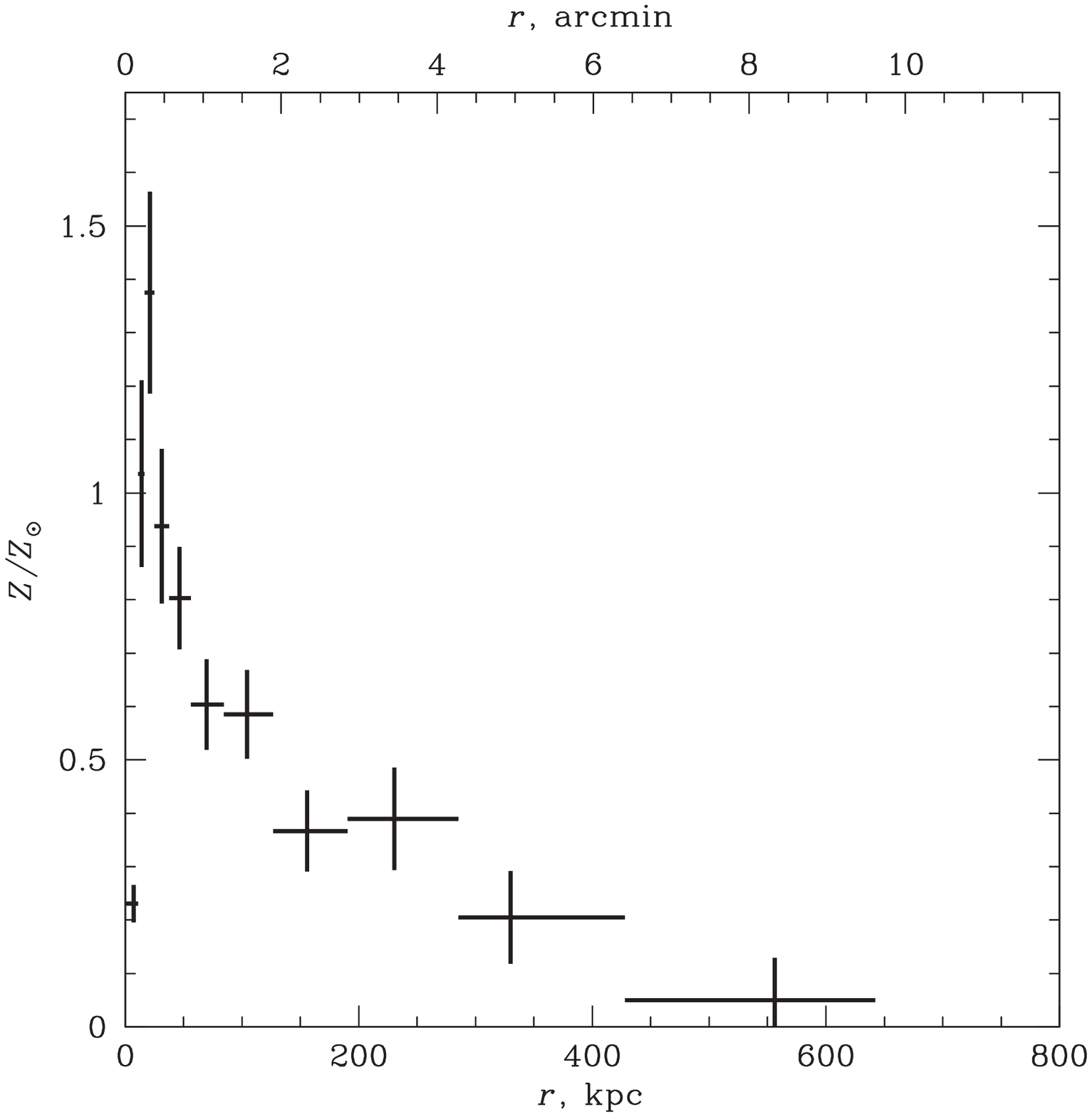}%
}
\vspace*{-0.2\baselineskip}
\caption{Temperature and metallicity profile for A1991. \emph{XMM-Newton} results from
  \cite{PrattArnaud2004} are shown in blue. Note that \emph{XMM-Newton} values were
  renormalized by $+10\%$ to account for cross-calibration (see
  \S~\ref{sec:a1413}).} 
\label{fig:a1991}
\end{figure*}

\subsection{Abell 262}

Abell~262 ($z=0.0162$, $T\simeq2$~keV) was observed in ACIS-S for 30~ksec. 
The X-ray image (Fig.~\ref{fig:mosaic:img}) on large scales in nearly
symmetric, but there is substructure in the very central region, probably
related to past activity of the central radio source \citep{Blanton2004}. We
include data from the S1 chip in this observation.  This CCD is not used for
temperature measurements in hot clusters, because its particle-induced
background is very high above 5~keV, but S1's performance is comparable to
that of S3 for clusters with $T\lesssim 2$~keV, such as A262. 

We do not find any problems with the soft background subtraction in the
A262 observation. The cluster brightness is sufficiently high everywhere
in the field of view and therefore any reasonable background variations
have negligible effect on the derived temperatures. The best-fit
Galactic absorption is constant with radius within the measurement
uncertainties but its average value,
$N_H=(8.1\pm0.3)\times10^{20}$~cm$^{-2}$ is significantly higher than
the radio value, $5.4\times10^{20}$~cm$^{-2}$. Excess absorption in this
cluster was also found in the \emph{ROSAT} data
\citep{1996ApJ...473..692D}. David et al.\ demonstrated that the excess
absorption can be fully explained by intervening Galactic cirrus visible
in the \emph{IRAS} $100\,\mu$ image. We use the best-fit X-ray value of
$N_H$ in the further analysis. 

The temperature and metallicity profiles of A262 are shown in
Fig.~\ref{fig:a262}. Note a strong central spike of metallicity, which
coincides with the temperature decrement within the inner 50~kpc. There
is a significant temperature decrease at $r>100-200$~kpc. Note that in this
case, we use a very different portion of the ACIS energy band for the
temperature determination and so systematics should be very different
from those in the hotter clusters. It is reassuring, therefore, that we
observe a qualitatively similar temperature structure.

\begin{figure*}
\vspace*{-\baselineskip}
\centerline{%
\includegraphics[width=0.485\linewidth]{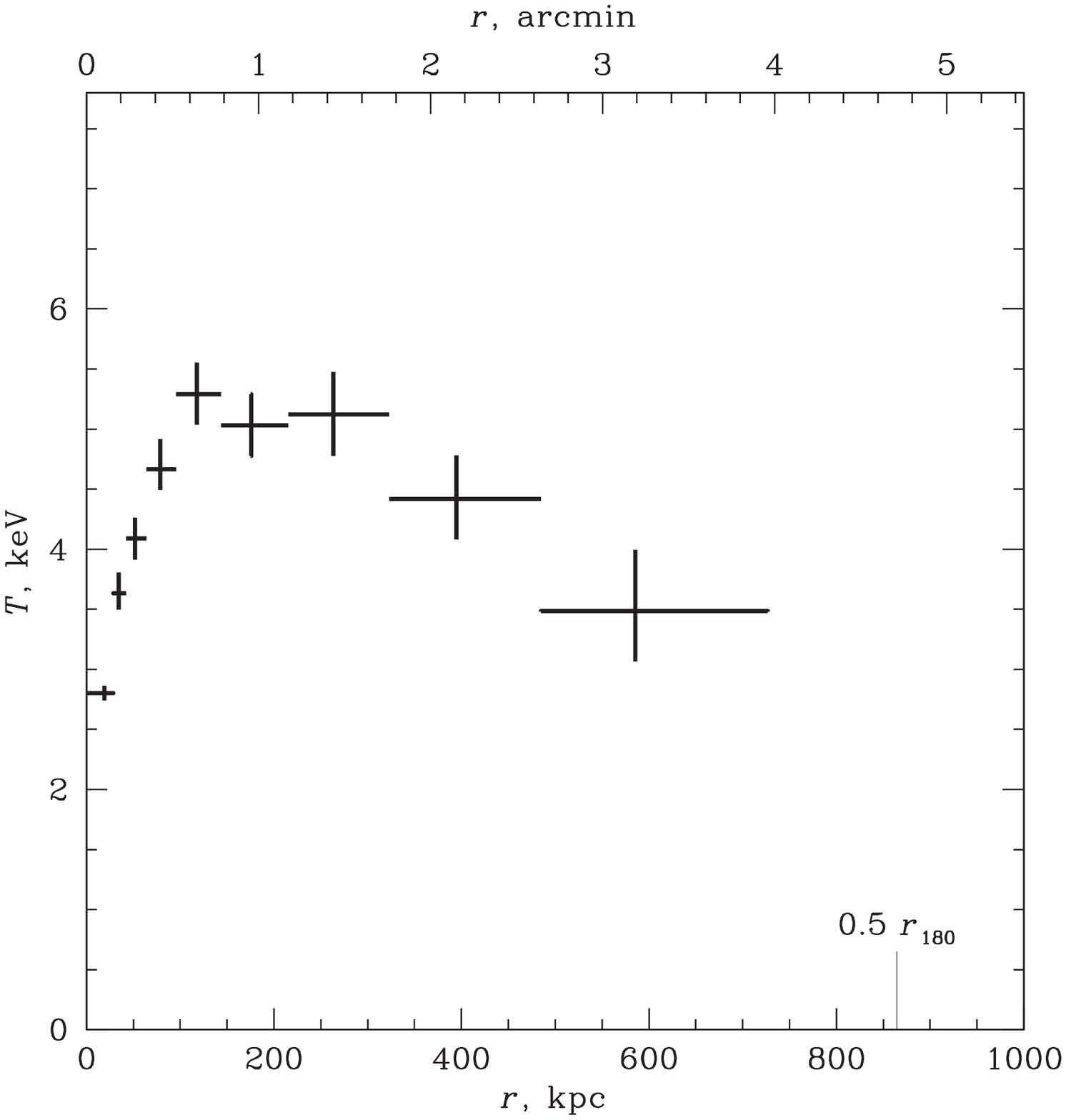}\hfill%
\includegraphics[width=0.485\linewidth]{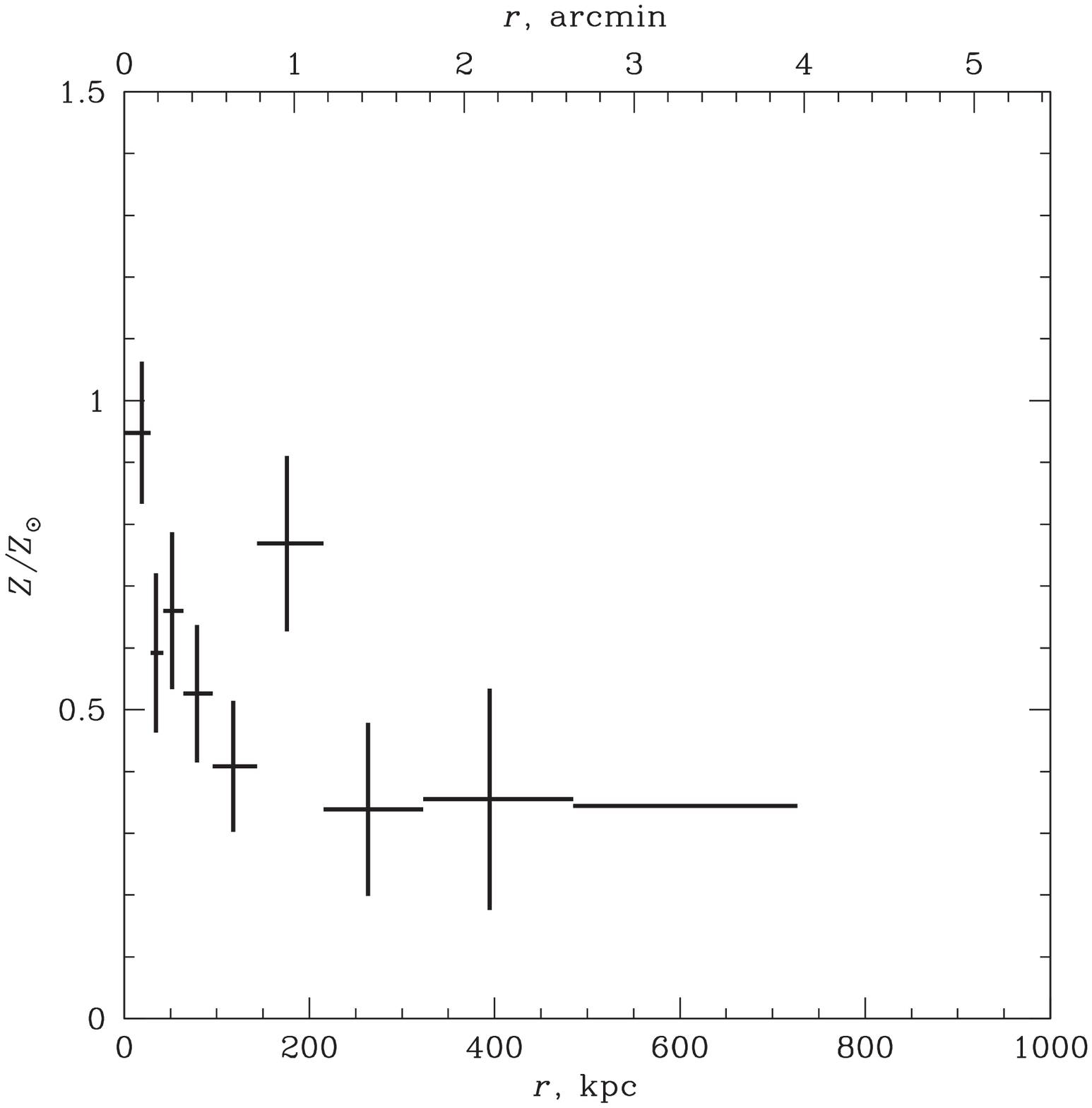}%
}
\vspace*{-0.2\baselineskip}
\caption{Temperature and metallicity profiles for A383.} 
\label{fig:a383}
\end{figure*}

\subsection{Abell 1991}

Abell~1991 is a $T\simeq 2.8$~keV cluster at $z=0.0592$ and its virial
radius fits almost entirely within the \emph{Chandra} field of
view. A1991 was observed for 40~ksec in ACIS-S
(Fig.~\ref{fig:mosaic:img}). \cite{2004ApJ...613..180S} report
substructure within the central $10''$ but otherwise, the X-ray image is
very symmetric. 

\begin{figure*}
\vspace*{-\baselineskip}
\centerline{%
  \includegraphics[width=0.485\linewidth]{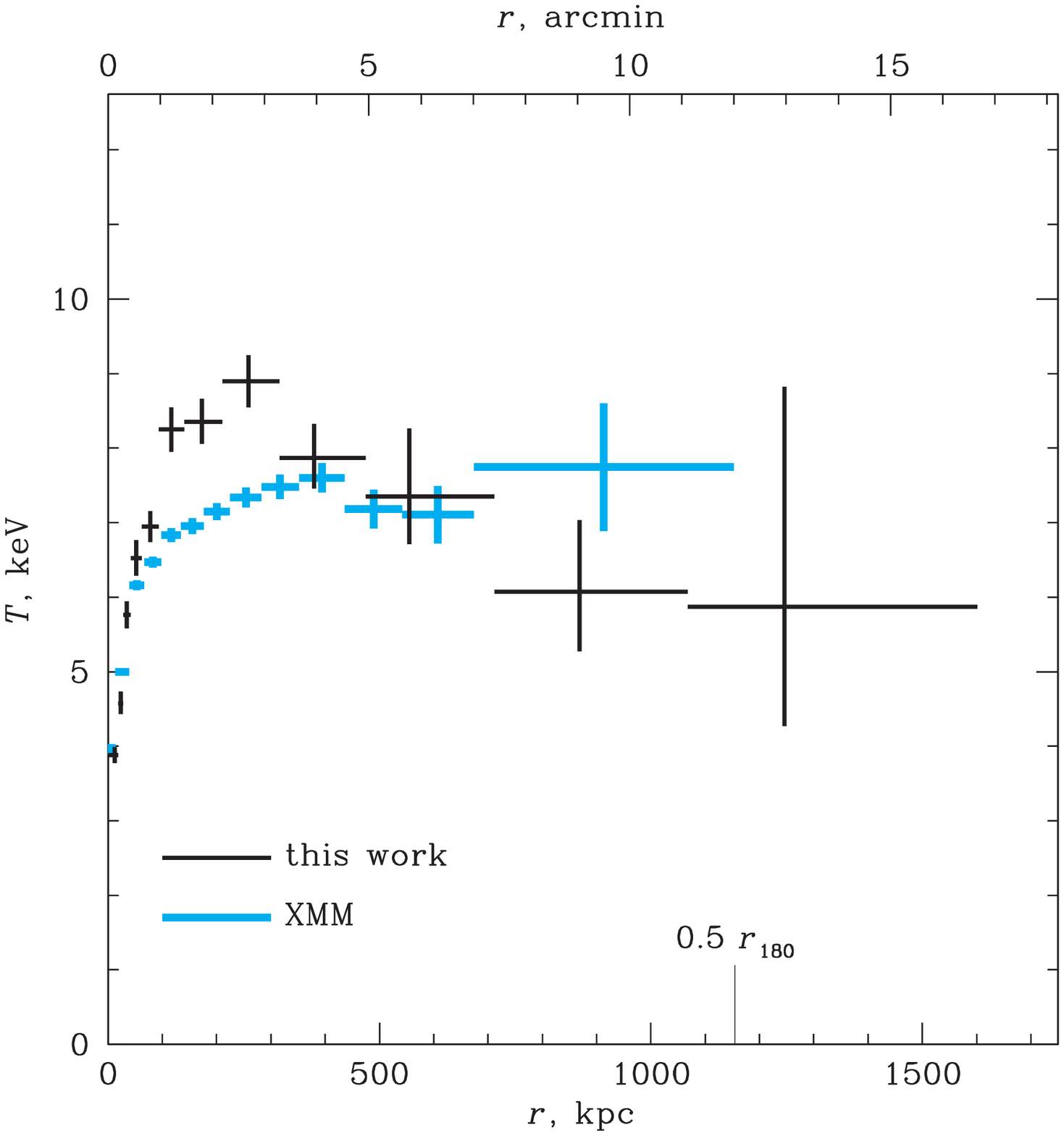}\hfill%
  \includegraphics[width=0.485\linewidth]{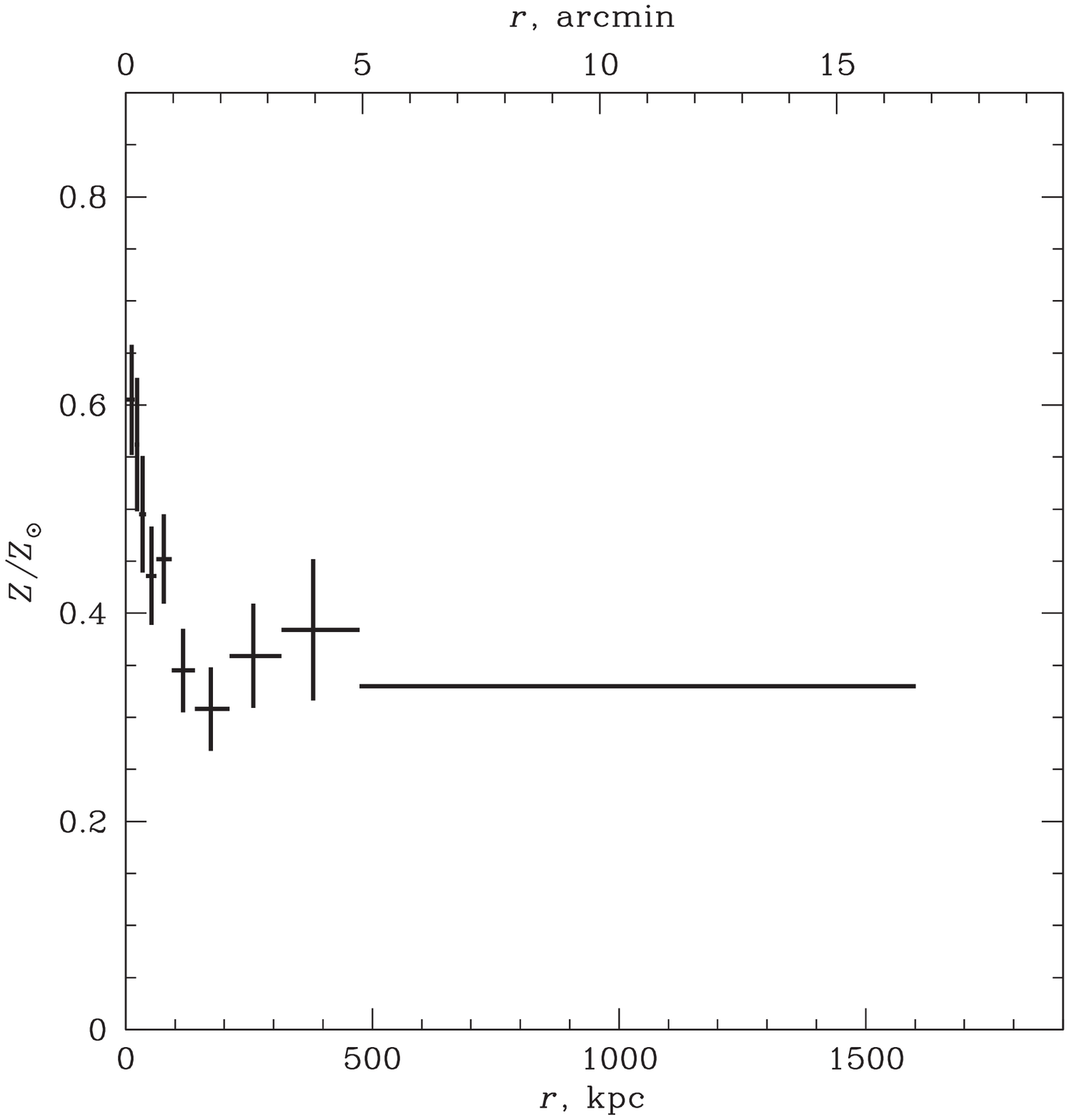}%
}
\vspace*{-0.2\baselineskip}
\caption{Temperature and metallicity profiles for A478. The \emph{XMM-Newton}
  measurements (renormalized by $+10\%$, see \S\,\ref{sec:a1413}) are shown
  in blue.} 
\label{fig:a478}
\end{figure*}

\begin{figure*}
\vspace*{-\baselineskip}
\centerline{%
\includegraphics[width=0.485\linewidth]{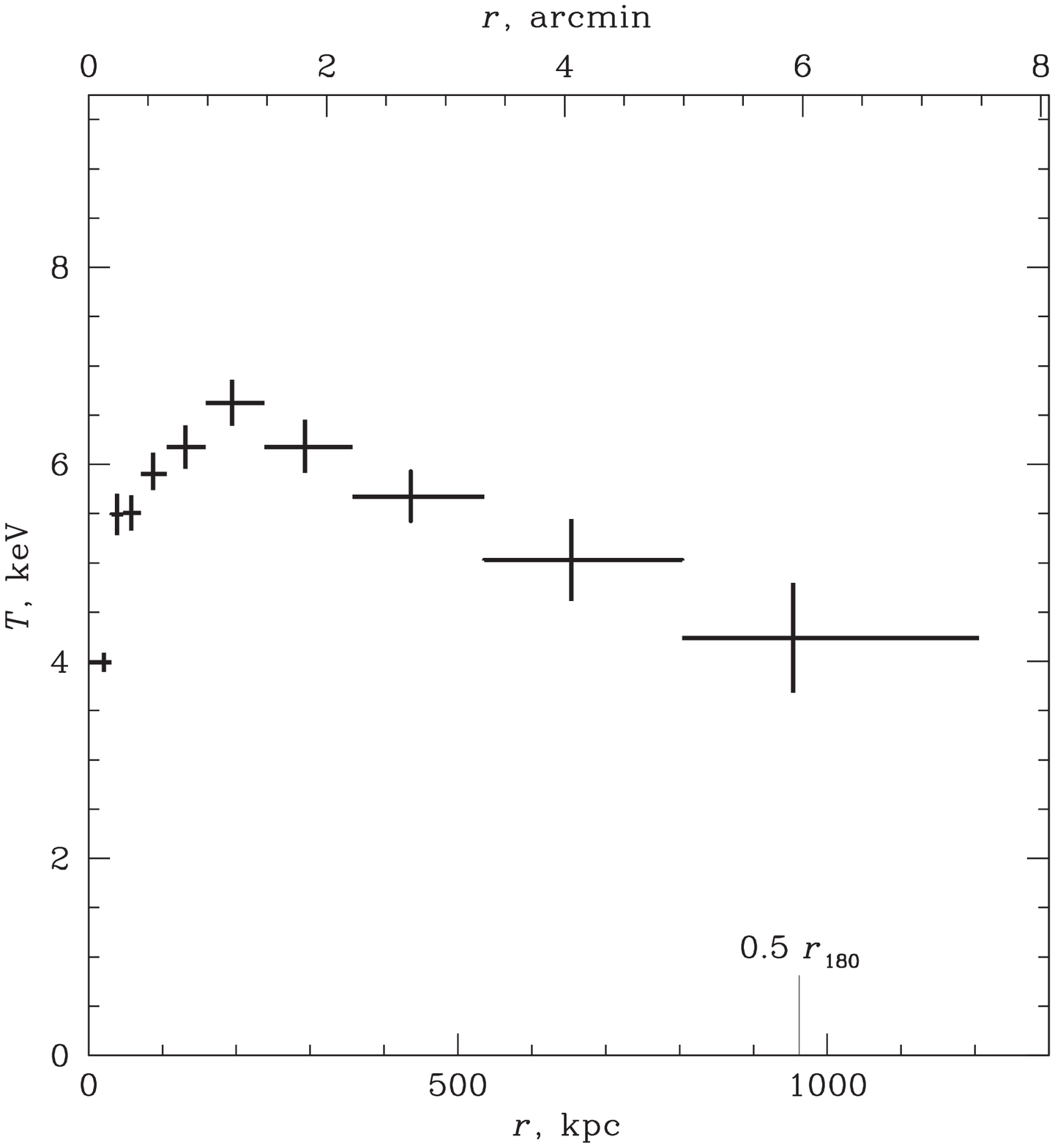}\hfill%
\includegraphics[width=0.485\linewidth]{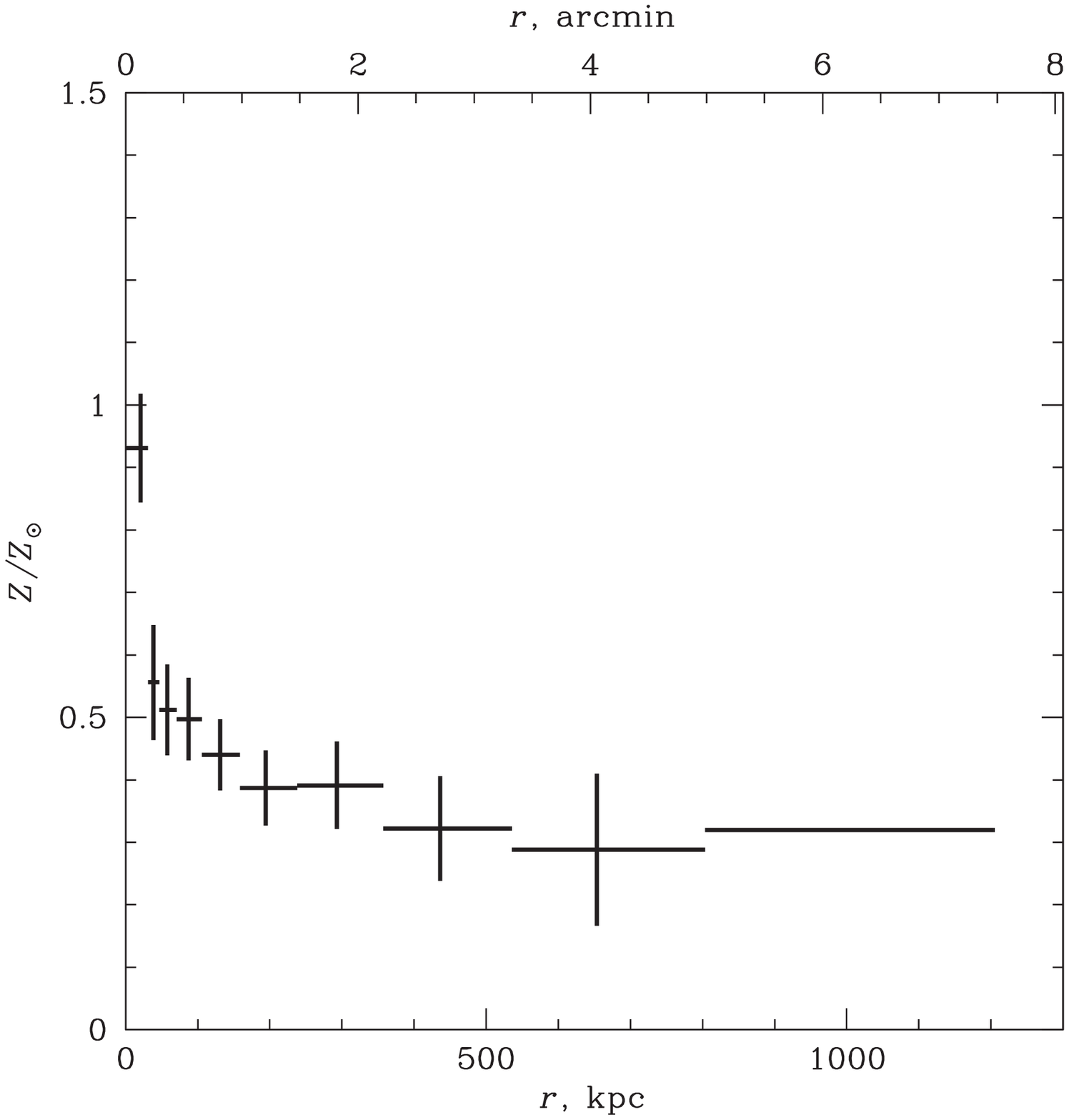}%
}
\vspace*{-0.2\baselineskip}
\caption{Temperature and metallicity profiles for A907.} 
\label{fig:a907}
\end{figure*}

Absorption, derived from the X-ray spectra, is constant with radius and
consistent with the radio value, $N_H=2.45\times10^{20}$~cm$^{-2}$. The
cluster is projected on the North Galactic Spur and in this field there
is a strong excess flux at low energies. Fortunately, the spectrum and
normalization of this soft background component can be determined from
the same observation because the \emph{Chandra}'s field of view extends
to $20'$ from the cluster center, where the object surface brightness is
negligible. Galactic foreground was measured independently from the
0.4--1.5~keV spectra in the $9.5'-14.5'$ and $14.5'-20'$ radial ranges
extracted in the S1 and ACIS-I chips. The excess flux requires two
components with different $T$, similar to results for the Chandra blank
fields in the Spur area \citep{2003ApJ...583...70M}. The best-fit
temperatures are $T_1=0.157\pm0.029$~keV and $T_2=0.51\pm0.15$~keV, and
relative emission measures are $K_2/K_1=0.285$. The normalization
corresponds to a flux of $2.45\times10^{-16}$~erg~s$^{-1}$~cm$^{-2}$ per
arcsec$^2$ in the 0.4--1.5~keV energy band. The normalizations are
consistent in all four spectra, indicating that this soft component
indeed represents a uniform foreground and is not related to the
cluster. 

The temperature and metallicity profiles of A1991 are shown in
Fig.~\ref{fig:a1991}. The spectrum in the innermost bin clearly requires
several temperature components and therefore the derived metallicity for
a single-temperature fit is artificially low. 

\emph{XMM-Newton} results for A1991 were recently published by
\cite{PrattArnaud2004}. They derive an essentially isothermal profile,
which disagrees with our results at $r>300$~kpc at a $\sim 3\sigma$
confidence level (blue points in Fig.~\ref{fig:a1991}). The difference
in modeling of the excess soft background in this field and background
flares are obvious suspects for the discrepancy between the
\emph{XMM-Newton} and \emph{Chandra} results; unfortunately, Pratt \&
Arnaud do not provide information necessary for a more detailed
comparison.

\subsection{Abell 383}

Abell~383 ($z=0.1883$) was observed with \emph{Chandra} in 3 pointings,
for 20 and 10 ksec in ACIS-I, and for 20~ksec in ACIS-S
(Fig.~\ref{fig:mosaic:img}). There are faint extended X-ray sources in
the X-ray surface brightness at $5.5'$ to the South-East and at $\sim
7'$ to the North-West, probably related to infalling subgroups. They
were excluded from the radial profile analysis. 

Galactic absorption derived from the X-ray spectra is constant with
radius and consistent with the radio value,
$3.92\times10^{20}$~cm$^{-2}$. The spectrum extracted in the outermost
regions of the field of view shows negative residuals near 0.6~keV,
similar to those in the A133 observation
(Fig.~\ref{fig:a133:soft:spec}). The residuals can be fit with the MEKAL
model with $T=0.18$~keV and a negative normalization corresponding to a
0.7--2~keV count rate of
$(-1.5\pm0.3)\times10^{-6}$~cnt~s$^{-1}$~arcmin$^{-2}$. This component
was included in the spectral fits. The temperature and abundance
profiles for A383 are shown in Fig.~\ref{fig:a383}. 

\subsection{Abell 478}

Abell~478, one of the highest-temperature clusters in our sample, was
observed in ACIS-S for 43~ksec (Fig.~\ref{fig:mosaic:img}). In the central
$30''$ there is substructure associated with activity of the central AGN but
otherwise, the cluster is symmetric.  \emph{Chandra} analysis of the central
$4'$ was presented by \cite{2003ApJ...587..619S}.  Our analysis supersedes
the temperature profile in that paper because we use updated calibrations. 
This cluster was also studied by \emph{XMM-Newton}
\citep{2004A&A...423...33P}. 

The spectra from the outermost regions of the field of view, which are
almost free from cluster emission, do not show any significant residuals
in the soft band. However, there are significant variations in the
Galactic absorption. The best-fit $N_H$ changes linearly with radius
from $(3.09\pm0.09)\times10^{21}$~cm$^{-2}$ at $r=0$ to
$(2.70\pm0.06)\times10^{21}$~cm$^{-2}$ at $r=4'$; these are
significantly in excess of the radio value,
$1.51\times10^{21}$~cm$^{-2}$. At $r>5'$, $N_H$ drops quickly, reaching
$1.5\times10^{21}$~cm$^{-2}$ at $r\approx 8'$ and stays constant at
larger radii.  Pointecouteau et al.\ obtained very similar results from
the \emph{XMM-Newton} data and argued that the excess absorption is most
likely caused by a Galactic dust cloud projected on one side of the
cluster, including the center. Since absorption is a strong function of
radius in this case, we leave it free in all spectral fits. The
absorption distribution may not be azimuthally symmetric relative to the
cluster center and using an average value in an annulus is not
accurate. We ignore this problem because $N_H$ variations are slow
inside the central $5'$ and at larger radii, the ACIS field of view
covers only the South-West sector (Fig.~\ref{fig:mosaic:img}).

The temperature and metallicity profiles are shown in
Fig.~\ref{fig:a478}. The abundance is unconstrained by the
\emph{Chandra} data in the individual annuli at $r>500$~kpc and we fixed
it at the average value in the $r=500-1600$~kpc range, $Z=0.33$. Any
reasonable variations of metallicity do not affect the temperature
determination in hot clusters such as A478. As in other clusters, we
observe a strong central temperature decrement, a peak at $r\sim
200-300$~kpc where $T\simeq8.5$~keV, and a gradual decline to $T\simeq
6$~keV at $r>700$~kpc. The \emph{XMM-Newton} temperature profile is
shown in blue in the left panel of Fig.~\ref{fig:a478}. It is
qualitatively different from \emph{Chandra} profile --- the temperature
gradually increases with radius to $r\sim 4'$ and then stays
approximately constant. The biggest difference between the two
instruments, however, is within the central~$3'$. \emph{Chandra} results
suggest a compact ($r<1'$) cold central region surrounded by a hotter
annulus, while \emph{XMM-Newton} observes a more distributed central
temperature decrement. Outside the central~$3'$, \emph{Chandra} and
\emph{XMM-Newton} temperatures are in reasonable agreement. We note that
A478 has one of the most centrally-peaked X-ray brightness profiles in
our sample. Therefore, is the prime suspect for the temperature
discrepancy in the central region is a combination of incomplete
correction for the PSF effects in the \emph{XMM-Newton} analysis and the
complex temperature structure of the central 100~kpc region in this
cluster \citep{astro-ph/0412316}.

\subsection{Abell 907}

Abell~907 ($z=0.1603$) was observed in 3 separate ACIS-I pointings for
49, 35, and 11 ksec, designed specifically for this measurement
(Fig.~\ref{fig:mosaic:img}). There are no significant radial variations
of the Galactic absorption, but the cluster-averaged value,
$N_H=(3.87\pm0.2)\times10^{20}$~cm$^{-2}$, is significantly lower than
that from the radio data, $5.4\times10^{20}$~cm$^{-2}$. We use the X-ray
value in our analysis. There are also negative residuals near 0.6~keV in
the spectra from the outermost regions. The corresponding background
correction can be modeled as a MEKAL spectrum with $T=0.278$~keV and
normalization corresponding to the 0.7--2~keV flux
$(-4.8\pm1.3)\times10^{-6}$~cnt~s$^{-1}$~arcmin$^{-2}$. The derived
temperature and metallicity profiles are shown in Fig.~\ref{fig:a907}.

\subsection{Abell 2029}

Abell 2029 ($T=8.5$~keV, $z=0.078$) was observed twice in ACIS-S for 10
and 80~ksec (both times in Faint mode) and in ACIS-I for 10~ksec (VFaint
mode). We use all three pointings. The X-ray morphology of this cluster
is very regular outside the central $\sim 50$ kpc.  \emph{Chandra} data
for the central region of A2029 was used by \cite{2004ApJ...616..178C}
to study interaction between X-ray and radio-emiting plasma, and by
Lewis et al.\ (2003) to derive the inner dark matter profile. 

\begin{figure}[t]
\vspace*{-1.3\baselineskip}
\centerline{\hspace*{5mm}\includegraphics[width=0.485\textwidth]{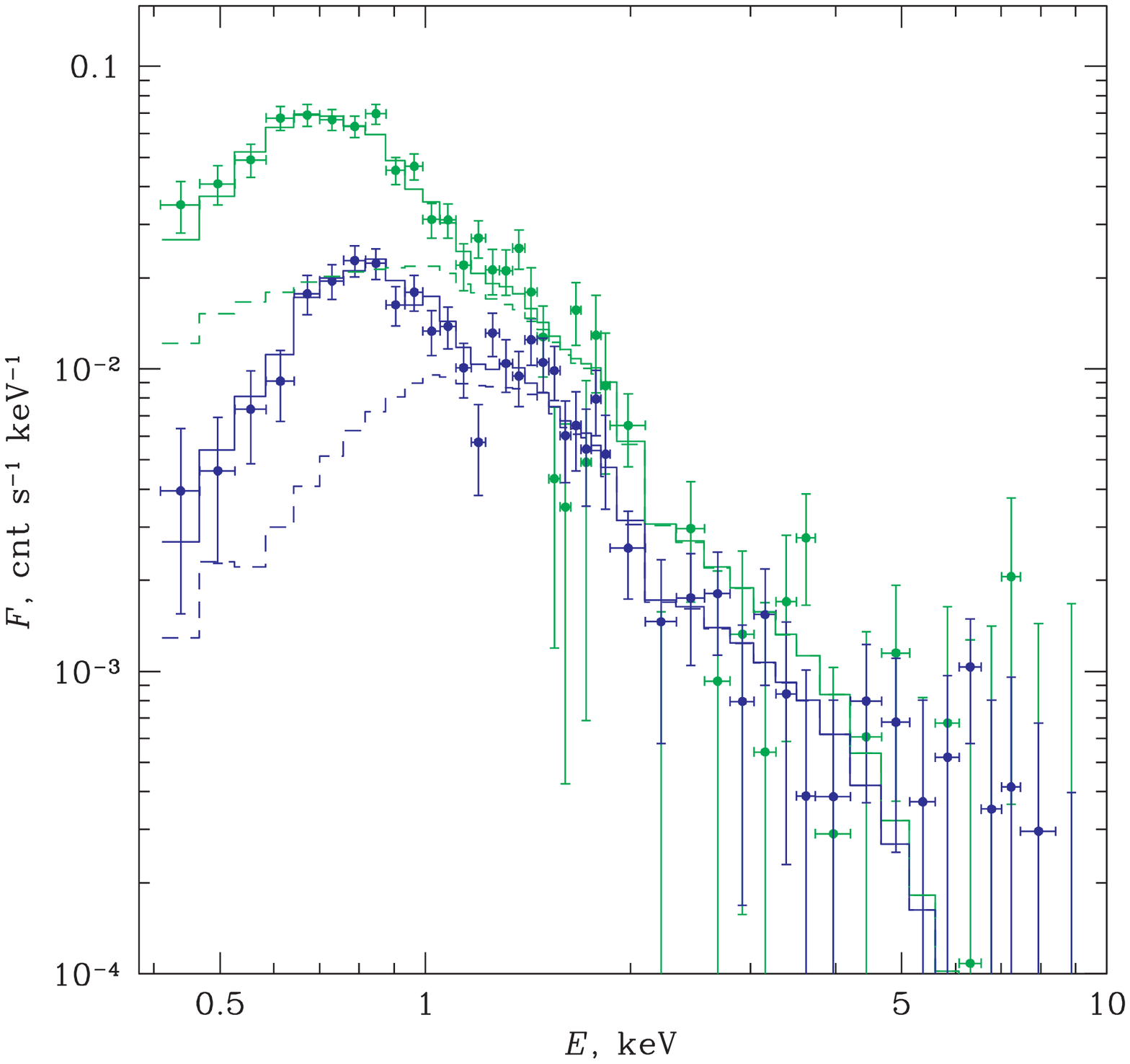}}
\vspace*{-1.0\baselineskip}
\caption{Observed spectra in the outermost region of A2029 ($r>14.5'$). 
  Data from S1 and FI chips are shown in green and blue, respectively. 
  The model includes the cluster emission and a 2-temperature Galactic
  component (see text for details). Total fit is shown by solid
  histograms. Dashed lines show contribution of the cluster component.} 
\label{fig:a2029:soft:spec}
\end{figure}

The best-fit Galactic absorption everywhere in the central region is
consistent with the radio value, $3.0\times10^{20}$~cm$^{-2}$. The major
complication for the \emph{Chandra} analysis of A2029 is high Galactic
soft X-ray foreground (A2029 is projected on the Northern Galactic
Spur). Because of the cluster proximity, its emission is non-negligible
everywhere in the ACIS field of view. Therefore, we had to fit the
Galactic flux jointly with the cluster emission using the spectrum
integrated at $r>14.5'$ from the cluster center. The cluster and
Galactic flux are easy to separate spectrally because the latter is much
softer (Fig.~\ref{fig:a2029:soft:spec}). We find that a 2-temperature
MEKAL model is required to represent the Galactic foreground. The
situation is similar to that in A1991, another cluster in our sample
projected on the Northern Galactic Spur. Its spectrum and normalization
is determined from the joint fit to spectra extracted in the S1 and FI
chips. The model includes an absorbed MEKAL spectrum to represent the
cluster, and two additional unabsorbed MEKAL components with the Solar
metallicity to represent the foreground. Spectral parameters for all
components are tied for S1 and FI datasets; the normalizations of the
cluster component are independent, while those of the Galactic
components are tied to be proportional to the area covered. The best fit
provides an excellent description of both datasets (solid lines in
Fig.~\ref{fig:a2029:soft:spec}). The best-fit temperatures of the
Galactic components are $T_1=0.22\pm0.05$~keV, and
$T_2=0.46\pm0.07$~keV, and the ratio of emission measures is
$K_2/K1=0.53$. These spectral parameters are similar to those derived
for A1991 and also for \emph{Chandra} blank field in the Spur region
\citep{2003ApJ...583...70M}, which gives us confidence that the Galactic
flux has been measured correctly. The best-fit temperature of the
cluster component in this region is $T=3.4^{+1.3}_{-0.9}$~keV (including
systematic uncertainties of the quiescent detector background). 

\begin{figure*}
\vspace*{-\baselineskip}
\centerline{%
\includegraphics[width=0.485\linewidth]{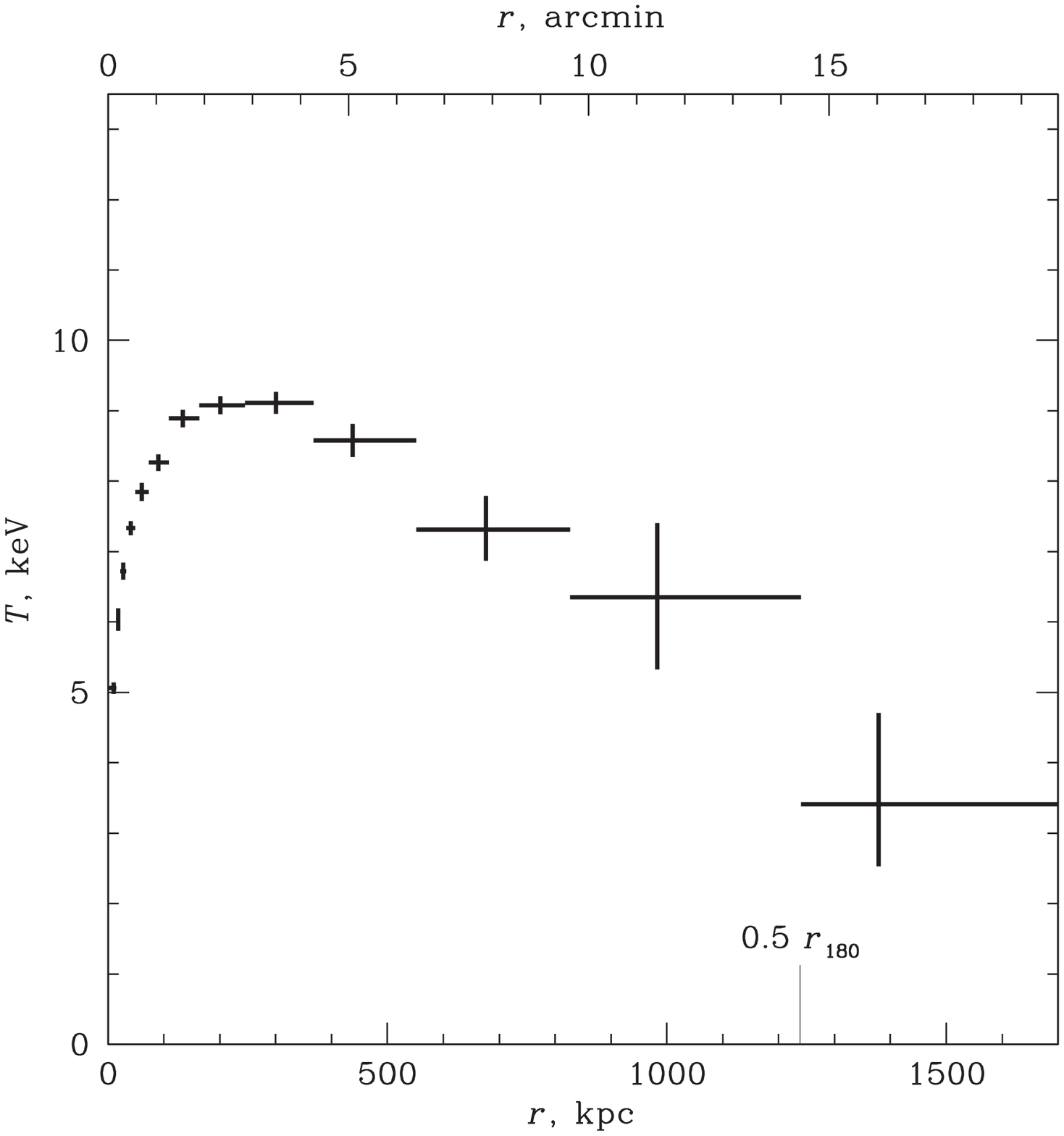}\hfill%
\includegraphics[width=0.485\linewidth]{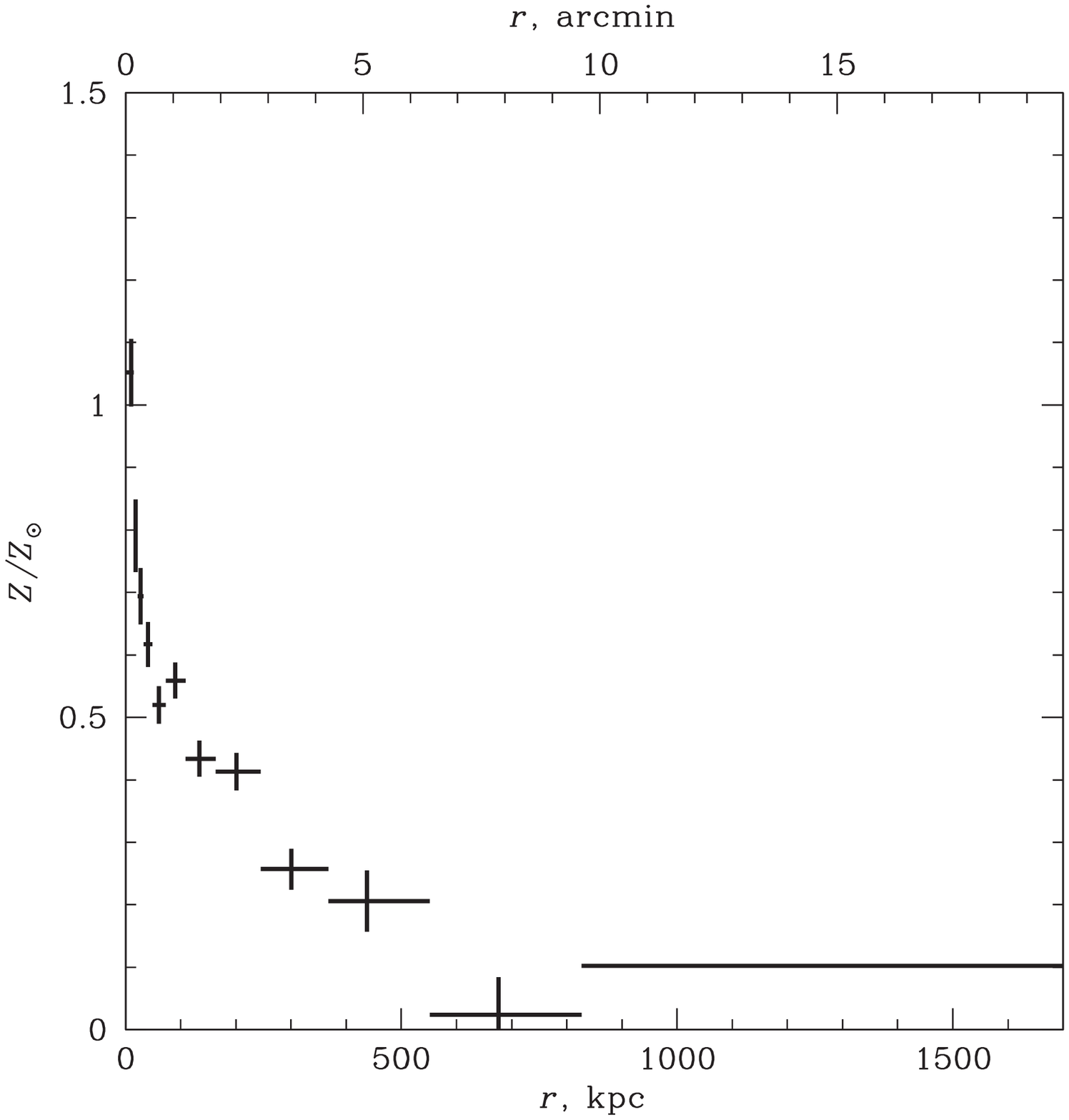}%
}
\vspace*{-0.2\baselineskip}
\caption{Temperature and metallicity profiles for A2029.} 
\label{fig:a2029}
\end{figure*}

Temperature fits in other regions are performed with normalization of
the Galactic component scaled by area. The temperature and abundance
profiles are shown in Fig.\ref{fig:a2029}. Despite difficulties with the
Galactic foreground modeling, A2029 has the most accurately measured
temperature profile among hot, $T\gtrsim 8$~keV, clusters in our
sample.

\begin{figure*}
\vspace*{-\baselineskip}
\centerline{%
\includegraphics[width=0.485\linewidth]{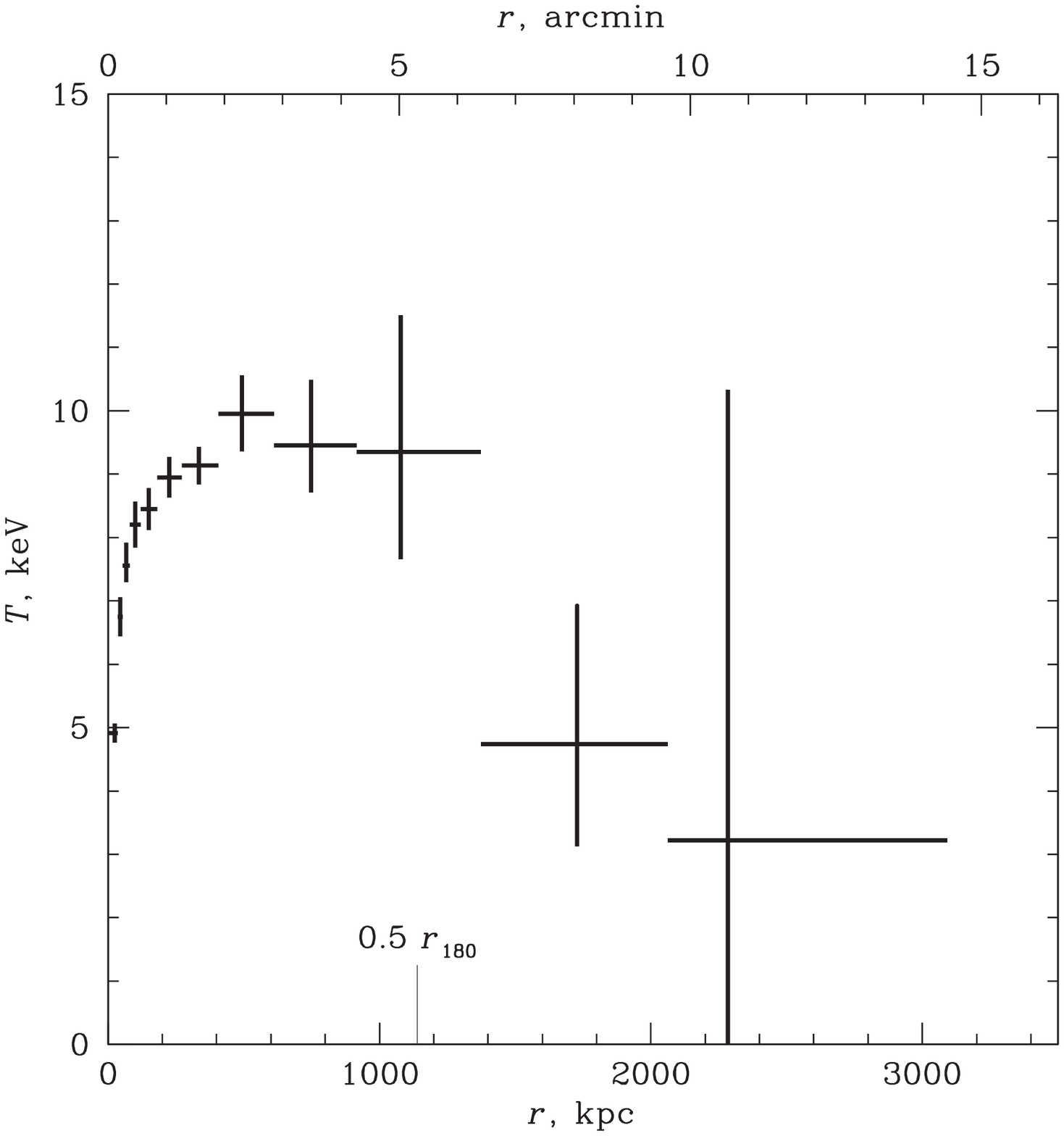}\hfill%
\includegraphics[width=0.485\linewidth]{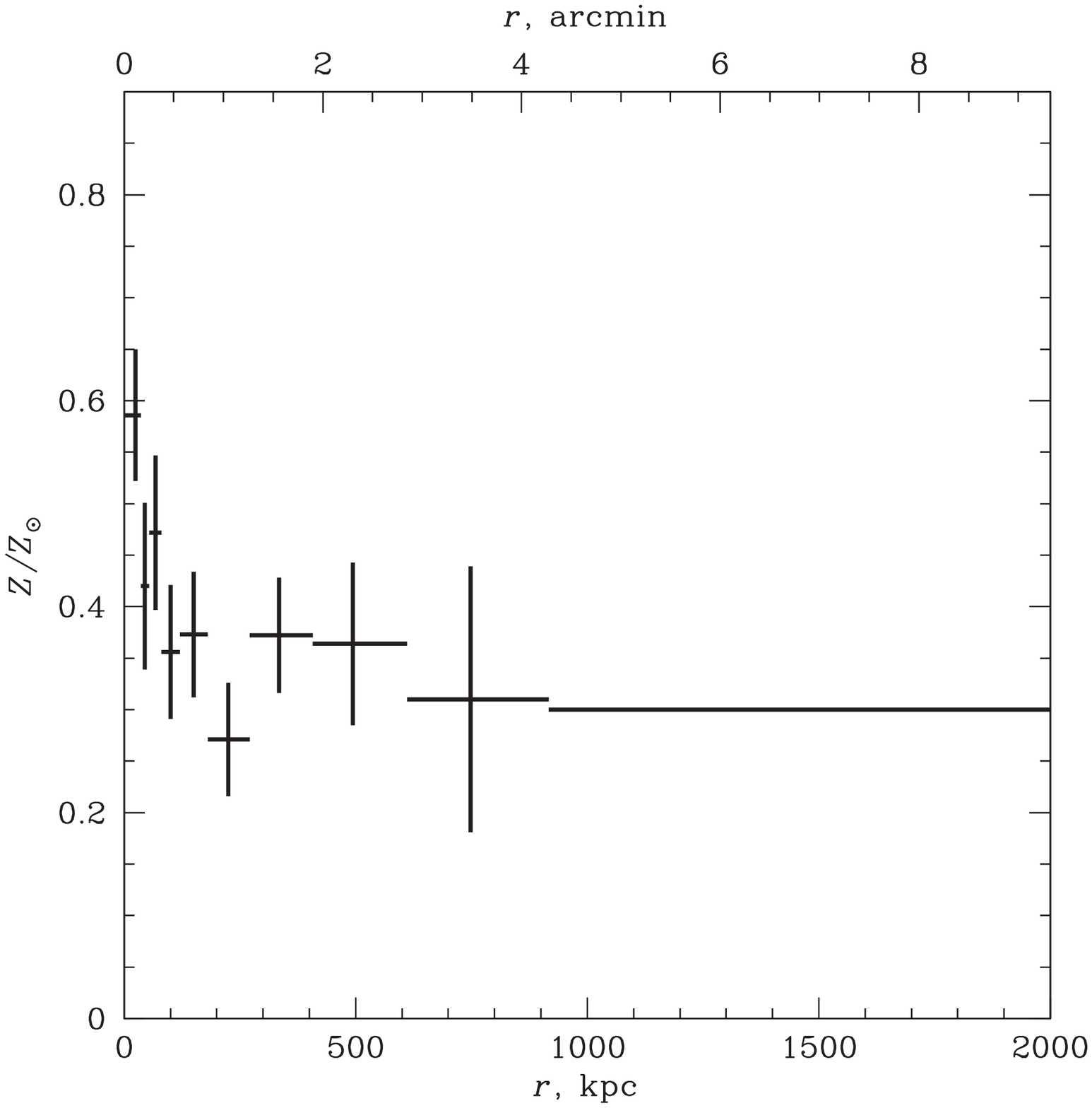}%
}
\vspace*{-0.2\baselineskip}
\caption{Temperature and metallicity profiles for A2390.} 
\label{fig:a2390}
\end{figure*}

\begin{figure*}
\vspace*{-\baselineskip}
\centerline{%
\includegraphics[width=0.485\linewidth]{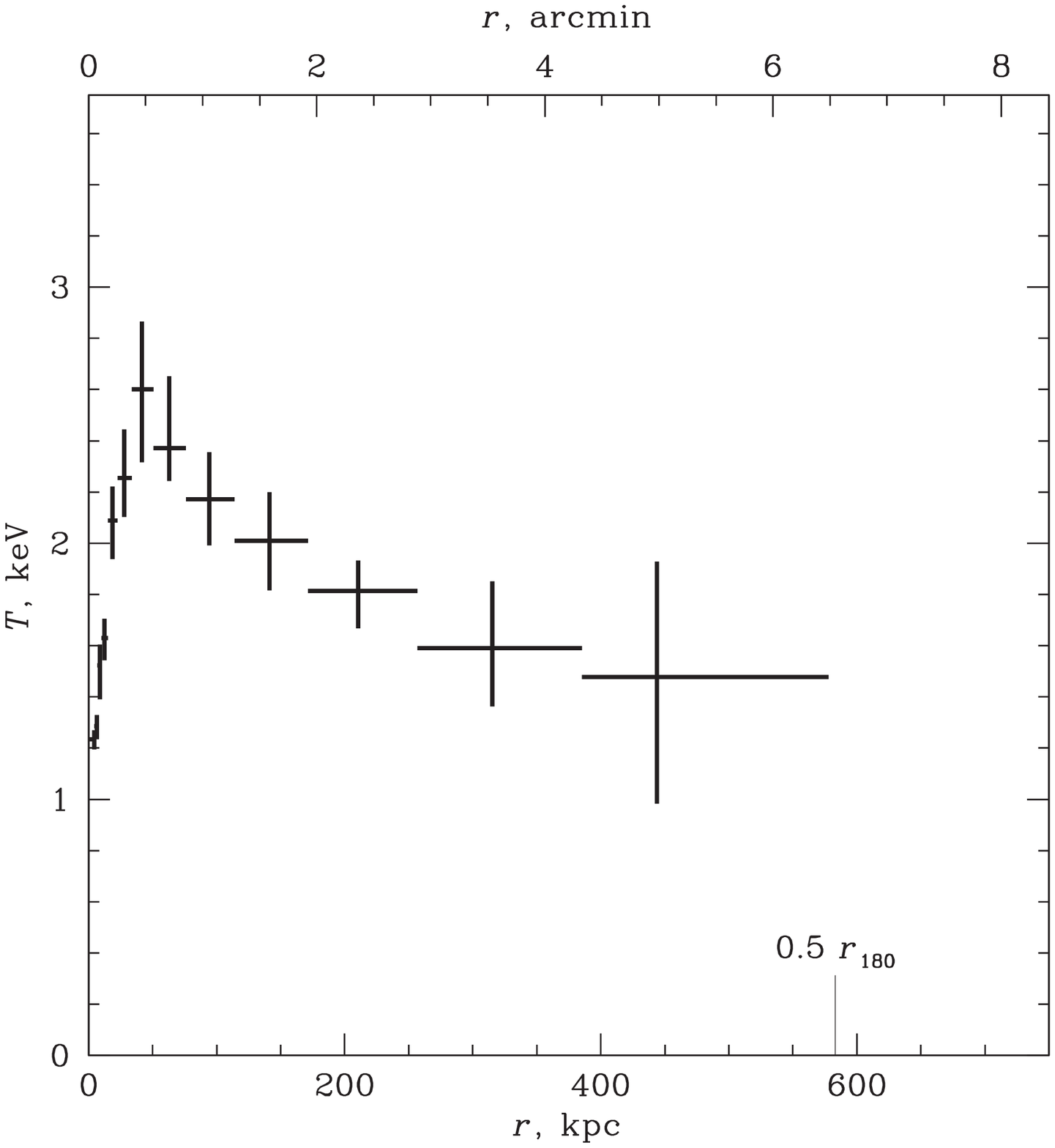}\hfill%
\includegraphics[width=0.485\linewidth]{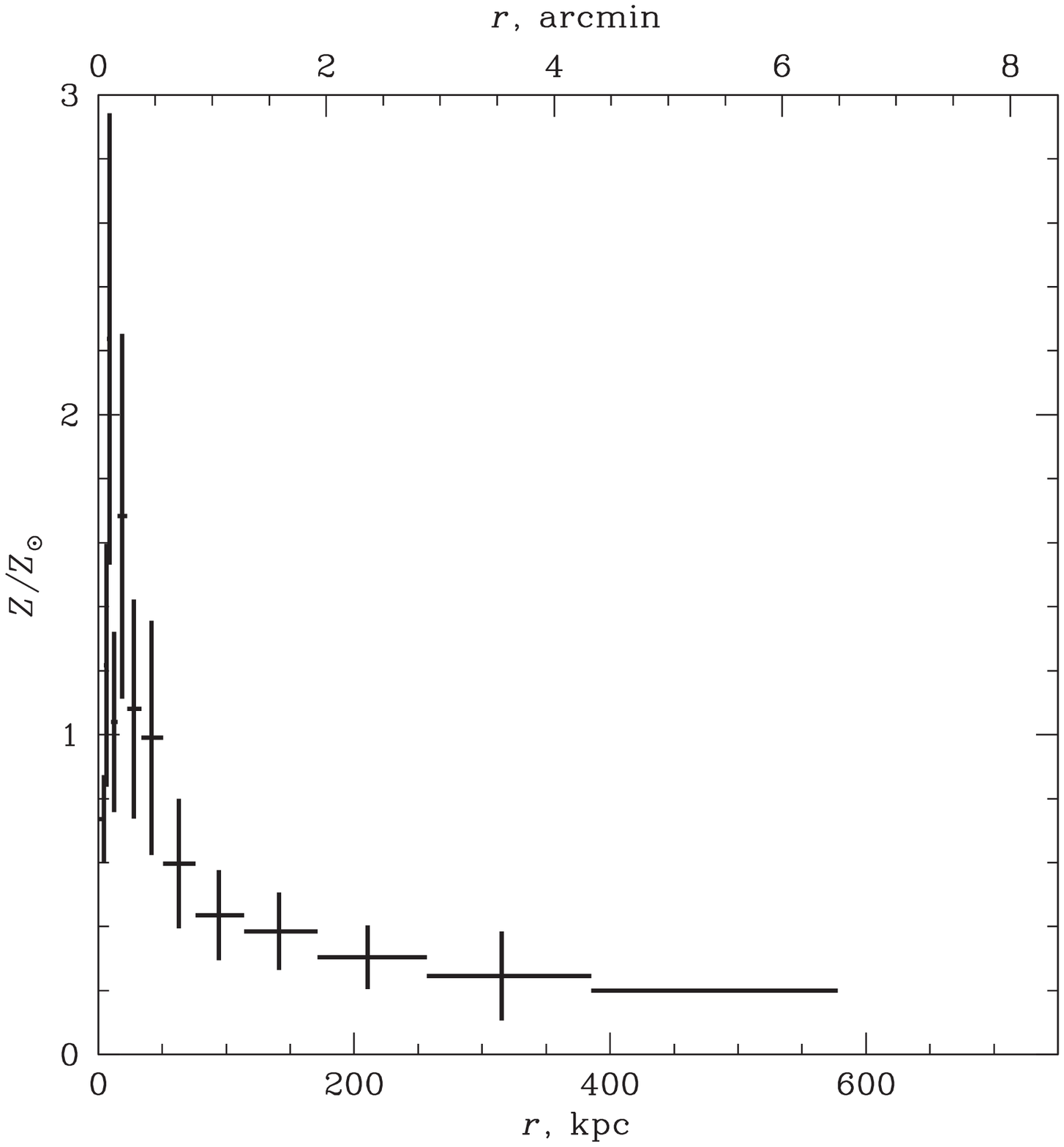}%
}
\vspace*{-0.2\baselineskip}
\caption{Temperature and metallicity profiles for RXJ~1159+5531}
\label{fig:cl1159}
\end{figure*}

\begin{figure*}
\vspace*{-\baselineskip}
\centerline{%
\includegraphics[width=0.485\linewidth]{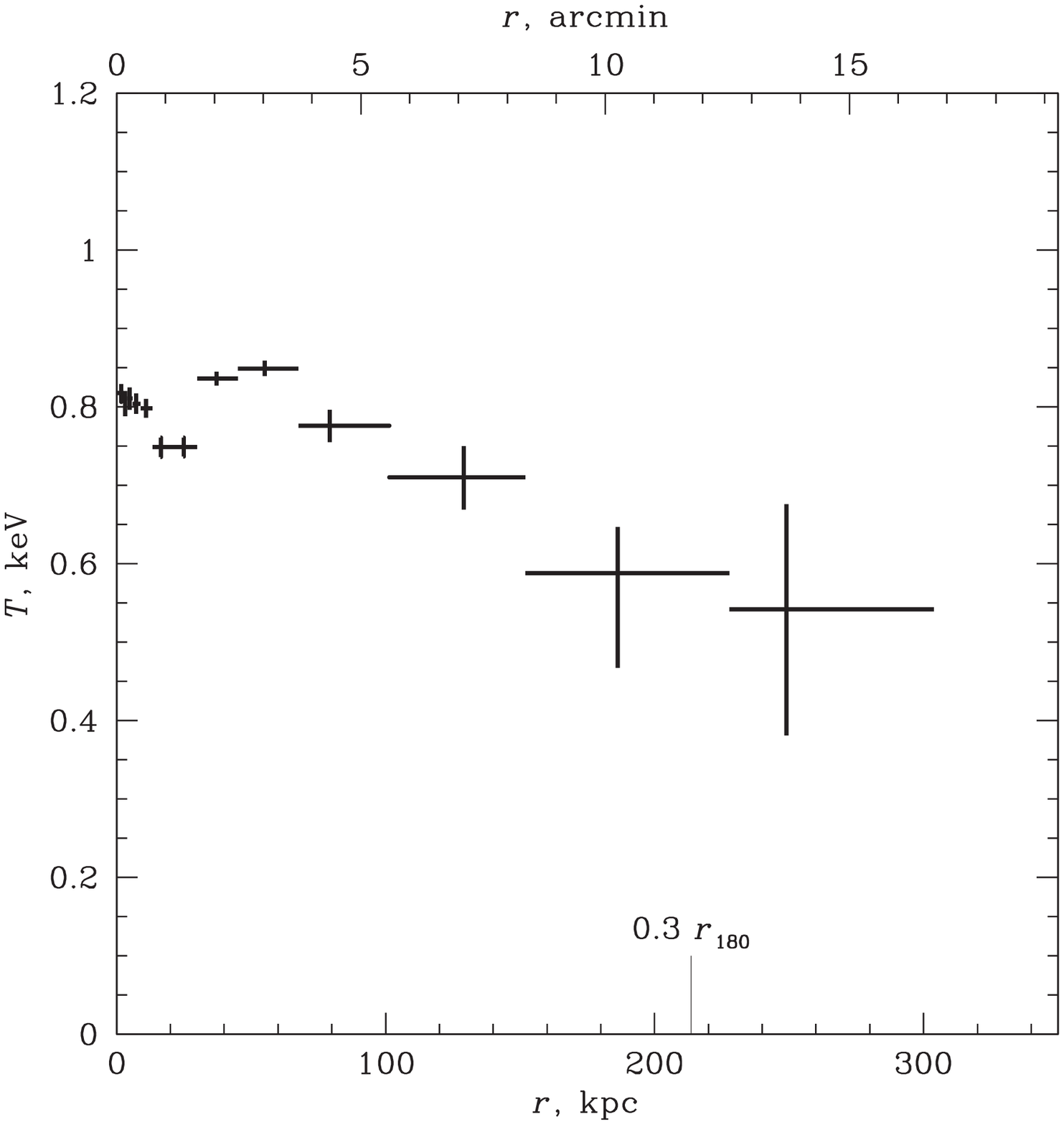}\hfill%
\includegraphics[width=0.485\linewidth]{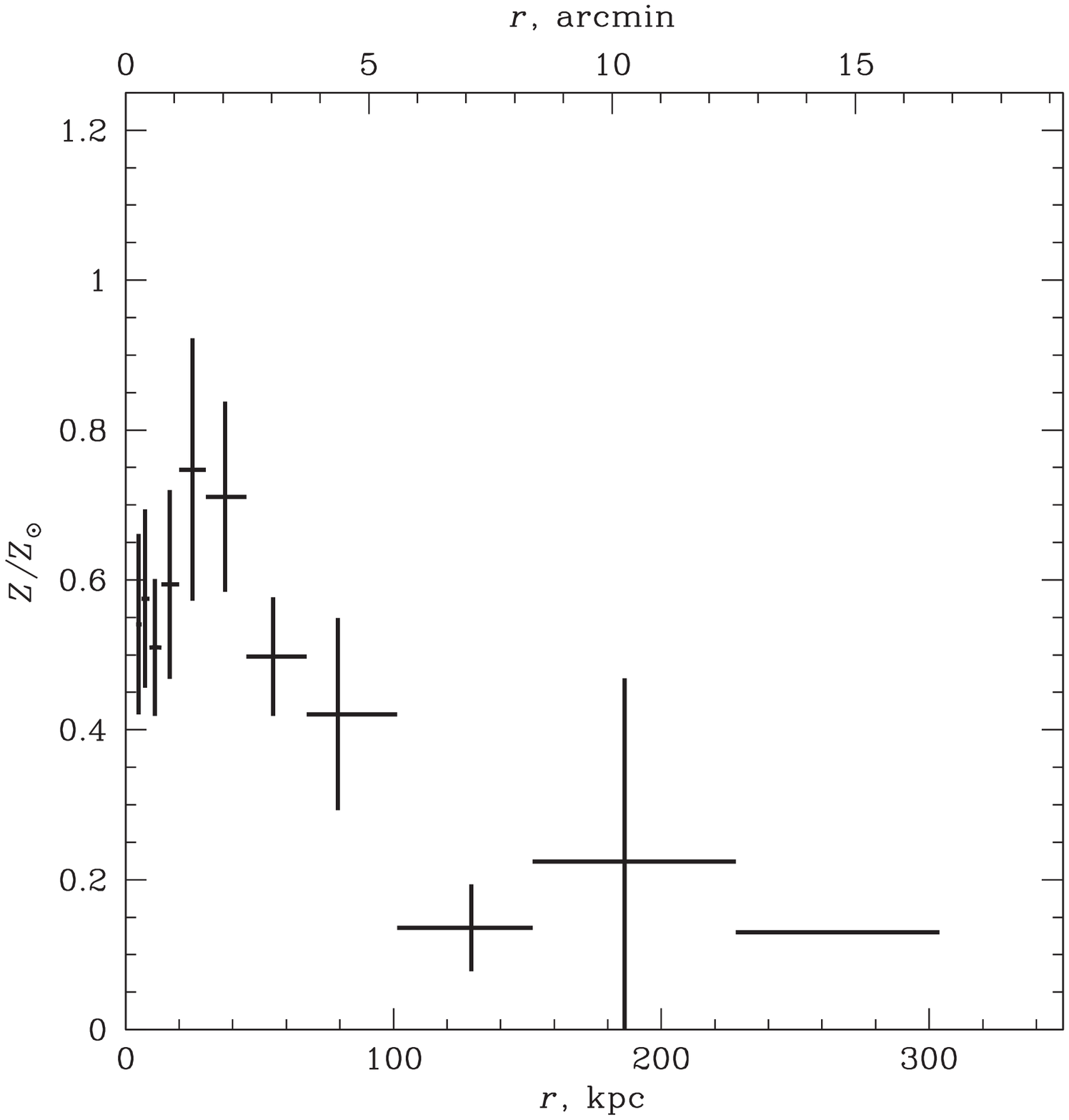}%
}
\vspace*{-0.2\baselineskip}
\caption{Temperature and metallicity profiles for USGC S152}
\label{fig:s152}
\end{figure*}

\begin{figure*}
\vspace*{-\baselineskip}
\centerline{%
\includegraphics[width=0.485\linewidth]{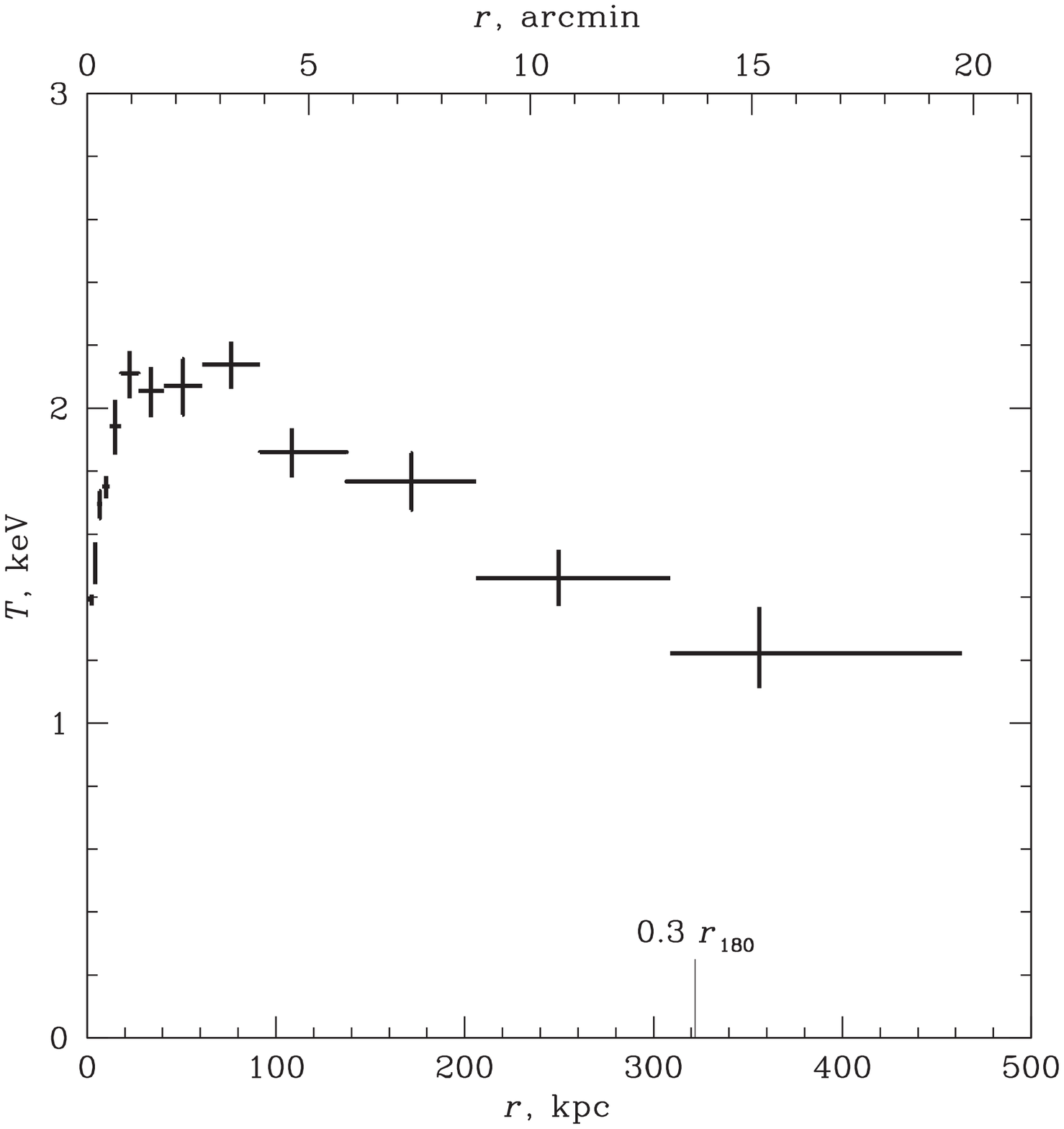}\hfill%
\includegraphics[width=0.485\linewidth]{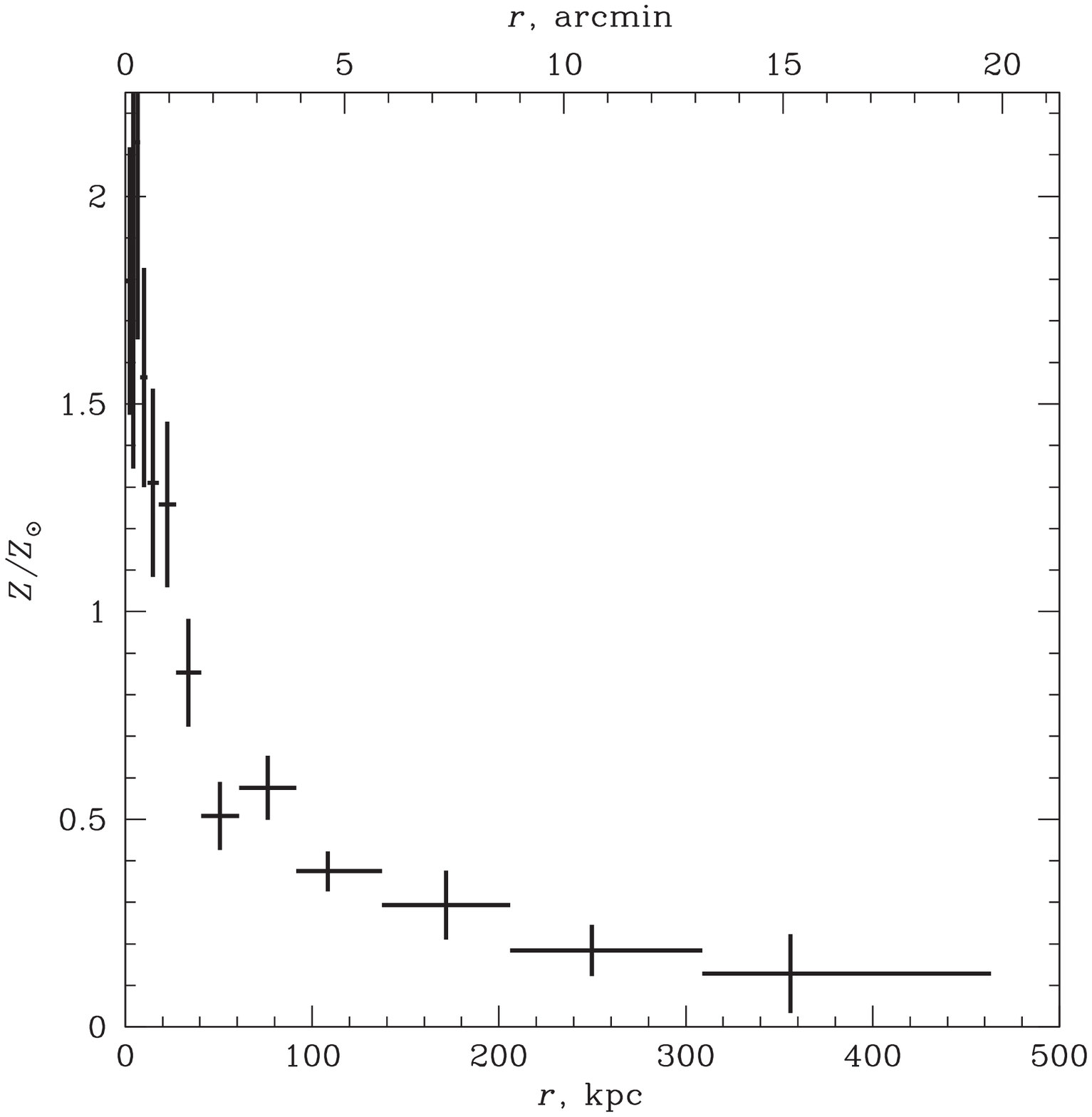}%
}
\vspace*{-0.2\baselineskip}
\caption{Temperature and metallicity profiles for MKW4}
\label{fig:mkw4}
\end{figure*}

\subsection{Abell 2390}

Abell 2390, the hottest and highest-redshift cluster in our sample
($z=0.2304$) was observed in 3 ACIS-S pointings for 10, 10, and
100~ksec. We discard the first two short observations. One of them was
performed in 1999 and accurate gain calibration for this period is
unavailable. The second short observation was telemetered in FAINT mode,
which results is poorer background rejection. The X-ray image shows some
substructure in the central $1.6'$ or $\sim 300$~kpc (cavities and
elongations) but is more regular on larger scales
(Fig.~\ref{fig:mosaic:img}). The best-fit absorption is constant with
radius, $N_H=(1.07\pm0.05)\times10^{21}$~cm$^{-2}$, but is higher than
the radio value $6.8\times10^{20}$~cm$^{-2}$; we use the X-ray value in
the further analysis. The spectra from the outermost regions show
positive soft-band residuals which can be fit with the MEKAL spectrum
with $T=0.24$~keV and normalization corresponding to the 0.7--2~keV flux
$(8.9\pm2.7)\times10^{-6}$~cnt~s$^{-1}$~arcmin$^{-2}$; this was used as
an additional soft background correction. The temperature and
metallicity profiles are shown in Fig.~\ref{fig:a2390}. There is a
marginally significant temperature drop from $T\simeq 9.5$~keV at $r\sim
500$~kpc to $4.7\pm2.3$~keV at $r>1200$~kpc.

\subsection{RXJ 1159+5531}

RXJ~1159+5531 ($T=1.9$~keV, $z=0.081$) is an ``X-ray Over-Luminous
Elliptical Galaxy'' selected from the 160~deg$^{2}$ \emph{ROSAT} PSPC
Survey \citep{1999ApJ...520L...1V}. Optically, this object appears as a
nearly isolated elliptical galaxy but its X-ray luminosity and extent is
typical of poor clusters.  It was observed in ACIS-I for 20~ksec in the
Fall of 1999 and in ACIS-S for 80~ksec in 2004. Accurate calibration is
still unavailable for the 1999 data and so we do not use this pointing in
the present analysis. The optical structure suggests that this object is
probably dynamically old and fully relaxed. The only detectable
substructure in the X-ray image is, indeed, confined to the very central
region of the cluster, $r<5$~kpc. The average X-ray absorption is
consistent with the radio column density,
$N_H=1.2\times10^{20}$~cm$^{-2}$. The soft background is clearly
over-subtracted. The required correction is well-fit by a MEKAL spectrum
with $T=0.17$~keV and negative normalization similar to most of other
similar cases. The temperature and metallicity profiles of RXJ~1159+5531
are shown in Fig.~\ref{fig:cl1159}.

\subsection{USGC S152}

The next two clusters are low-redshift, low-temperature systems. 
USGC~S152 ($z=0.0153$) was observed in a single ACIS-S pointing for
30~ksec (Fig.~\ref{fig:mosaic:img}). This is the lowest-temperature
($T\sim0.8$~keV) cluster in our sample. The outermost region of the
field of view ($r>16.5'$) is virtually free from the cluster emission
and the spectrum extracted at these radii shows no problems with the
background subtraction in the soft band. The best-fit absorption is
$N_H=(1.55\pm0.10)\times10^{21}$~cm$^{-2}$, which is significantly
higher than the radio value, $4.34\times10^{20}$~cm$^{-2}$; there are no
statistically significant variations of absorption with radius. The
temperature and metallicity profiles are shown in Fig.~\ref{fig:s152}. 
This is the only cluster in our sample without a strong central
temperature decrement, probably because the energy output from a recent
AGN outburst or from the stellar winds of the central cD galaxy is
non-negligible in this low-temperature system.

\subsection{MKW4}

MKW4 ($z=0.0199$) was observed in a single ACIS-S pointing for 30~ksec
(Fig.~\ref{fig:mosaic:img}). Faint cluster emission is detectable to the
very edge of the ACIS field of view. We find no indication of problems
with the soft background subtraction. However, the best-fit absorption
is significantly higher than the radio value
($1.89\times10^{20}$~cm$^{-2}$) and varies with radius from
$(6.4\pm0.8)\times10^{20}$~cm$^{-2}$ near the cluster center to
$(2.6\pm0.8)\times10^{20}$~cm$^{-2}$ in the outermost regions. We,
therefore, allowed $N_H$ to be a free parameter in the spectral fits.
The results for this cluster are shown in Fig.~\ref{fig:mkw4}. Note a
remarkably strong metallicity gradient --- $Z$ changes (for a
single-temperature model) from $\sim 2$ Solar in the center to $<0.3$
Solar outside the central 200~kpc.

\section{Discussion of systematic effects}

To the best of our knowledge, all calibration effects which could affect
the temperature profile measurements have been accounted for in our
analysis. It is useful, nevertheless, to discuss independent indicators
of systematic errors. The main possible sources of bias are the
following. (i) \emph{Statistical}. Our spectral fits use $\chi^2$
minimization. The cluster emission is fainter at large radii and the
photon counting statistics are increasingly non-Gaussian, which in
principle can lead to underestimation of the temperature. The
statistical bias can be avoided by adequately grouping the spectra, so
that there is a sufficient number of counts in each bin prior to
background subtraction. The binning we used --- into 26 channels of
approximately equal log width --- resulted in $>100$ counts per channel
in all cases. Therefore, $\chi^2$ statistics are fully adequate for
spectral fitting; this was verified in several cases by fitting the
spectra with a factor of 2 coarser binning.

\begin{figure*}
\vspace*{-2.5\baselineskip}
\centerline{\includegraphics[width=0.5\linewidth]{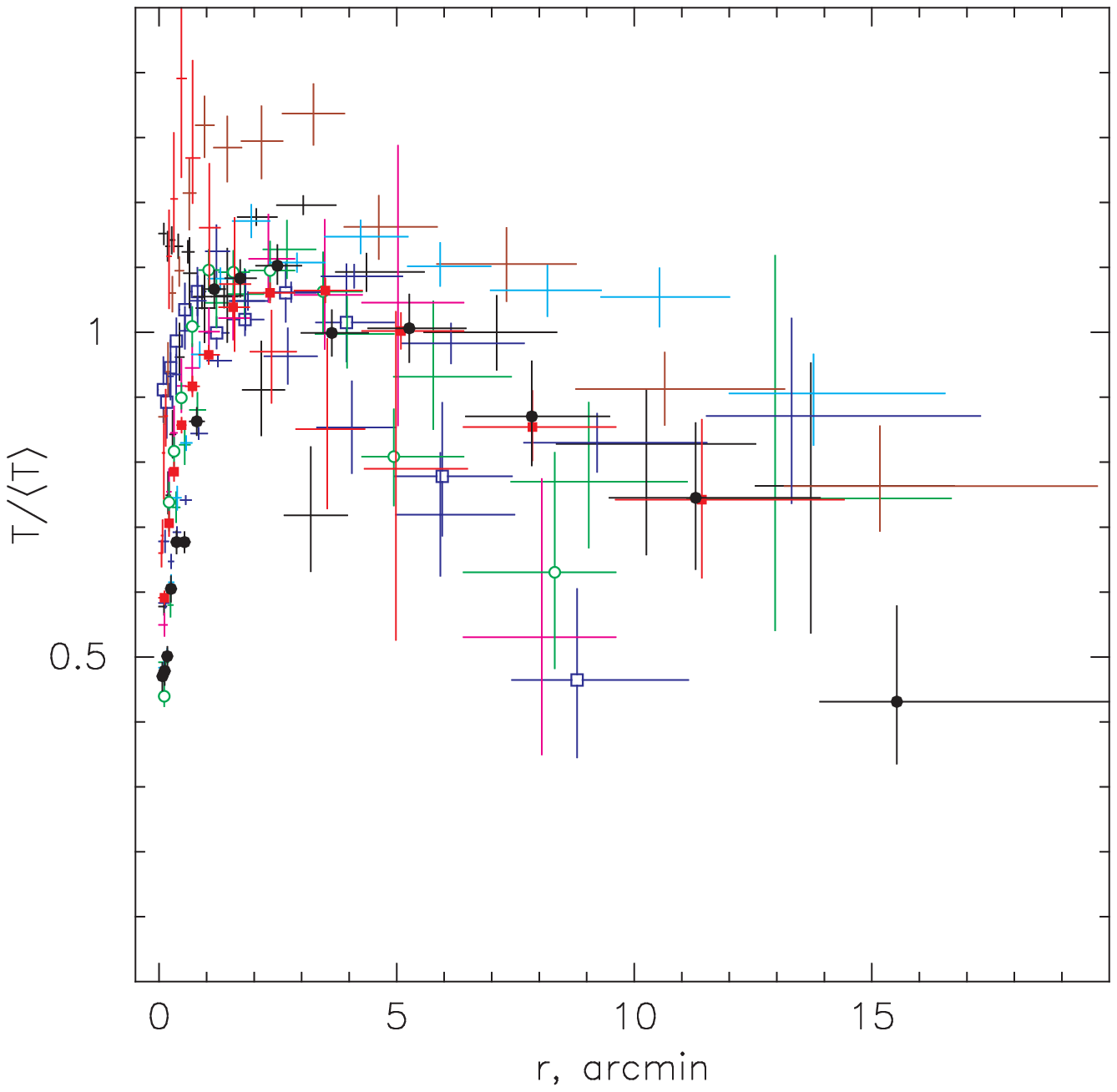}\hfill%
           \includegraphics[width=0.5\linewidth]{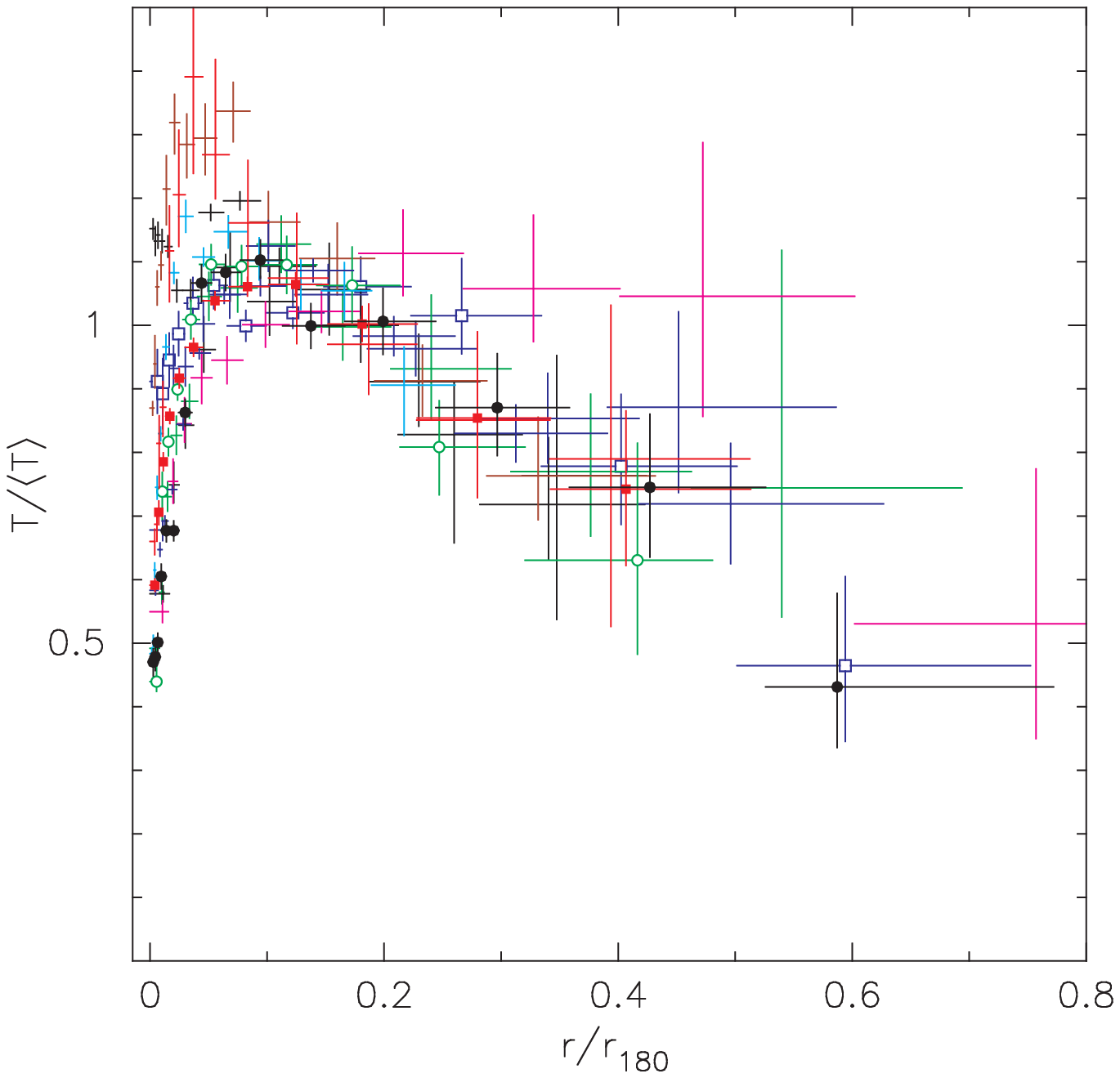}%
}
\caption{Temperature profiles for all clusters plotted as a function of
  angular distance from the center and in units of the cluster virial
  radius. The temperatures were scaled to the cluster emission-weighted
  temperature excluding the central 70~kpc regions affected by cooling.
  The virial radii were estimated from these average temperatures using
  a relation from Evrard et al.\ (1996), $r_{180} =
  2.74\,\text{Mpc}\;(\langle T\rangle/10\,\text{keV})^{1/2}$. Our best
  measurements, for A1991 ($T=2.6$~keV), A133 ($T=4.2$~keV), A1413
  ($T=7.3$~keV), and A2029 ($T=8.5$~keV), are open circles, filled
  circles, open squares, and filled squares, respectively. The strongest
  outliers in the right panel are MKW~4 (brown) and RXJ~1159+5531 (red)
  whose temperature profile peak at $r<70$~kpc, and A2390 (magenta)
  whose central cool region extends to $r\sim 400$~kpc.  }
\label{fig:comp:prof}
\end{figure*}

(ii) \emph{Effective area}. Miscalibration of the off-axis effective
area can result in artificial temperature gradients. However, effective
area calibration can be verified internally from our data. The most
sensitive test is provided by different pointings of A1795, with the
cluster placed at different detector locations and off-axis angles in
both ACIS-S and ACIS-I. If miscalibration of effective area were
responsible for an observed factor of $\sim 2$ temperature decrease at
large radii we would derive very different temperatures for the central
region of A1795 from on-axis and off-axis pointings. Instead, we find
excellent agreement between these pointings (see \S\ref{sec:a1795} and
full report in Vikhlinin et al.\ 2004a\footnote{ACIS calibration memo,
\url{http://cxc.harvard.edu/contrib/alexey/contmap.pdf}}). Further
evidence against effective area-related bias is provided by the off-axis
ACIS-I observation of A133 which resulted in one of our most accurate
temperature profiles (Fig.~\ref{fig:a133}). Unlike most other cases, the
cluster center in this pointing was placed off-axis and the temperature
decrease is observed on-axis. Finally, miscalibration of effective area
would tend to cause all temperature profiles to be similar when plotted
as a function of \emph{angular} distance from the center. Instead, we
observe a very large scatter in such a plot (Fig.~\ref{fig:comp:prof}a). 

(iii) \emph{Background}. Incorrect background subtraction can affect
temperature measurements at large radii where the surface brightness is
low.  If incorrect background subtraction is the cause for temperature
decrements at large radii, the background normalization must be
overestimated in all cases we analyzed. This is unlikely given that we
used a procedure that was demonstrated to work very well for a large
number of \emph{Chandra} observations (Markevitch et al.\ 2003). Typical
scatter of the background normalization is included in our temperature
uncertainties. The accuracy of the background subtraction can be
verified directly for several clusters with small angular size that we
observed in ACIS-I, such as A383 and A907, because a sufficient fraction
of the ACIS field of view is free from cluster emission in these
cases. There are no background residuals in these areas, and the
temperature profiles for these two clusters are typical of the entire
sample. 

To summarize, we cannot identify any obvious source of systematic error
which could be responsible for the observed temperature decrease at
large radii. An implicit indication that our measurements are correct is
given by the similarity of all temperature profiles when expressed in
units of the virial radius, as discussed below.

\section{Self-similarity of temperature profiles}
\label{sec:self:sim}

Theory predicts that clusters should be approximately self-similar because
they form from scale-free density perturbations, and their dynamics are
governed by the scale-free gravitational force. Self-similarity implies, in
particular, that cluster temperature (and density etc.) profiles should be
similar when radii are scaled to the cluster virial radius, which can be
estimated from the average temperature, $r_{180}\propto \langle
T\rangle^{1/2}$ (for detailed discussion see, e.g., Bryan \& Norman 1998). 
This prediction is strongly confirmed by our measurements. 

The scaled temperature profiles for all clusters are shown in
Fig.~\ref{fig:comp:prof}b. The temperatures were scaled to the
integrated emission-weighted temperature, excluding the central $70$~kpc
region usually affected by radiative cooling. The same average
temperature was used to estimate the cluster virial radius, $r_{180}$
using a relation from Evrard, Metzler \& Navarro (1996), $r_{180} =
1.95\,h^{-1}\,\text{Mpc} (\langle T\rangle/10\,\text{keV})^{1/2}$. The
scaled profiles are almost identical at $r\gtrsim0.15 r_{180}$ and the
general trend can be represented with the following functional form,
\begin{equation}
  T/\langle T\rangle = 
  \begin{cases}
    1.07, & \text{$0.035 < r/r_{180} < 0.125$}\\
    1.22 - 1.2\,r/r_{180},& \text{$0.125 < r/r_{180} < 0.6$},
  \end{cases}
\end{equation}
with a 15--20\% scatter. The only significant outlier is A2390 (shown in
magenta). However, this cluster is unusual in that its central cool
region in this cluster extends to $r\sim 400$~kpc, probably because the
cold gas is pushed out from the center by radio lobes. 

The strongest scatter of the temperature profiles is observed in the
central cooling regions. This is not unexpected, because in these
regions, non-gravitational processes such as radiative cooling and
energy output from the central AGNs are important, thus breaking
self-similarity. The largest outliers in the central region are MKW~4
and RXJ~1159+5531 whose cooling regions are very compact and the
temperature profiles peak near $r\simeq 50$~kpc (thus our fixed exclusion
radius of 70~kpc is too big).

\begin{figure}[b]
\vspace*{-2.5\baselineskip}
\centerline{\includegraphics[width=1.0\linewidth]{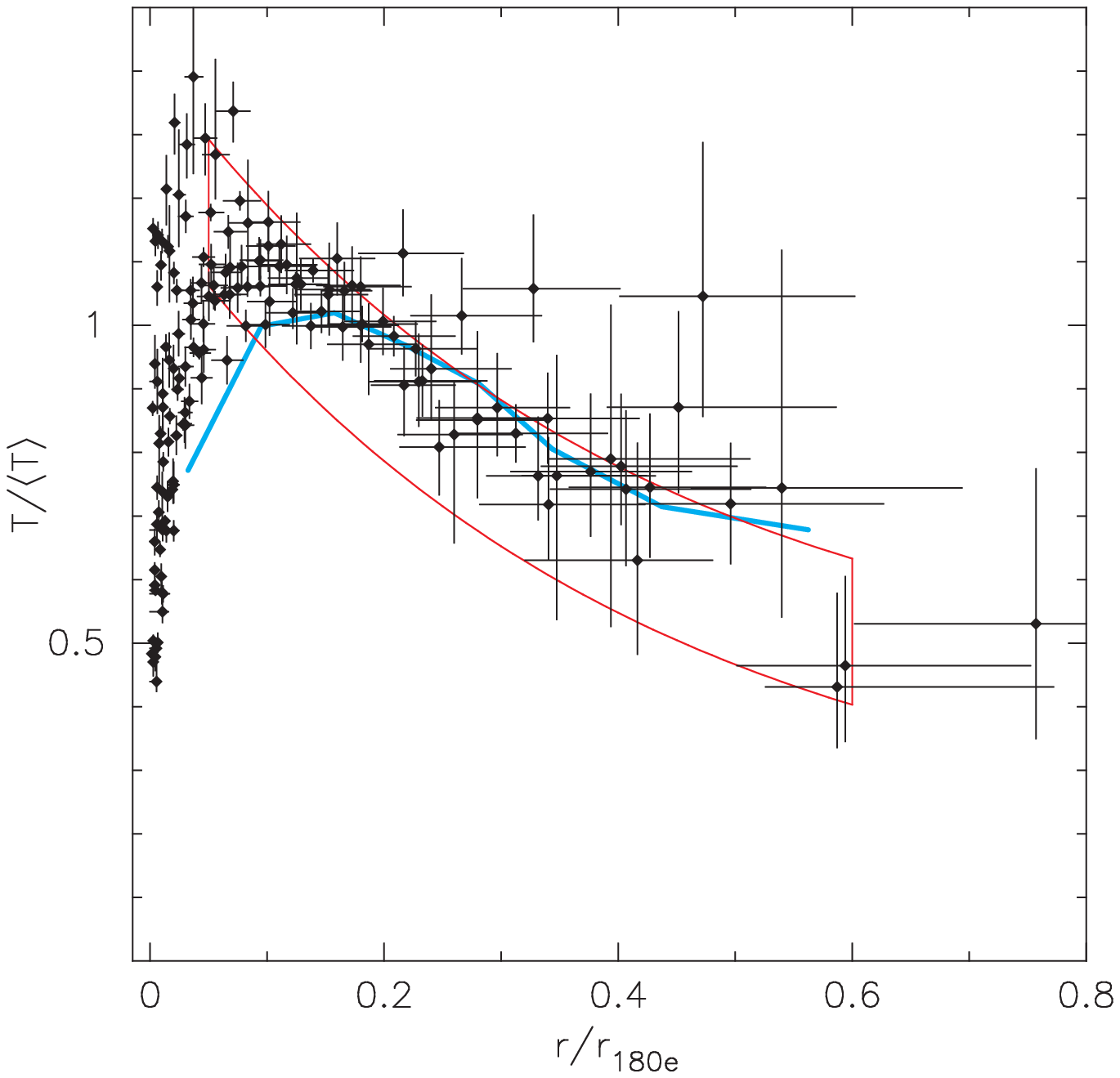}}
\caption{Comparison of the \emph{Chandra} temperature profiles and the
  average profiles from \emph{ASCA} \citep{1998ApJ...503...77M} and
  \emph{Beppo-SAX} \citep{2002ApJ...567..163D}. \emph{ASCA} results are
  shown as the red band with the width equal to the scatter of the
  best-fit values in a sample of 30 clusters. \emph{Beppo-SAX} results
  (blue line) represent the average temperature profile in a sample of
  11 ``cooling flow'' clusters.} 
\label{fig:tprof:asca:sax}
\end{figure}

Some of the previous studies of large cluster samples with \emph{ASCA} and
\emph{Beppo-SAX} have already uncovered the similarity of the cluster
temperature profiles at large radii
\citep{1998ApJ...503...77M,2002ApJ...567..163D}. Our measurements are
consistent with these earlier results both qualitatively and quantitatively. 
Red band in Fig.~\ref{fig:tprof:asca:sax} represents the typical scatter of
individual profiles in the M98 sample. Our radial trend at $r>0.1\,r_{180}$
is in good agreement with their profile, even though our sample has only two
clusters in common with M98 (A478 and A1795). A difference in the
qualitative trend around $0.1\,r_{180}$ is fully expected, considering that
the \asca\ measurement was at the limit of that instrument's angular
resolution, and given the difference in spectral modeling (we use
single-temperature fits whereas M98 used a then-common model with ambient +
cooling flow components for the cluster centers, and included only the
ambient hot phase in the final profiles). 

The average \emph{Beppo-SAX} profile for the subsample of clusters with
cool cores (such as ours) from \cite{2002ApJ...567..163D} is shown by
blue line in Fig.~\ref{fig:tprof:asca:sax}. The results are remarkably
close at $r>0.2\,r_{180}$. At smaller radii, the \emph{Beppo-SAX}
profile does not recover the continuation of a temperature increase
towards smaller radii, which is consistent with the poorer angular
resolution of that telescope.

A declining temperature profile has been recently reported from an XMM
analysis of 16 low-redshift clusters \citep{astro-ph/0412233}. On average,
temperature drops by $\sim 30\%$ from its peak value near
$r=0.4r_{180}$ (the outer radius for most clusters in Piffaretti et
al.\ sample), fully consistent with our results.

Declining temperature profiles are generally reproduced in recent
cosmological numerical simulations
\citep{2002ApJ...579..571L,2003MNRAS.346..731A,2003MNRAS.339.1117V,2004MNRAS.348.1078B,2004MNRAS.354..111E,2004MNRAS.tmp..504K,2004rcfg.procE..29M},
although earlier simulations were more discrepant and produced both
nearly isothermal profiles \citep{1996ApJ...469..494E} and strong
temperature gradients \citep{1993ApJ...412..455K}. The best agreement
with observed temperature profile seems to be reached by Eulerian codes
(Loken et al.\ 2002 and discussion therein) and by the
entropy-conserving versions of the SPH code \citep{2002MNRAS.333..649S}
used, e.g., by Ascasibar et al.\ and Borgani et al. A more detailed
comparison with observations requires folding the simulation output
through the detector response \citep{2004MNRAS.354...10M}. This is
beyond the scope of this paper. The general agreement of numerical and
observational results suggests that the declining temperature profiles
is a natural product of gravitational heating of ICM in the process of
cluster formation. This conclusion is unaffected by the presence of heat
conductivity of up to 1/3 of Spitzer value \citep{2004ApJ...606L..97D}. 
% 
% The presence of heat conductivity of up to 1/3 of
% Spitzer value does not change this conclusion
% \citep{2004ApJ...606L..97D}. 

\section{Summary and conclusions}

\emph{Chandra} is well-suited for measurements of the temperature
profiles to 0.5--0.6 of the virial radius thanks to its stable detector
background, and fine angular resolution. Our analysis of high-quality
observations of 11 low-redshift clusters reveals an almost universal
behavior of temperature profiles in the systems spanning the range of
average temperatures from 1 to 10~keV. Projected temperature as a
function of radius reaches a maximum at $r\sim 0.1-0.2 r_{180}$,
just outside the central cooling region. The temperature gradually
declines at larger radii, reaching 0.5 of its peak value near
$r\sim0.5-0.6r_{180}$, the outer boundary of our measurements in
most cases. 

In the future papers, we will use these \emph{Chandra} observations for
accurate measurements of the baryon and total mass profiles at large
radii. We will also carry out a consistent comparison of clusters from
high-resolution numerical simulations with their real-world
counterparts. 

\acknowledgements

We thank A.~Kravtsov and P.~Mazzotta for useful discussion. This work
was supported by NASA grant NAG5-9217 and contract NAS8-39073.

\end{document}